%% file: main.tex
\documentclass[11pt,a4paper]{article}
\usepackage[margin=2.5cm]{geometry}
\usepackage[T1]{fontenc}\usepackage{lmodern}
\usepackage{setspace}
\newif\ifssbuild\ssbuildfalse
\usepackage{amsmath,amssymb,amsthm,mathtools,booktabs,graphicx,enumitem,xcolor}
\usepackage[round,authoryear]{natbib}
\usepackage{url}
\usepackage[hypertexnames=false,colorlinks=true,linkcolor=blue!50!black,citecolor=blue!50!black]{hyperref}
\newtheorem{theorem}{Theorem}\newtheorem{proposition}{Proposition}
\newtheorem{corollary}{Corollary}\newtheorem{lemma}{Lemma}
\theoremstyle{definition}\newtheorem{example}{Example}\newtheorem{remark}{Remark}
\newcommand{\E}{\mathbb{E}}\newcommand{\R}{\mathbb{R}}


\title{Heavy tails in Gaussian-process surrogates:\\
process-level profiling degeneracy versus\\ observation-level identifiability}
\author{Neelesh S.\ Upadhye and Rohit Kumar Mani\\[2pt]
\normalsize Department of Mathematics, Indian Institute of Technology Madras,\\ \normalsize Chennai 600036, India}
\date{}

\makeatletter
\let\arxivmaketitle\@maketitle
\makeatother
\begin{document}
\maketitle

\begin{flushleft}\small
\textit{Running head:} Heavy tails in Gaussian-process surrogates\\
\textit{Address for correspondence:} Neelesh S.\ Upadhye, Department of Mathematics, Indian Institute of Technology Madras, Chennai 600036, India. E-mail: \texttt{neelesh@iitm.ac.in}
\end{flushleft}

\begin{abstract}
In this article, we study heavy-tailed alternatives to Gaussian-process surrogates, namely Student-$t$ processes, $\alpha$-stable processes and Gaussian processes with heavy-tailed noise. These are Gaussian scale mixtures, and the position of the mixing variable decides whether the tails can be learned from a single realization. For process-level mixtures with a free amplitude, the profile likelihood of the mixing law is data-free and is maximized by the Gaussian law, the plug-in predictive is a data-free transformation of the Gaussian-process predictive, and no estimator can be consistent for the mixing law over the model along one path. For observation-level mixtures with symmetric $\alpha$-stable noise, the tail index is identifiable. Further, for one-dimensional inputs under fixed-domain asymptotics with $\alpha_0<2$ and compact hyperparameter sets, a cross-fitted plug-in estimator and the Laplace marginal-likelihood estimate computed at the global posterior mode attain the efficiency bound of the model with a known regression function, also for Mat\'ern-$5/2$ sample paths. We also characterize small-sample failures of the Laplace approximation as folds of a posterior-mode branch and give a diagnostic. Finally, in a repeated Bayesian optimization study, observation-level tails reduced regret under gross outliers, as did Student-$t$ noise, and no degradation was detected under Gaussian noise.
\end{abstract}


\section{Introduction}
Gaussian-process (GP) surrogates are fundamental to Bayesian optimization (BO) \citep{Jones1998,Srinivas2012,Rasmussen2006}. However, many objectives of practical interest are heavy-tailed; for example, returns of trading rules, simulation outputs with rare failures and measurements with gross errors. Two kinds of remedy are in use. The first replaces the process, for example by the Student-$t$ process (TP) of \citet{Shah2014} or by an $\alpha$-stable process with sub-Gaussian (scale-mixture) representation \citep{SamorodnitskyTaqqu1994}. The second replaces the noise. Observation-level $t$ errors are classical in robust modeling \citep{Lange1989}, where the degrees of freedom are estimated by maximum likelihood. For GP regression the analog is the GP with Student-$t$ likelihood \citep{Vanhatalo2009,Jylanki2011}, used in BO by \citet{MartinezCantin2018}; GPs with other robust likelihoods are also used.
Heavy-tailed bandits and BO under heavy-tailed payoffs are also studied from the regret side \citep{Bubeck2013,Chowdhury2019}, as is GP bandit optimization under adversarial corruptions \citep{Bogunovic2020}, which is close to our outlier experiment. Robust conjugate GP regression (RCGP; \citealp{Altamirano2024}) robustifies the GP posterior by generalized Bayes with outlier-dependent weights, and \citet{Ezzerg2026} combine it with UCB to obtain regret bounds under corruptions of unbounded magnitude but bounded frequency. We use an RCGP-type surrogate as a robust baseline in Section~\ref{sec:bostudy}.

This article is motivated by a recurring empirical puzzle. When the TP's degrees of freedom $\nu$ are fitted by maximum likelihood (ML), they run to the upper bound, even on data with outliers or strongly heavy-tailed noise. In our experiments $\hat\nu$ reached the bound in all 90 synthetic data sets and all 60 NIFTY~50 data sets (Section~\ref{sec:bo}). \citet{Shah2014} (\S4.2) state that the TP predictive mean and covariance differ from the GP's after the kernel hyperparameters are learned. \citet{Tang2016} showed that, for kernels with a suitable scaling property, the predictive mean is in fact the GP's. In this article, we show that, with a free amplitude and for any mixing law, the predictive \emph{scale} is also the GP's. For the TP the covariance differs only by the factor $(\nu+n)/(\nu+n-2)$, the variance of a standard $t_{\nu+n}$ law.

The explanation is structural. Every model above is a Gaussian scale mixture, and what matters is whether one mixing variable is shared by the whole sample (\emph{process level}, M1) or each observation has its own (\emph{observation level}, M2). For M1 the phenomenon has a long history in finite dimensions. \citet{Zellner1976} showed that, with multivariate-$t$ errors, ML and posterior inference for the regression coefficients coincide with the normal case, and \citet{Breusch1997} argued that the two models are inferentially indistinguishable; \citet{Ng2002} gave the Bayesian analog for elliptical errors, and \citet{Chib1991} gave general conditions on the prior under which the prior of $\nu$ in multivariate-$t$ regression is not updated by the data, together with an example in which learning about $\nu$ is not precluded. For a possibly nonlinear regression under any elliptical sampling density, \citet{OsiewalskiSteel1993} showed that posterior and predictive inference is unaffected by the choice of density within the elliptical class when an improper prior is placed on a precision factor. \citet{FernandezSteel1999} studied the pathologies of estimating $\nu$, and \citet{WangIp2003} showed that a moment estimator of $\nu$ for multivariate-$t$ regression errors \citep{Singh1988} has a distribution that does not depend on the true $\nu$ and diverges in probability as $n$ grows; Corollary~\ref{cor:tp} is the ML counterpart for the TP. Further, \citet{Kano1994} showed that elliptical processes consistent under marginalization must be normal scale mixtures. For TPs, \citet{Tang2016} showed that for kernels satisfying a scaling condition (their ``$\beta$-property'') the ML hyperparameters do not depend on $\nu$ and the predictive mean equals the GP's; see also \citet{Tang2017}. \citet{Bankestad2023} noted that the mixing law of an elliptical process cannot be inferred from a single path, and moved to per-observation mixing. Finally, \citet{Andrade2023} used regular variation to show that, under conditions that include a light-tailed prior on the scale, the posterior of a Student-$t$ process model rejects outlying observations as they grow. Note that this is a Bayesian statement under a prior. Under ML with a free amplitude the TP's plug-in predictive mean is the GP's (Theorem~\ref{thm:pred}), so robustness of that kind must come from the prior, not from the likelihood (Remark~\ref{rem:scope}).

Hence neither the distinction between a mixing variable shared by the sample and one per observation, nor invariance over the mixing law in Bayesian regression with an improper prior on the scale \citep{OsiewalskiSteel1993}, is new, and we do not claim them. What is new are the corresponding likelihood results for GP surrogates with a free amplitude, namely that the Gaussian law maximizes the profile likelihood among all mixing laws and the exact form of the plug-in predictive and of its acquisitions, the impossibility of consistent estimation along one path, and precise efficiency and computational results for stable noise. The main contributions of this article are as follows. Our main result, items (a)--(d) below, is that a tail parameter shared by the whole sample cannot be learned from one realization of a GP surrogate with a free amplitude. Next, the constructive alternative, item (e), puts the tails in the observation noise, where they can be learned efficiently. Further, item (f) is a specific failure of the Laplace approximation for such noise, with a diagnostic. Finally, item (g) is a repeated BO study.
\begin{enumerate}[label=(\alph*),leftmargin=2.2em]
\item A factorization of the likelihood of any process-level scale mixture with free amplitude (Theorem~\ref{thm:factor}): the profile likelihood of the mixing law is data-free, and the kernel hyperparameters equal the GP's. No estimator can be consistent for the mixing law over the model along one path (Proposition~\ref{prop:noconsist}).
\item Over \emph{all} mixing laws, the Gaussian law strictly maximizes the profile likelihood (Theorem~\ref{thm:gauss}); for the TP, $\hat\nu$ runs to the upper bound (Corollary~\ref{cor:tp}).
\item The plug-in predictive is the GP mean plus the GP standard deviation times a data-free shape (Theorem~\ref{thm:pred}), with a unique fitted amplitude whenever $\log V$ is unimodal (Proposition~\ref{prop:shat}), which includes all positive stable laws (Lemma~\ref{lem:kanter}).
\item Probability of improvement ranks candidates exactly as under the GP; UCB and EI differ only through fixed functions (Corollary~\ref{cor:acq}).
\item For symmetric $\alpha$-stable noise: identifiability (Proposition~\ref{prop:m2id}); a $\sqrt n$-consistent difference-based estimator (Theorem~\ref{thm:ecf}); efficiency of a cross-fitted plug-in estimator (Theorem~\ref{thm:eff}, with a moment-free running-median pilot or the GP posterior mode, Theorem~\ref{thm:gppilot}) and of the global-mode Laplace marginal-likelihood estimate (Theorem~\ref{thm:lapeff}), also for Mat\'ern-$5/2$ sample paths (Theorem~\ref{thm:rough}); Wald intervals (Corollary~\ref{cor:wald}); and an $O(n)$ fitting algorithm for one input.
\item For the Laplace approximation: a comparison with importance sampling and a variational bound (Section~\ref{sec:exact}); its gross failures as folds of a posterior-mode branch, a sufficient condition for no fold, the eventual uniqueness and nondegeneracy of the global mode under infill (Proposition~\ref{prop:infill}), and an eigenvalue diagnostic (Section~\ref{sec:fold}); and a bootstrap calibration of the test of $\alpha=2$.
\item A repeated sequential BO study with five surrogates and nine problems (Section~\ref{sec:bostudy}).
\end{enumerate}
The message for practice is simple: from a single realization, heavy tails in a GP surrogate can be learned only where the data can see them, in the noise.

The organization of this article is as follows. In Section~\ref{sec:setup}, we introduce the process-level and observation-level scale-mixture models. In Section~\ref{sec:m1}, we study process-level mixtures and prove that the profile likelihood of the mixing law is data-free. In Section~\ref{sec:m2}, we show that, for observation-level symmetric $\alpha$-stable noise, the tail index is identifiable and, for one-dimensional inputs, can be estimated efficiently, and we examine the accuracy of the Laplace approximation and its failures. In Section~\ref{sec:bo}, we discuss the consequences for BO and report a repeated sequential BO study. Finally, Section~\ref{sec:disc} contains concluding remarks and open problems. \ifssbuild Proofs not given in the text, further remarks and simulation results, the comparison of the Laplace approximation with importance sampling, the evidence on folds and the full design of the BO study are in the Supplementary Material.\else The proofs of the results of Section~\ref{sec:eff} and further results of the BO study are given in the Supplementary Material.\fi

\section{Two ways to put heavy tails into a GP surrogate}\label{sec:setup}
In this section, we introduce the two models studied in this article. Let $x_1,\dots,x_n$ be inputs, $K_0=K_0(\kappa)$ a positive-definite matrix built from a correlation kernel $R_\ell$ plus a nugget ($K_0=R_\ell+\rho I$), and $\kappa=(\ell,\rho)$.

\textbf{M1 (process level).} $y\mid V \sim N_n(0,\,cV K_0)$ with $V\sim F_\theta$ on $(0,\infty)$ and amplitude $c>0$. The whole vector shares one mixing variable.

\textbf{M2 (observation level).} $f\sim \mathrm{GP}(0,\sigma_f^2R_\ell)$ and $y_i = f(x_i) + \sigma\sqrt{S_i}\,Z_i$, with $S_i\stackrel{iid}{\sim}F_\theta$ and $Z_i\stackrel{iid}{\sim}N(0,1)$. For stable noise we take $S_i=2P_i$ with $\E \exp\{-\lambda P_i\}=\exp\{-\lambda^{\alpha/2}\}$, so that $\sqrt{S_i}Z_i$ has characteristic function $\exp\{-|t|^\alpha\}$ (at $\alpha=2$, variance 2).

\begin{example}[Mixing laws]\label{ex:mix}
(a) $V\sim$ inverse-gamma: M1 is the Student-$t$ process; M2 is the GP with Student-$t$ likelihood.
(b) $V$ positive $(\alpha/2)$-stable: M1 is the sub-Gaussian $\alpha$-stable process; M2 is the GP with symmetric $\alpha$-stable noise.
(c) $V$ generalized inverse Gaussian: generalized hyperbolic.
(d) $V^{-1/2}\sim U(0,1)^{1/q}$, $q>0$: slash (the classical slash law is $q=1$).
\end{example}

\paragraph{Three notions.} We separate three properties of a tail parameter $\theta$, because they behave differently in M1 and M2.
\begin{enumerate}[label=(\Roman*),leftmargin=2.2em]
\item \emph{Distributional identifiability:} for fixed $n$ and design, distinct parameter values give distinct laws of $y$.
\item \emph{Single-path estimability:} some estimator computed from one realization is consistent as $n\to\infty$ with the inputs filling a fixed domain.
\item \emph{Profile-likelihood learnability:} after nuisance parameters (here the amplitude and kernel hyperparameters) are profiled out, the likelihood of $\theta$ depends on the data, so the profile ML estimate can respond to them.
\end{enumerate}
In M1, (I) holds under a mild condition (Remark~\ref{rem:m1id}), while (II) fails (Proposition~\ref{prop:noconsist}) and (III) fails (Corollary~\ref{cor:profile}). In M2 all three hold (Proposition~\ref{prop:m2id}, Theorem~\ref{thm:ecf}, Section~\ref{sec:cons}). The title refers to (III) for M1 and to (I)--(II) for M2.

\section{Process-level mixtures: the profile likelihood of the mixing law is data-free}\label{sec:m1}
In this section, we consider process-level mixtures (M1) and show that, with a free amplitude, the profile likelihood of the mixing law does not depend on the data.\ifssbuild{} Proofs not given here are in Supplementary Section~\ref{supp:m1}.\fi{} Let $\beta_0 = y^\top K_0^{-1}y$ and $g_n(u) = u^{-n/2}\exp\{-1/(2u)\}$, $u>0$.

\begin{theorem}[Factorization]\label{thm:factor}
Under M1, $L(\theta,c,\kappa) = C(y,\kappa)\,M_n(\theta,c/\beta_0)$, where
\[
C(y,\kappa) = (2\pi)^{-n/2}|K_0|^{-1/2}\beta_0^{-n/2},\qquad M_n(\theta,s)=\E_\theta\big[g_n(sV)\big].
\]
\end{theorem}
\begin{proof}
For $u>0$, \[\varphi_n(y;0,uK_0) = (2\pi)^{-n/2}|K_0|^{-1/2}u^{-n/2}\exp\{-\beta_0/(2u)\} = C(y,\kappa)\,g_n(u/\beta_0).\] Put $u=cV$ and take expectations. This proves the result.
\end{proof}

\begin{corollary}\label{cor:profile}
(i) The profile likelihood $\sup_c L = C(y,\kappa)\,H_n(\theta)$ with $H_n(\theta)=\sup_{s>0}M_n(\theta,s)$, which depends on $(n,\theta)$ only, not on $y$ or $\kappa$ (for $y\ne0$).
(ii) The ML estimate of $\kappa$ maximizes $C(y,\kappa)$ and so equals the GP ML estimate for every $\theta$.
(iii) $\hat\theta$ maximizes the data-free function $H_n$.
\end{corollary}

Corollary~\ref{cor:profile} concerns the profile likelihood. The next result shows that no other estimator does better along one path: the free amplitude makes the single draw of $V$ uninformative about $\theta$, whatever the estimator.

\begin{proposition}[No single-path consistency]\label{prop:noconsist}
Let $K_n$ ($n\ge1$) be any known positive-definite $n\times n$ matrices, for nested or non-nested designs, and let $y_n\mid V\sim N_n(0,cVK_n)$ with $V\sim F_\theta$, where $\theta$ ranges over a metric space $(\Theta,d)$ and each $F_\theta$ is a law on $(0,\infty)$, degenerate laws included. Let $\theta_1\ne\theta_2$, and let $T_n=T_n(y_n)$ be any estimators.
\begin{enumerate}[label=(\alph*),leftmargin=2em]
\item There is no sequence $T_n$ such that, for each $\theta\in\{\theta_1,\theta_2\}$, $d(T_n,\theta)\to0$ in probability under $(\theta,c)$ for Lebesgue-almost every $c>0$.
\item If $F_{\theta_1}$ and $F_{\theta_2}$ are not mutually singular, then for every $c>0$ no $T_n$ is consistent at both $(\theta_1,c)$ and $(\theta_2,c)$, so $\theta$ cannot be estimated consistently even when $c$ is known. This covers any two non-degenerate members of each family of Example~\ref{ex:mix}: inverse-gamma, positive stable laws with $\alpha<2$ and generalized inverse Gaussian laws have positive densities on $(0,\infty)$, and slash laws on $(1,\infty)$. It does not cover the Gaussian member ($V\equiv1$).
\end{enumerate}
The result holds a fortiori when the kernel hyperparameters are unknown.
\end{proposition}
\begin{proof}[Proof of Proposition~\ref{prop:noconsist}]
Let $\|P-Q\|_{\rm TV}=\sup_A|P(A)-Q(A)|$, which is $<1$ unless $P\perp Q$. For a law $\mu$ on $(0,\infty)$ let $P^n_\mu=\int N_n(0,sK_n)\,\mu(ds)$; the law $P^n_{\theta,c}$ of $y_n$ under $(\theta,c)$ is $P^n_{\mu}$ with $\mu$ the law of $cV$. A Markov kernel cannot increase total variation, so $\|P^n_{\mu_1}-P^n_{\mu_2}\|_{\rm TV}\le\|\mu_1-\mu_2\|_{\rm TV}$ for every $n$. If $T_n$ were consistent for $\theta_1$ under $P^n_{\mu_1}$ and for $\theta_2$ under $P^n_{\mu_2}$, then with $A_n=\{d(T_n,\theta_1)<\varepsilon\}$ and $2\varepsilon<d(\theta_1,\theta_2)$, $\|P^n_{\mu_1}-P^n_{\mu_2}\|_{\rm TV}\ge P^n_{\mu_1}(A_n)-P^n_{\mu_2}(A_n)\to1$; this is impossible when $\mu_1$ and $\mu_2$ are not mutually singular. (b) follows with $\mu_j$ the law of $cV_j$, $V_j\sim F_{\theta_j}$. For (a), let $\Pi$ be the law of $\tilde c=\exp(Z)$, $Z\sim N(0,1)$, independent of $V$. Consistency at almost every $c$ gives, by bounded convergence (the map $c\mapsto P^n_{\theta,c}(A)$ is continuous for Borel $A$, by Scheff\'e's lemma), $\int P^n_{\theta,c}\{d(T_n,\theta)\ge\varepsilon\}\,\Pi(dc)\to0$, that is, consistency under $P^n_{\mu_\theta}$ with $\mu_\theta$ the law of $\tilde cV$. Since $\log(\tilde cV)=Z+\log V$ has a positive density on $\R$ for any law of $V$, $\mu_{\theta_1}$ and $\mu_{\theta_2}$ are both equivalent to Lebesgue measure on $(0,\infty)$, hence not mutually singular. This proves the result.
\end{proof}
For each fixed $n$ the law of $y_n$ determines $\theta$ under a mild condition (Remark~\ref{rem:m1id}). But by the proof, for a common known $c$ the total-variation distance between the laws of $y_n$ under $\theta_1$ and $\theta_2$ is at most $\|F_{\theta_1}-F_{\theta_2}\|_{\rm TV}$ for every $n$, and with $c$ averaged over $\Pi$ it is at most $\|\mu_{\theta_1}-\mu_{\theta_2}\|_{\rm TV}<1$: information about $\theta$ does not accumulate along the path. Part (a) includes $F_{\theta_1}$ degenerate, so with a free amplitude not even a Gaussian process can be told apart consistently from a Student-$t$ process. With $c$ known the Gaussian member can be separated from a continuous mixing law, since $\beta_0/n=cV\chi^2_n/n\to cV$ in probability whatever $K_n$; this is why (b) needs mixing laws that are not mutually singular.

\begin{theorem}[The Gaussian law maximizes the profile likelihood]\label{thm:gauss}
For every law $F$ on $(0,\infty)$, $H_n(F)\le g_n(1/n)=n^{n/2}\exp\{-n/2\}=H_n(\delta_1)$, with equality if and only if $F$ is degenerate.
\end{theorem}
\begin{proof}
$g_n$ is bounded, continuous, has a unique maximizer at $u=1/n$, and tends to $0$ as $u\to0$ and $u\to\infty$. By dominated convergence $s\mapsto \E[g_n(sV)]$ is continuous and vanishes at $0$ and $\infty$, so its supremum is attained at some $s^*$. Then $\E[g_n(s^*V)]\le g_n(1/n)$, with equality iff $s^*V=1/n$ a.s. This proves the result.
\end{proof}

\begin{remark}[Parametric families]\label{rem:families}
Theorem~\ref{thm:gauss} compares mixing laws; it does not by itself say where the ML estimate lies within a given parametric family $\{F_\theta\}$. What it does give is this: no non-degenerate $F_\theta$ attains $\sup_\theta H_n(\theta)$ when the family contains a degenerate law, or has one as a weak limit along which $H_n(\theta)\to H_n(\delta_1)$, since then the supremum is $H_n(\delta_1)$. So the profile ML estimate is either the Gaussian member, if there is one, or does not exist in the interior. For the named families the Gaussian is represented as follows. Student-$t$: $\nu\to\infty$, and Corollary~\ref{cor:tp} proves that $H_n$ is strictly increasing, so $\hat\nu$ runs to the upper bound. Sub-Gaussian $\alpha$-stable: $\alpha=2$ gives $V\equiv1$, a member of the family, so $\hat\alpha=2$. Slash: $q\to\infty$. Generalized inverse Gaussian (generalized hyperbolic processes): the law concentrates at a point as $\chi,\psi\to\infty$ with $\chi/\psi$ fixed. For a family whose closure does not contain a degenerate law, such as the slash law with fixed $q$, Theorem~\ref{thm:gauss} says that the GP has a strictly higher profile likelihood than every member.
\end{remark}

\begin{corollary}[Student-$t$ process]\label{cor:tp}
For $\nu>2$ and $V$ inverse-gamma with shape $\nu/2$ and scale $(\nu-2)/2$ (so $\E V=1$), $\log H_n(\nu) - \log H_n(\infty) = h_n(\nu)-h_n(\infty)$ with
$h_n(\nu)=\log\Gamma(\tfrac{\nu+n}{2})-\log\Gamma(\tfrac\nu2)-\tfrac n2\log\nu-\tfrac{\nu+n}{2}\log(1+\tfrac n\nu)$ and $h_n(\infty)=-\tfrac n2(1+\log2)$.
Moreover $h_n$ is strictly increasing in $\nu$, so $\hat\nu$ always runs to the upper bound of any search interval.
\end{corollary}
\begin{proof}[Proof of Corollary~\ref{cor:tp} (monotonicity)]
Differentiating, $h_n'(\nu)=\tfrac12\{r(\tfrac{\nu+n}2)-r(\tfrac\nu2)\}$ with $r(x)=\psi_0(x)-\log x$, where $\psi_0$ is the digamma function. The function $r$ is strictly increasing because $\psi_0'(x)=\sum_{k\ge0}(x+k)^{-2}>\int_0^\infty(x+t)^{-2}\,dt=1/x$.
\end{proof}

\begin{remark}[Sub-Gaussian $\alpha$-stable]
Numerically, $\log H_n(\alpha)-\log H_n(2)$ is increasing in $\alpha$. For example, at $n=25$ it is $-1.62, -0.95, -0.19, -0.02$ at $\alpha=1.1,1.5,1.9,1.99$. It is identical across data sets, as Theorem~\ref{thm:factor} requires. Monotonicity in $\alpha$ is not needed for the conclusion, since Theorem~\ref{thm:gauss} already gives $\hat\alpha=2$.
\end{remark}

For a fixed mixing law $F_\theta$, write $\mathcal S_n(\theta)=\arg\max_{s>0}M_n(\theta,s)$. It is non-empty by the proof of Theorem~\ref{thm:gauss}, and it depends on $(n,\theta)$ only.

\begin{proposition}[Uniqueness of $\hat s$]\label{prop:shat}
Put $T=\log V$ and $k_n(t)=\log g_n\{\exp(t)\}=-\tfrac n2 t-\tfrac12\exp\{-t\}$. If $T$ has a unimodal law, then $\mathcal S_n(\theta)$ is a single point $\hat s_n(\theta)$. This holds for inverse-gamma, generalized inverse Gaussian and slash mixing, where $T$ has a log-concave density, and for positive stable mixing (Corollary~\ref{cor:stableshat}).
\end{proposition}
\begin{proof}[Proof of Proposition~\ref{prop:shat}]
With $s=\exp(\tau)$, $M_n\{\theta,\exp(\tau)\}=\E\,\exp\{k_n(\tau+T)\}=(\exp\{k_n\}\ast F_{-T})(\tau)$. The function $\exp\{k_n\}$ is integrable, since $\int \exp\{k_n\}=2^{n/2}\Gamma(n/2)$, and it is log-concave, since $k_n''(t)=-\tfrac12\exp\{-t\}<0$. By Ibragimov's theorem \citep{Ibragimov1956}, the convolution of a log-concave density with a unimodal law is unimodal. Since this convolution has the continuous density $\tau\mapsto M_n\{\theta,\exp(\tau)\}/\int \exp\{k_n\}$, that function is non-decreasing and then non-increasing, and its set of maximizers is a closed interval.

It is also real-analytic. For $\tau=a+ib$ with $|b|<\pi/2$, we have $|\exp\{k_n(\tau+u)\}|=\exp\{-\tfrac n2(a+u)-\tfrac12\exp\{-(a+u)\}\cos b\}$. This is at most $\exp\{\tfrac n2\log(n/\cos b)-\tfrac n2\}$ uniformly in $u$, so the integral is holomorphic in the strip. If the interval of maximizers were non-trivial, the function would be constant on it, hence constant everywhere by the identity theorem. That contradicts $M_n>0$ and $M_n\to0$ at $\pm\infty$. So the maximizer is unique.

For the examples: for inverse-gamma $V$, $T$ has density $\propto\exp\{-at-b\exp(-t)\}$; for generalized inverse Gaussian $V$, $\propto\exp(\lambda t-\tfrac12(\chi \exp\{-t\}+\omega \exp\{t\}))$; for slash mixing, $T=-2\log U$ is exponential with rate $1/2$. All three are log-concave.
\end{proof}

\begin{lemma}[Kanter's function is log-convex]\label{lem:kanter}
For $\gamma\in(0,1)$ let \[A_\gamma(u)=\frac{\sin(\gamma u)^{\gamma/(1-\gamma)}\,\sin((1-\gamma)u)}{\sin(u)^{1/(1-\gamma)}},\qquad u\in(0,\pi).\] Then $\log A_\gamma$ is strictly convex and strictly increasing, and $(\log A_\gamma)'(0+)=0$.
\end{lemma}
\begin{proof}
Write $a=\gamma$ and $b=1-\gamma$. Since $(\log\sin cu)''=-c^2/\sin^2(cu)$,
\[
b\,(\log A_\gamma)''(u)=\frac1{\sin^2u}-\frac{a^3}{\sin^2(au)}-\frac{b^3}{\sin^2(bu)} .
\]
Let $q(x)=\sin^{-2}x-x^{-2}=\sum_{k\neq0}(x-k\pi)^{-2}$. Pairing the terms $\pm k$ gives $q'(x)=2\sum_{k\ge1}\{(k\pi-x)^{-3}-(k\pi+x)^{-3}\}>0$ on $(0,\pi)$, and $q(0+)=\tfrac13$. So $q$ is positive and increasing on $(0,\pi)$.

Because $a^3/(au)^2+b^3/(bu)^2=(a+b)/u^2=u^{-2}$, the pole terms cancel. With $au,bu\in(0,u)$ this leaves
\[
b\,(\log A_\gamma)''(u)=q(u)-a^3q(au)-b^3q(bu)\ \ge\ (1-a^3-b^3)\,q(u)>0 .
\]
Finally, $(\log A_\gamma)'(u)=\tfrac{a^2}{b}\cot(au)+b\cot(bu)-\tfrac1b\cot u$. Its $1/u$ terms have total coefficient $\tfrac ab+1-\tfrac1b=0$, so $(\log A_\gamma)'(0+)=0$, and strict convexity makes it positive on $(0,\pi)$.
\end{proof}

\begin{corollary}[Stable mixing]\label{cor:stableshat}
If $V$ is positive $\gamma$-stable ($\E\,\exp\{-\lambda V\}=\exp\{-\lambda^\gamma\}$, $\gamma=\alpha/2\in(0,1)$), then $\log V$ is unimodal \citep[Theorem~4.1]{Kanter1975}. So $\hat s$ is unique for sub-Gaussian $\alpha$-stable mixing for every $\alpha\in(0,2)$.
\end{corollary}
\begin{proof}[Self-contained proof]
By Kanter's representation \citep{Kanter1975}, $V\stackrel d=(A_\gamma(U)/\mathcal E)^{(1-\gamma)/\gamma}$ with $U\sim U(0,\pi)$ and $\mathcal E\sim{\rm Exp}(1)$ independent. So $\log V=\tfrac{1-\gamma}{\gamma}\{B+\Lambda\}$ with $B=\log A_\gamma(U)$ and $\Lambda=-\log\mathcal E$.

$\Lambda$ is standard Gumbel, with log-concave density $\exp\{-g-\exp\{-g\}\}$. By Lemma~\ref{lem:kanter}, $h=\log A_\gamma$ is increasing, so $B$ has density $\{\pi\,h'(h^{-1}(w))\}^{-1}$ on $(h(0+),\infty)$. This density is non-increasing because $h'$ is increasing, so $B$ is unimodal with its mode at the left endpoint. By Ibragimov's theorem $B+\Lambda$ is unimodal, and hence so is $\log V$. Proposition~\ref{prop:shat} applies.
\end{proof}

\begin{remark}
Log-concavity would be too strong a hypothesis here. By \citet{Simon2011}, a positive $\gamma$-stable law is multiplicatively strongly unimodal, equivalently $\log V$ has a log-concave density \citep{CuculescuTheodorescu1998}, if and only if $\gamma\le\tfrac12$, that is $\alpha\le1$. (Correspondingly, $\log A_\gamma(U)$ has a non-increasing but not log-concave density in general; Simon notes this at $\gamma=\tfrac12$.) So for $1<\alpha<2$ the unimodal hypothesis of Proposition~\ref{prop:shat} is needed. The deterministic quadrature in \texttt{check\_shat.py} agrees: $M_n$ has a single maximum for $n\in\{10,25,100\}$ and $\alpha\in\{0.6,1,1.3,1.6,1.9\}$. At $\alpha=1.99$ the curve is too flat near its peak for the grid, which reports spurious local maxima. It also reproduces the profile values of the preceding Remark, e.g.\ $\log H_{25}(1.9)-\log H_{25}(2)=-0.19$.
\end{remark}

\begin{theorem}[Predictive distribution]\label{thm:pred}
Fix $\theta$, and let $(\hat c,\hat\kappa)$ maximize the M1 likelihood given $\theta$. Let $\mu_{\rm GP}(x_\ast)$ and $\sigma_{\rm GP}(x_\ast)$ be the plug-in predictive mean and standard deviation of the GP fitted by ML, for the same target. For a new response this includes the nugget; for the latent value it does not. Here M1 is read as $y=\sqrt{cV}\,(f_0+\sqrt\rho\,\varepsilon)$, so the noise shares $V$. Then:
\begin{enumerate}[label=(\roman*),leftmargin=2em]
\item $\hat\kappa=\hat\kappa_{\rm GP}$, and $\hat c=\hat s\,\beta_0$ with $\hat s\in\mathcal S_n(\theta)$.
\item For a new response or latent value $y_\ast$ at $x_\ast$, the plug-in predictive at $(\hat c,\hat\kappa,\theta)$ is
\[
y_\ast\mid y \;\stackrel{d}{=}\; \mu_{\rm GP}(x_\ast)+\sigma_{\rm GP}(x_\ast)\,\xi,\qquad \xi=\sqrt{n\hat s\,\tilde V}\,Z,
\]
where $Z\sim N(0,1)$ is independent of $\tilde V$, and $\tilde V$ has law $\propto g_n(\hat s v)\,dF_\theta(v)$.
\item The law $G_{n,\theta,\hat s}$ of $\xi$ is symmetric, with a continuous, strictly positive density. For each $\hat s\in\mathcal S_n(\theta)$ it depends on $(n,\theta,\hat s)$ only, not on $y$, the inputs or $x_\ast$. Under the hypothesis of Proposition~\ref{prop:shat}, $\hat s$ is unique, and we write $G_{n,\theta}$.
\end{enumerate}
\end{theorem}
\begin{proof}[Proof of Theorem~\ref{thm:pred}]
(i) is Corollary~\ref{cor:profile}(ii) together with the definition of $\mathcal S_n$.

(ii) Given $V=v$, the vector $(y,y_\ast)$ is Gaussian with covariance $cv K_0^+$, where $K_0^+$ is the $(n+1)\times(n+1)$ matrix of $K_0$ extended to $x_\ast$. For the latent value, the new diagonal entry omits the nugget. Hence $y_\ast\mid y,v\sim N(\mu,\,cv\tau^2)$ with $\mu=k_\ast^\top K_0^{-1}y$ and $\tau^2=k_{\ast\ast}-k_\ast^\top K_0^{-1}k_\ast$. Neither depends on $(c,v)$.

By Bayes' rule and the proof of Theorem~\ref{thm:factor}, \[\pi(v\mid y)\propto\varphi_n(y;0,cvK_0)\,dF_\theta(v)\propto g_n(cv/\beta_0)\,dF_\theta(v).\] At $c=\hat c=\hat s\beta_0$ this is $\propto g_n(\hat s v)\,dF_\theta(v)$, the law of $\tilde V$. Therefore $y_\ast\mid y\stackrel{d}{=}\mu+\sqrt{\hat s\beta_0\tilde V}\,\tau Z$.

For the GP ($F=\delta_1$), maximizing $g_n(c/\beta_0)$ gives $\hat c_{\rm GP}=\beta_0/n$, so $\sigma^2_{\rm GP}=\beta_0\tau^2/n$ and $\mu_{\rm GP}=\mu$ at $\hat\kappa_{\rm GP}=\hat\kappa$. Substituting gives $\sqrt{\hat s\beta_0\tilde V}\,\tau=\sqrt{n\hat s\tilde V}\,\sigma_{\rm GP}$.

(iii) The normalizer of $\tilde V$ is $M_n(\theta,\hat s)=H_n(\theta)\in(0,\infty)$, and $n\hat s\tilde V>0$ almost surely, with $\E\tilde V^{-1/2}<\infty$ because $u^{-1/2}g_n(u)$ is bounded. So $\xi$ is a scale mixture of $N(0,1)$ with a strictly positive continuous density, and by (ii) its mixing law involves only $(n,\theta,\hat s)$. This proves the result.
\end{proof}

\begin{corollary}[Student-$t$ process]\label{cor:tppred}
For $\nu>2$ and $V\sim$ inverse-gamma$(\nu/2,(\nu-2)/2)$, $\hat s=\nu/\{n(\nu-2)\}$, $\tilde V$ is inverse-gamma with shape $\tfrac{\nu+n}2$ and scale $\tfrac{(\nu-2)(\nu+n)}{2\nu}$, and $G_{n,\nu}$ is the standard $t_{\nu+n}$ law. So the TP predictive is $\mu_{\rm GP}+\sigma_{\rm GP}\,t_{\nu+n}$. In the parametrization of \citet{Shah2014}, the fitted TP has $\beta_{\rm S}=y^\top K_y^{-1}y=n(\nu-2)/\nu$ exactly. Substituting $\hat s$ into $M_n$ gives $\log H_n(\nu)=\tfrac n2\log(2n)+h_n(\nu)$, the expression in Corollary~\ref{cor:tp}.
\end{corollary}
\begin{proof}[Proof of Corollary~\ref{cor:tppred}]
$M_n(s)\propto s^{-n/2}\{b+1/(2s)\}^{-(a+n/2)}$ with $a=\nu/2$ and $b=(\nu-2)/2$, which is maximized at $s=a/(nb)$. The posterior of $V$ is inverse-gamma with shape $a+n/2$ and scale $b+1/(2\hat s)$. If $W\sim$ inverse-gamma$(a',b')$, then $\sqrt W Z\sim\sqrt{b'/a'}\,t_{2a'}$. Here $n\hat s\,b'/a'=1$.
\end{proof}
Numerically, a TP fitted with fixed $\nu\in\{3,5,20\}$ reproduces the GP's length-scale and noise ratio to four digits, and the GP's predictive mean and scale to $10^{-7}$. It has $\nu+n$ degrees of freedom and $\beta_{\rm S}=n(\nu-2)/\nu$, on both Gaussian and $t_{1.5}$-noise data (\texttt{check\_pred.py}).

\begin{corollary}[Acquisition functions]\label{cor:acq}
Consider maximization. Let $z(x)=\{\mu_{\rm GP}(x)-y^+\}/\sigma_{\rm GP}(x)$ for an incumbent $y^+$, wherever $\sigma_{\rm GP}(x)>0$, and assume $\hat s$ is unique. Under M1 at the ML:
\begin{enumerate}[label=(\alph*),leftmargin=2em]
\item Probability of improvement is $G_{n,\theta}(z(x))$, a strictly increasing function of the GP's $z(x)$. It ranks every candidate exactly as GP-PI does, and has the same maximizer.
\item The upper confidence bound at level $q$ is $\mu_{\rm GP}+G_{n,\theta}^{-1}(q)\,\sigma_{\rm GP}$. This is GP-UCB with $\Phi^{-1}(q)$ replaced by $G_{n,\theta}^{-1}(q)$, which is constant in $x$ but changes with $n$, i.e.\ a particular $n$-dependent multiplier.
\item Expected improvement is $\sigma_{\rm GP}(x)\,e_G(z(x))$ with $e_G(z)=\E(z+\xi)^+=\E[\sigma_{\tilde V}\,e_\Phi(z/\sigma_{\tilde V})]$, where $\sigma_{\tilde V}=\sqrt{n\hat s\tilde V}$ and $e_\Phi(z)=z\Phi(z)+\phi(z)$. This is finite because $\sqrt u\,g_n(u)$ is bounded. It differs from GP-EI only through the fixed function $e_G$. Since $e_G/e_\Phi$ is not constant, EI rankings, unlike PI rankings, can differ from the GP's.
\end{enumerate}
\end{corollary}
By part (a), a fitted TP and the GP rank candidates identically under PI.

\begin{remark}[How far from Gaussian is \texorpdfstring{$G_{n,\theta}$}{G}?]\label{rem:gshape}
For the principal families considered here, $G_{n,\theta}$ approaches $N(0,1)$ as $n$ grows. For the TP this is exact: $t_{\nu+n}\to N(0,1)$. For sub-Gaussian $\alpha$-stable mixing we have checked it numerically, and the effect is similar in size%
:

\begin{center}\small
\begin{tabular}{@{}lccc@{}}\toprule
& $n=10$ & $n=25$ & $n=100$\\\midrule
97.5\% quantile of $G$, $\alpha=1.0$ & 2.20 & 2.06 & 1.98\\
97.5\% quantile of $G$, $\alpha=1.6$ & 2.09 & 2.03 & 1.98\\
$e_G(-2)/e_\Phi(-2)$, $\alpha=1.0$ & 2.38 & 1.49 & 1.11\\
$e_G(-2)/e_\Phi(-2)$, $\alpha=1.6$ & 1.75 & 1.35 & 1.10\\
TP $\nu=3$: 97.5\% quantile; $e_G(-2)/e_\Phi(-2)$ & 2.16; 2.11 & 2.05; 1.45 & 1.98; 1.11\\\bottomrule
\end{tabular}
\end{center}
The GP values are $1.96$ and $1$. The $\alpha=1$ rows are exact: positive $\tfrac12$-stable mixing is the L\'evy law, an inverse gamma with shape $\tfrac12$, so M1 at $\alpha=1$ is the multivariate Cauchy process (the TP with $\nu=1$) and $G=t_{n+1}$. More generally, tilting by $g_n(\hat s v)\sim v^{-n/2}$ gives $G_{n,\alpha}$ a power tail of index $\alpha+n$, the analog of $t_{\nu+n}$. So even a Cauchy-like process ($\alpha=1$) has a nearly Gaussian predictive once $n=100$. Heavy tails enter M1 only through data-free changes of the acquisitions, such as an inflation of EI in low-$z$ regions, and for these families the change vanishes as $n$ grows (\texttt{check\_shape.py}).
\end{remark}

\begin{remark}[Scope]\label{rem:scope}
The result depends on (a) a single realization and (b) a free amplitude estimated by ML. With $m$ independent realizations sharing $\theta$ but with independent mixing variables, as in multi-task settings, $\theta$ becomes consistently estimable as $m\to\infty$ from the $m$ values $\beta_0^{(j)}$, as in iid scale-mixture estimation; for fixed $m$, the argument of Proposition~\ref{prop:noconsist} applied to the $m$-fold product of mixing laws still excludes consistency as $n\to\infty$. A proper prior on $c$ also makes the posterior of $\theta$ depend on the data, but only through a single draw of $cV$; the robustness results of \citet{Andrade2023} for the TP posterior, which use a light-tailed prior on the scale, are of this kind. Under M1 no estimator can be consistent for $\theta$ over the model along one path (Proposition~\ref{prop:noconsist}).
\end{remark}

\section{Observation-level mixtures: the tail index is identifiable and estimable from one realization}\label{sec:m2}
In this section, we consider observation-level mixtures (M2).\ifssbuild{} Proofs not given here are in Supplementary Sections~\ref{app:proofs}--\ref{app:lap}, \ref{supp:m2} and~\ref{supp:fold}.\fi{} Note that, in M2, each observation carries its own mixing variable, so a sample of size $n$ contains $n$ independent draws of it. We treat symmetric $\alpha$-stable noise: $\varepsilon_i$ has characteristic function $\exp\{-|t|^\alpha\}$, and $y=f+\sigma\varepsilon$ with $f\sim N_n(0,\Sigma_f)$, $\Sigma_f=\sigma_f^2R_\ell$, and $R_\ell$ a correlation matrix.

\begin{proposition}[Identifiability]\label{prop:m2id}
The law of $y$ determines $\alpha\in(0,2]$. If $\alpha<2$ it also determines $\sigma$ and $\Sigma_f$, and hence $(\sigma_f,\ell)$ whenever $\ell\mapsto R_\ell$ is injective. If $\alpha=2$ it determines $\Sigma_f+2\sigma^2I$.
\end{proposition}
\begin{proof}[Proof of Proposition~\ref{prop:m2id}]
The log characteristic function is $\Psi(t)=-\tfrac12t^\top\Sigma_ft-\sigma^\alpha\sum_i|t_i|^\alpha$.

Along $t=se_i$, $-\Psi(se_i)=\sigma^\alpha|s|^\alpha+\tfrac12(\Sigma_f)_{ii}s^2$. In all cases \[\lim_{s\to0}\log\{-\Psi(se_i)\}/\log|s|=\alpha.\] For $\alpha<2$ this gives $\alpha$, then $\sigma^\alpha=\lim_{s\to0}-\Psi(se_i)/|s|^\alpha$, and then $(\Sigma_f)_{ii}$.

For the off-diagonal entries, $\Psi(s(e_i+e_j))-\Psi(se_i)-\Psi(se_j)=-s^2(\Sigma_f)_{ij}$. For $\alpha=2$, $\Psi$ is the quadratic form of $\Sigma_f+2\sigma^2I$.
\end{proof}

\begin{remark}[Identifiability in M1]\label{rem:m1id}
Proposition~\ref{prop:m2id} concerns notion (I). The M1 parameters are identifiable in this sense too, under a mild condition. Under M1, $y=\sqrt{\beta_0}\,K_0^{1/2}U$ with $U$ uniform on the unit sphere and independent of $\beta_0=y^\top K_0^{-1}y\stackrel d=cV\chi^2_n$, where $V$ and $\chi^2_n$ are independent. The law of the direction $y/\|y\|$ (an angular central Gaussian law) determines $K_0$ up to a positive multiple, which is absorbed in $c$; then the law of $y$ determines the law of $\beta_0$. Since $\E(\chi^2_n)^{is}$ never vanishes (it is a ratio of Gamma functions), the law of $\beta_0$ determines the characteristic function of $\log(cV)$, and hence the law of $cV$. Thus $(\theta,c)$ is identifiable whenever $(\theta,c)\mapsto\text{law of }cV$ is injective, as for the Student-$t$ family with $\E V=1$. What fails in M1 is (II) and (III). A single path contains a single draw of $V$: no estimator can be consistent for $\theta$ over the model along it (Proposition~\ref{prop:noconsist}), and by Theorem~\ref{thm:factor} it carries no information about $\theta$ once $c$ is profiled out. Under M2, by contrast, the residuals $y_i-f(x_i)$ are $n$ independent draws of the noise.
\end{remark}

If $f$ were known, $\hat\alpha$ would be the iid stable MLE, which is $\sqrt n$-consistent and asymptotically normal \citep{DuMouchel1973}. With $f$ unknown, a single path still identifies $\alpha$ consistently and at the $\sqrt n$ rate. Differences of neighboring observations remove the smooth signal, as in difference-based variance estimation \citep{Rice1984,Gasser1986,HallKayTitterington1990}. Press's two-point empirical characteristic function (ECF) estimator \citep{Press1972,Koutrouvelis1980}, applied to the differences, then estimates $\alpha$. For iid samples, \citet{Krutto2018} proved asymptotic normality of this two-point log-ECF (``cumulant'') estimator and gave a sample-based rule for choosing the two arguments; see also \citet{LemberKrutto2022}. What is new here is the differencing construction, the bound on the signal remainder, and the treatment of a data-driven choice of the arguments (Remark~\ref{rem:ecf}(c)).

Let $x_{n,1}<\dots<x_{n,n}$ be the inputs in a compact interval, $m=\lfloor n/2\rfloor$, and
\[
D_j=y_{2j}-y_{2j-1}=\sigma\eta_j+\Delta_j,\qquad \eta_j=\varepsilon_{2j}-\varepsilon_{2j-1},
\]
with $\Delta_j=f(x_{n,2j})-f(x_{n,2j-1})$ for $j=1,\dots,m$, and let $h_n=\max_j(x_{n,2j}-x_{n,2j-1})$. The $\eta_j$ are iid, and $\psi(t)=\E\cos(t\sigma\eta_j)=\exp(-2\sigma^\alpha|t|^\alpha)$. Hence $\log\psi(t_1)/\log\psi(t_2)=(t_1/t_2)^\alpha$ for any $0<t_1<t_2$. With $\hat\phi(t)=m^{-1}\sum_j\cos(tD_j)$, define
\[
\tilde\alpha_n=g\big(\hat\phi(t_1),\hat\phi(t_2)\big),\qquad g(u_1,u_2)=\frac{\log\{\log u_1/\log u_2\}}{\log(t_1/t_2)} .
\]

\begin{theorem}[A $\sqrt n$-consistent estimator of the tail index]\label{thm:ecf}
Let $\alpha\in(0,2]$ and $\sigma>0$. Suppose $f$ is Lipschitz with constant $L$ on the input interval and $h_n\to0$. Conditionally on $f$:
\begin{enumerate}[label=(\roman*),leftmargin=2em]
\item $\tilde\alpha_n\to\alpha$ almost surely, and $-\log\hat\phi(t_1)/(2t_1^{\tilde\alpha_n})\to\sigma^\alpha$ almost surely;
\item if $\sqrt n\,h_n^2\to0$, then $\sqrt m(\tilde\alpha_n-\alpha)\Rightarrow N(0,v)$ with $v=\nabla g^\top\Sigma\nabla g$. Here $\nabla g$ is evaluated at $(\psi(t_1),\psi(t_2))$, and $\Sigma_{kl}=\tfrac12\{\psi(t_k-t_l)+\psi(t_k+t_l)\}-\psi(t_k)\psi(t_l)$.
\end{enumerate}
\end{theorem}
\begin{proof}[Proof of Theorem~\ref{thm:ecf}]
Let $\bar\phi(t)=m^{-1}\sum_j\cos(t\sigma\eta_j)$. Since $|\cos a-\cos b|\le|a-b|$, we have $|\hat\phi(t)-\bar\phi(t)|\le t\max_j|\Delta_j|\le tLh_n\to0$.

The summands of $\bar\phi(t)$ are iid, take values in $[-1,1]$ and have mean $\psi(t)$. By Hoeffding's inequality, $P\{|\bar\phi(t)-\psi(t)|>\epsilon\}\le2\exp\{-m\epsilon^2/2\}$, which is summable in $n$, so $\bar\phi(t)\to\psi(t)$ almost surely by Borel--Cantelli. This holds for any construction of the noise across $n$. Since $\psi(t_k)\in(0,1)$ and $g$ is continuous there, (i) follows; the statement for $\sigma^\alpha$ is the same argument.

For (ii), the symmetry of $\eta_j$ gives $\E\cos(tD_j)=\psi(t)\cos(t\Delta_j)$. Hence $\hat\phi(t)-\psi(t)=A_n(t)+B_n(t)$, where
\begin{gather*}
A_n(t)=\frac1m\sum_j\{\cos(tD_j)-\psi(t)\cos(t\Delta_j)\},\\ B_n(t)=\frac{\psi(t)}m\sum_j\{\cos(t\Delta_j)-1\}.
\end{gather*}
The summands of $A_n$ are independent, bounded and centered. As $\max_j|\Delta_j|\to0$, their covariances converge to $\Sigma$; this uses $\cos a\cos b=\tfrac12\{\cos(a-b)+\cos(a+b)\}$. By the Lindeberg--Feller CLT, $\sqrt m\,A_n\Rightarrow N(0,\Sigma)$. Moreover $|B_n(t)|\le\tfrac12t^2L^2h_n^2$, so $\sqrt m\,B_n\to0$. The delta method completes the proof.
\end{proof}

\begin{remark}[Scope and practical choices]\label{rem:ecf}\leavevmode
\begin{enumerate}[label=(\alph*),leftmargin=2em]
\item \emph{Smoothness of $f$.} The theorem conditions on $f$, so it needs only the realized $f$ to be Lipschitz, with noise independent of $f$ and of the design. Mat\'ern sample paths with $\nu>1$ admit a continuously differentiable modification, which is Lipschitz on a compact interval. If $f$ is only $\gamma$-H\"older, the same proof gives (ii) under $\sqrt n\,h_n^{2\gamma}\to0$. Stratified and equispaced designs have $h_n=O(1/n)$, so the condition holds for every $\gamma>1/4$.
\item \emph{Second differences.} On an equispaced design with $f\in C^2$, the second differences $y_{3j}-2y_{3j-1}+y_{3j-2}$ have noise ch.f.\ $\exp\{-(2+2^\alpha)\sigma^\alpha|t|^\alpha\}$. The ratio identity still holds, the signal remainder is $O(h_n^2)$, and the bias term is $O(h_n^4)$, with $m=\lfloor n/3\rfloor$. The price is efficiency: per observation, the standard deviation is larger by about $\sqrt{3/2}$ (RMSE $0.034$ against $0.026$ at $n=7{,}680$, $\alpha_0=1.3$).
\item \emph{Choice of $t$.} We use $t_k=c_k/s_n$ with $s_n=\operatorname{median}|D_j|$ and $(c_1,c_2)=(0.2,1.0)$, which minimized the average asymptotic standard deviation over $\alpha\in\{1,1.3,1.7,1.95\}$ among the values tried. Since $s_n\to s^\ast=\sigma\,\mathrm{median}|\eta|>0$, Theorem~\ref{thm:ecf} holds with $v$ evaluated at $t_k^\ast=c_k/s^\ast$ (Lemma~\ref{lem:datat}). In outline: because $s\mapsto g(\psi(c_1/s),\psi(c_2/s))\equiv\alpha$ is constant, the error in $s_n$ enters only through the modulus of continuity of the ECF process. That process is uniformly consistent on compacts \citep{FeuervergerMureika1977} and asymptotically equicontinuous \citep{Csorgo1981}; Lemma~\ref{lem:datat} gives a self-contained bracketing proof that also handles the signal remainder. On pure noise ($m=20{,}000$, 1{,}000 replicates) the Monte Carlo standard deviation of $\sqrt m(\tilde\alpha_n-\alpha)$ was within 5\% of $\sqrt v$ at $\alpha\in\{1.3,1.7,1.95\}$ (\texttt{mc\_check.py}).
\item \emph{Contrast with M1.} Under M1 a single path carries one draw of $V$, and no estimator can be consistent for the mixing law over the model (Proposition~\ref{prop:noconsist}). Under M2 the differences carry $m$ independent draws of the noise, and Theorem~\ref{thm:ecf} turns this into a consistent estimator.
\item \emph{Clipping.} Part (ii) is for the unclipped $\tilde\alpha_n$, which is defined with probability tending to one. In practice $\tilde\alpha_n$ is clipped at 2. At $\alpha=2$, $\sqrt m\{\min(\tilde\alpha_n,2)-2\}\Rightarrow\min(Z,0)$ with $Z\sim N(0,v)$, which halves the mean squared error. For iid stable samples, \citet{Krutto2018} and \citet{LemberKrutto2022} study the log-ECF estimator of $\alpha$ and the choice of its arguments.
\item \emph{The ML estimate.} Theorem~\ref{thm:ecf} is not efficient. Section~\ref{sec:eff} shows that two ML estimators are efficient: cross-fitted plug-in ML (Theorem~\ref{thm:eff}), and the Laplace marginal-likelihood estimate at the global posterior mode (Theorem~\ref{thm:lapeff}), for $f\in H^3$, one-dimensional inputs and hyperparameters in a compact set. The IS marginal-likelihood estimate is not covered. $\tilde\alpha_n$ is a consistent starting value for all of them. Started there, the root-mean-squared error of the Laplace ML estimate is close to the known-$f$ bound over the simulated range $n\le7{,}680$ (Section~\ref{sec:cons}).
\end{enumerate}
\end{remark}

\subsection{Efficient estimation of the tail index from one realization}\label{sec:eff}
Theorem~\ref{thm:ecf} gives a simple $\sqrt n$-consistent estimator, but not an efficient one. We now show that, with $f$ unknown, $\alpha$ can be estimated from one realization as efficiently as if $f$ were known. The estimator is a plug-in maximum-likelihood estimator with cross-fitting. Let $\theta=(\alpha,\log\sigma)$ and $\ell_\theta(r)=\log\{\sigma^{-1}p_\alpha(r/\sigma)\}$, with score $s_\theta=\nabla_\theta\ell_\theta$. Let $I(\theta)=\E\,s_\theta s_\theta^\top(\sigma\varepsilon)$ be the Fisher information of the model in which $f$ is known, so that $[I(\theta)^{-1}]_{\alpha\alpha}^{1/2}$ is the known-$f$ bound used in Section~\ref{sec:cons}.

\emph{Cross-fitting.} Split the ordered inputs into two folds by the parity of their rank. For $i$ in one fold, let $\hat f_i$ be an estimate of $f(x_i)$ computed only from the observations in the other fold, and write $\Delta_i=\hat f_i-f(x_i)$ and $\|\Delta\|_n=(n^{-1}\sum_i\Delta_i^2)^{1/2}$. The estimator is
\[
\hat\theta=\arg\max_{\theta\in\Theta}\ \sum_{i=1}^n\ell_\theta(y_i-\hat f_i),
\]
for a compact set $\Theta\subset(0,2)\times\R$.

\emph{Assumptions.}
\begin{enumerate}[label=(A\arabic*),leftmargin=2.6em]
\item\label{A:design} The inputs $x_{n,1}<\dots<x_{n,n}$ lie in a fixed compact interval, and all spacings (including those to the endpoints) are at most $C_D/n$.
\item\label{A:f} $f$ is Lipschitz with constant $L$. (As in Theorem~\ref{thm:ecf}, we condition on the realized $f$.)
\item\label{A:noise} $y_i=f(x_i)+\sigma_0\varepsilon_i$ with $\varepsilon_i$ iid S$\alpha$S with characteristic function $\exp\{-|t|^{\alpha_0}\}$, $\alpha_0\in(0,2)$, and $\theta_0=(\alpha_0,\log\sigma_0)$ is an interior point of $\Theta$.
\item\label{A:reg} \emph{Regularity of the stable score.} $\sup_{\theta\in\Theta,\,r\in\R}|\partial_r\ell_\theta(r)|<\infty$, and $\E\sup_\Theta|\ell_\theta(\sigma_0\varepsilon)|<\infty$. On a neighborhood $U$ of $\theta_0$, the functions $\partial_r^j\partial_\theta^\beta\ell_\theta(r)$ with $j\in\{1,2\}$ and $|\beta|\in\{1,2\}$ are bounded uniformly in $(\theta,r)$; $\E\sup_U\|\partial_\theta^\beta\ell_\theta(\sigma_0\varepsilon)\|^2<\infty$ for $|\beta|\le2$; $\E\,s_{\theta_0}=0$, $\E\,\partial_\theta s_{\theta_0}=-I(\theta_0)$; and $I(\theta_0)$ is nonsingular.
\item\label{A:pilot} \emph{Pilot.} Each $\hat f_i$ is a measurable function of the other fold's data, and $\|\Delta\|_n=o_p(n^{-1/4})$.
\end{enumerate}
Assumption~\ref{A:reg} concerns only the S$\alpha$S family, and it holds for every compact $\Theta$.

\begin{proposition}[Regularity of the S$\alpha$S log-density]\label{prop:reg}
Let $\Theta\subset(0,2)\times\R$ be compact. Then:
\begin{enumerate}[label=(\roman*),leftmargin=2em]
\item for every $j\ge1$ and every multi-index $\beta$, \[\sup_{\theta\in\Theta,\,r\in\R}(1+|r|)^j\,|\partial_r^j\partial_\theta^\beta\ell_\theta(r)|<\infty;\]
\item for every multi-index $\beta$ (including $\beta=0$), $|\partial_\theta^\beta\ell_\theta(r)|\le C_\beta\{1+\log(1+|r|)\}$ on $\Theta\times\R$;
\item for every $\theta\in(0,2)\times\R$, $\E_\theta s_\theta=0$, $\E_\theta\,\partial_\theta s_\theta=-I(\theta)$, and $I(\theta)$ is finite, nonsingular and continuous.
\end{enumerate}
In particular, \ref{A:reg} holds whenever $\theta_0$ is interior to $\Theta$, with $U$ any compact neighborhood of $\theta_0$.
\end{proposition}
The proof (Supplementary Section~\ref{app:reg}) rests on a tail expansion of $p_\alpha$ that is uniform in $\alpha$ on compacts and may be differentiated in both $x$ and $\alpha$ (Lemma~\ref{lem:tail}): $p_\alpha(x)=c_\alpha|x|^{-1-\alpha}\{1+O(|x|^{-\alpha})\}$ with $c_\alpha=\Gamma(1+\alpha)\sin(\pi\alpha/2)/\pi$. So $\partial_\alpha\ell_\theta(r)$ grows like $-\log|r|$, every derivative in $r$ decays, and the constants degenerate only at the ends of $(0,2)$, where $c_\alpha\to0$; at $\alpha=2$ this is the non-regularity noted in Section~\ref{sec:exact}. The numerical check on the density tables (\texttt{check\_regularity.py}) agrees with the proposition.

\begin{theorem}[Efficient estimation with $f$ unknown]\label{thm:eff}
Under \ref{A:design}--\ref{A:pilot}, $\hat\theta\to\theta_0$ in probability and
\[
\sqrt n\,(\hat\theta-\theta_0)\Rightarrow N\big(0,I(\theta_0)^{-1}\big).
\]
In particular $\sqrt n(\hat\alpha-\alpha_0)\Rightarrow N(0,[I(\theta_0)^{-1}]_{\alpha\alpha})$. This is the limit of the maximum-likelihood estimator when $f$ is known \citep{DuMouchel1973}, so $\hat\alpha$ attains the known-$f$ bound.
\end{theorem}
The proof is in Supplementary Section~\ref{app:proofs}. The key step is an orthogonality. Because $p_\alpha$ is symmetric, the scores for $\alpha$ and $\sigma$ are even functions of the residual, and their derivatives in the residual are odd. An error $\Delta_i$ in $f$ therefore enters the estimating equation only through a term that has conditional mean zero under cross-fitting, and through a second-order term of size $\|\Delta\|_n^2$. This is Neyman orthogonality in the sense of double/debiased machine learning \citep{Chernozhukov2018}. Here it needs no correction term, because symmetry of the noise makes the score orthogonal to $f$. Cross-fitting removes the dependence between $\Delta_i$ and $\varepsilon_i$ without any rate condition on the complexity of $\hat f$.

\begin{corollary}[Confidence intervals]\label{cor:wald}
Under the conditions of Theorem~\ref{thm:eff}, the interval
\[
\hat\alpha\pm z_{1-\gamma/2}\,[I(\hat\theta)^{-1}]_{\alpha\alpha}^{1/2}/\sqrt n
\]
has asymptotic coverage $1-\gamma$.
\end{corollary}
\begin{proof}[Proof of Corollary~\ref{cor:wald}]
$I(\theta)$ depends only on $\alpha$, since the model is a scale family in $\sigma$. It is continuous in $\alpha$ under~\ref{A:reg}, so $I(\hat\theta)\to I(\theta_0)$ in probability. Slutsky's lemma completes the proof.
\end{proof}

A pilot that provably satisfies~\ref{A:pilot} under heavy tails, with no moments of the noise, is the running median.

\begin{lemma}[Running-median pilot]\label{lem:median}
Assume \ref{A:design}--\ref{A:noise}. For $i$ in one fold, let $\hat f_i$ be the median of the observations at the $k_n$ inputs of the other fold that are nearest to $x_i$ in rank, with $k_n$ odd. Then, for $k_n\ge k_0(\alpha_0)$,
\[
\E\|\Delta\|_n^2\ \le\ C\Big(\frac{k_n^2}{n^2}+\frac1{k_n}\Big),
\]
with $C$ depending only on $L$, $C_D$, $\alpha_0$ and $\sigma_0$. Hence \ref{A:pilot} holds whenever $n^{1/2}/k_n\to0$ and $k_n/n^{3/4}\to0$, for example $k_n\asymp n^{2/3}$.
\end{lemma}

\paragraph{The GP posterior mode as pilot.} In practice the pilot is the posterior mode of the GP with S$\alpha$S noise. Let $\kappa=(s,\ell)$ be the amplitude and length-scale of the Mat\'ern-$5/2$ kernel $k_\kappa$, with reproducing-kernel Hilbert space (RKHS) $\mathcal H_\kappa$, let $c\in\R$ be a prior mean, and write $\psi=(\theta,\kappa,c)$. For a set $J$ of $N$ design points, let $\hat f_\psi$ minimize over $g\in\mathcal H_\kappa$
\[
Q_\psi(g)=-\sum_{i\in J}\ell_\theta\{y_i-g(x_i)\}+\tfrac12\|g-c\|_{\mathcal H_\kappa}^2 .
\]
By the representer theorem, $(\hat f_\psi(x_i))_{i\in J}$ is the posterior mode under the prior $N(c\mathbf 1,K_\kappa)$ and the S$\alpha$S likelihood with parameter $\theta$ (Section~\ref{sec:fold}). We use two further assumptions.
\begin{enumerate}[label=(A\arabic*),leftmargin=2.6em]\setcounter{enumi}{5}
\item\label{A:sob} $f\in H^3[0,T]$, the Sobolev space on the design interval $[0,T]$. It equals $\mathcal H_\kappa$ as a set, with norms equivalent uniformly over compact sets of $\kappa$ \citep[Ch.~10]{Wendland2005}.
\item\label{A:box} $\psi$ ranges over a compact set $\Psi=\Theta'\times\mathcal K\times[-M,M]$, with $\Theta'\subset(0,2)\times\R$ and $\mathcal K\subset(0,\infty)^2$ compact rectangles.
\end{enumerate}
The working parameter $\theta\in\Theta'$ need not equal $\theta_0$: the pilot may use a wrong tail index or scale.

\begin{theorem}[GP posterior-mode pilot]\label{thm:gppilot}
Assume \ref{A:design}, \ref{A:noise}, \ref{A:sob} and \ref{A:box}, and let $J$ be a set of $N\ge n/3$ design points whose spacings, including those to the endpoints, are at most $2C_D/n$ (for instance, one fold). Then, uniformly over $\psi\in\Psi$ and over all minimizers $\hat f_\psi$,
\begin{gather*}
\|\hat f_\psi-f\|_N=O_p\big(n^{-2/5}\log^6n\big),\qquad \|\hat f_\psi\|_{H^3}=O_p\big(n^{1/10}\log^6n\big),\\
\sup_{x}|\hat f_\psi(x)-f(x)|=O_p\big(n^{-7/30}\log^{6}n\big),
\end{gather*}
where $\|h\|_N^2=N^{-1}\sum_{i\in J}h(x_i)^2$. The same holds for any $\hat f_\psi$ with $\sup_{\psi\in\Psi}\{Q_\psi(\hat f_\psi)-Q_\psi(f)\}=O_p(1)$.
\end{theorem}
\emph{Outline of the proof of Theorem~\ref{thm:gppilot}.} The proof (Supplementary Section~\ref{app:gp}) needs no moments of the noise. The loss $-\ell_\theta$ is not convex, but it is Lipschitz (Proposition~\ref{prop:reg}(i)), its increments grow only logarithmically, and its excess risk $\E\{\ell_\theta(\tilde\varepsilon)-\ell_\theta(\tilde\varepsilon-t)\}$ is bounded below by a multiple of $\min(t^2,1)$ for every working $\theta$, because the noise is symmetric and unimodal. Chaining in a metric truncated at the scale of the loss first gives consistency in the empirical norm, then in the sup norm, and then the rate. The rate is not optimized; the logarithmic factors come from peeling.

\begin{corollary}[The GP pilot satisfies~\ref{A:pilot}]\label{cor:gppilot}
Under \ref{A:design}, \ref{A:noise}, \ref{A:sob} and \ref{A:box}, let $\hat\psi\in\Psi$ be computed from one fold only (including any standardization of the data), and let $\hat f_i$, for $x_i$ in the other fold, be the linear interpolation at $x_i$ of $\hat f_{\hat\psi}$ at the first fold's inputs (constant beyond the end points). Then $\|\Delta\|_n=O_p(n^{-2/5}\log^6n)$, so~\ref{A:pilot} holds and Theorem~\ref{thm:eff} applies.
\end{corollary}
The pilot of Table~\ref{tab:cf} is of this form: $\hat\psi$ is the box-constrained Laplace ML estimate on that fold's data, centered by their median and scaled by their median absolute deviation, which converge to finite non-zero limits (Remark~\ref{rem:effscope}(b)).

\paragraph{Without cross-fitting: the marginal-likelihood estimate.} Now take $J=\{1,\dots,n\}$, let $c=\hat c$ be any statistic with $|\hat c|\le M$ with probability tending to one, and let $W_\psi=\mathrm{diag}\{-\partial_r^2\ell_\theta(y_i-\hat f_\psi(x_i))\}$. The Laplace log marginal likelihood of Section~\ref{sec:fold} is, up to a constant,
\[
L(\theta,\kappa)=\sum_{i=1}^n\ell_\theta\{y_i-\hat f_\psi(x_i)\}-\tfrac12\|\hat f_\psi-c\|_{\mathcal H_\kappa}^2-\tfrac12\log\big|I+K_\kappa^{1/2}W_\psi K_\kappa^{1/2}\big| .
\]
\begin{proposition}[The global posterior mode under infill]\label{prop:infill}
Under \ref{A:design}, \ref{A:noise}, \ref{A:sob} and \ref{A:box}, there is a constant $\eta_0>0$ such that, with probability tending to one, simultaneously for all $\psi\in\Psi$: $Q_\psi$ has a unique global minimizer $\hat f_\psi$; it is nondegenerate, with $\lambda_{\min}(I+K_\kappa^{1/2}W_\psi K_\kappa^{1/2})\ge\tfrac12$; $\psi\mapsto\hat f_\psi$ is continuously differentiable; and $Q_\psi$ has no other stationary point $g$ with $\max_i|g(x_i)-f(x_i)|\le\eta_0$. So the branch of global modes has no fold in $\Psi$, $L$ is finite and continuously differentiable on $\Psi$, and the guard $\lambda_{\min}\ge\tau$ of Section~\ref{sec:fold} is inactive at the global mode for $\tau\le\tfrac12$.
\end{proposition}
The proposition does not exclude local stationary points of $Q_\psi$ farther from $f$, or folds of branches through them, and it does not show that a particular mode-finding algorithm selects the global branch; see Remark~\ref{rem:effscope}(a).
\begin{theorem}[Efficiency of the Laplace ML estimate]\label{thm:lapeff}
Assume \ref{A:design}, \ref{A:noise}, \ref{A:sob} and \ref{A:box}, and let $\theta_0$ be interior to a compact $\Theta\subset\Theta'$. Let $L$ be evaluated at the global mode $\hat f_\psi$ (unique with probability tending to one by Proposition~\ref{prop:infill}), and let $(\hat\theta,\hat\kappa)$ be any maximizer of $L$ over $\Theta\times\mathcal K$. Then $\hat\theta\to\theta_0$ in probability, whatever the behavior of $\hat\kappa$, and
\[
\sqrt n\,(\hat\theta-\theta_0)\Rightarrow N\big(0,I(\theta_0)^{-1}\big).
\]
So the Laplace ML estimate of $\alpha$ attains the known-$f$ bound, and the intervals of Corollary~\ref{cor:wald} are valid for it.
\end{theorem}
The theorem concerns the global-mode criterion. An implementation is covered on the event that the computed mode equals the global mode at every hyperparameter value the optimizer evaluates and the optimizer returns a global maximizer of $L$ over $\Theta\times\mathcal K$; we prove neither (Remark~\ref{rem:effscope}(a)).
\emph{Outline of the proof of Theorem~\ref{thm:lapeff}.} The proof is in Supplementary Section~\ref{app:lap}. By the envelope theorem, the $\theta$-score of $L$ is $\sum_is_\theta(y_i-\hat f_\psi(x_i))$ minus half the $\theta$-derivative of the log-determinant, and the latter is $O_p(n^{1/3})$ because posterior variances are $O(n^{-2/3})$. Without cross-fitting, the first-order term $n^{-1}\sum_i\dot s(\tilde\varepsilon_i)\Delta_i$ no longer has conditional mean zero. But $\E\dot s_\theta(\tilde\varepsilon)=0$ still holds for every $\theta$, and by Theorem~\ref{thm:gppilot} the errors $\Delta$ lie in a class of functions that is small in the empirical norm and not too large in $H^3$. Stochastic equicontinuity over that class then makes the term $o_p(n^{-1/2})$. Consistency of $\hat\theta$ does not need $\hat\kappa$ to converge, which matters because some kernel parameters are not consistently estimable under infill \citep{Zhang2004}: the criterion $n^{-1}L$ converges to $\E\,\ell_\theta(\tilde\varepsilon_1)$ uniformly in $\kappa$. The same argument shows that the plug-in estimator with an uncross-fitted GP pilot, at any $\hat\psi\in\Psi$, is efficient.

Proposition~\ref{prop:infill} proves the heuristic of Section~\ref{sec:fold} for the global mode with hyperparameters in a compact set. Since $\Psi$ is any compact set, for every compact $\Psi$ the probability of a fold of the global-mode branch in $\Psi$ tends to zero; folds of that branch at large $n$ need hyperparameters that approach the boundary of the parameter space, such as length-scales or noise scales tending to zero, or $\alpha$ tending to $0$ or $2$. Folds of other branches are not excluded.

\paragraph{Rough regression functions.} Assumption~\ref{A:sob} excludes the sample paths of the Mat\'ern-$5/2$ prior itself, which are almost surely not in its RKHS. For $\beta\in(1,3]$ let $C^\beta[0,T]$ be the functions whose derivative of order $k=\lceil\beta\rceil-1$ is $(\beta-k)$-H\"older.
\begin{enumerate}[label=(A\arabic*$'$),leftmargin=2.6em]\setcounter{enumi}{5}
\item\label{A:hol} $f\in C^\beta[0,T]$ for some $\beta\in(3/2,3]$.
\end{enumerate}
\begin{theorem}[Rough regression functions]\label{thm:rough}
Assume \ref{A:design}, \ref{A:noise}, \ref{A:hol} and \ref{A:box}, with $J$ as in Theorem~\ref{thm:gppilot}. Then, uniformly over $\psi\in\Psi$ and over all minimizers (or any $\hat f_\psi$ with $\sup_\psi\{Q_\psi(\hat f_\psi)-\inf_gQ_\psi(g)\}=O_p(1)$),
\begin{gather*}
\|\hat f_\psi-f\|_N=O_p\big(n^{-\min(\beta/6,\,2/5)}\log^6n\big),\\ \|\hat f_\psi\|_{H^3}=O_p\big(n^{\max\{(3-\beta)/6,\,1/10\}}\log^6n\big),\\
\sup_x|\hat f_\psi-f|=O_p\big(n^{-\min\{(\beta-1)/6,\,7/30\}}\log^6n\big),
\end{gather*}
Corollary~\ref{cor:gppilot} holds with the rate $n^{-\min(\beta/6,\,2/5)}\log^6n$, so~\ref{A:pilot} holds, and Proposition~\ref{prop:infill} and Theorem~\ref{thm:lapeff} hold with~\ref{A:hol} in place of~\ref{A:sob}. A centered Mat\'ern-$5/2$ process on $[0,T]$ has a modification whose paths lie in $C^\beta[0,T]$ for every $\beta<5/2$. So under model~M2 with a Mat\'ern-$5/2$ correlation (any amplitude and length-scale), conditionally on the realized $f$, which is independent of the noise, the GP pilot satisfies~\ref{A:pilot}, the global-mode branch has no folds, and the Laplace ML estimate of $\alpha$ is efficient. Given $f$ the noise is still iid S$\alpha$S, so each statement holds for almost every $f$, with constants depending on $\|f\|_{C^\beta}<\infty$. The same statements then hold unconditionally, by integrating the conditional probabilities over $f$ (Lemma~\ref{lem:uncond}).
\end{theorem}
\emph{Outline of the proof of Theorem~\ref{thm:rough}.} The proof (Supplementary Section~\ref{app:gp}) compares $\hat f_\psi$ with a mollification of $f$ at bandwidth $n^{-1/6}$, which balances the approximation error $b^\beta$ against the growth $b^{\beta-3}$ of the $H^3$ norm. The threshold $\beta>3/2$ is exactly what makes $\sqrt n\,\|\Delta\|_n^2\to0$.

\begin{remark}[Scope of Theorems~\ref{thm:eff}, \ref{thm:gppilot}, \ref{thm:lapeff} and~\ref{thm:rough}]\label{rem:effscope}\leavevmode
\begin{enumerate}[label=(\alph*),leftmargin=2em]
\item \emph{Global and computed modes.} The theorems are about global minimizers of $Q_\psi$, or approximate ones in the sense of Theorem~\ref{thm:gppilot}. The algorithm computes a local mode by Newton's method from an IRLS start (Section~\ref{sec:alg}). By Proposition~\ref{prop:infill}, with probability tending to one $Q_\psi$ is strictly convex on a sup-norm neighborhood of $f$ that contains every global minimizer. This rules out folds of the global-mode branch, not of other local branches; the finite-sample failures of Section~\ref{sec:fold} are folds of the branch that the solver tracks. In the GP-pilot fits of Table~\ref{tab:cf}, Newton's method started from the true $f$ reached the same mode as the algorithm, to within $10^{-6}$ in sup-norm, in every fit with $n\ge1{,}920$; at $n\le480$ it reached a different mode (sup-distance above $10^{-4}$) in 12 of 800 fits (Section~\ref{sec:cons}).
\item \emph{Implementation.} \ref{A:box} excludes $\alpha=2$ from the working likelihood: there the loss is quadratic, not Lipschitz, and the mode is a linear smoother of infinite-variance noise, which our argument does not control. \ref{A:box} also needs boxes that are compact in the original units. Standardizing $y$ by its sample standard deviation violates this asymptotically, since for $\alpha_0<2$ that deviation diverges like $n^{1/\alpha_0-1/2}$; centring by the median and scaling by the median absolute deviation does not. All fits in Tables~\ref{tab:cons}--\ref{tab:exact} use the latter and $\alpha\le1.98$, with the Gaussian model fitted separately.
\item \emph{The Gaussian boundary.} All results exclude $\alpha_0=2$, where the stable family is non-regular (Section~\ref{sec:exact}).
\end{enumerate}
\end{remark}
\emph{Smoothness of $f$ (addition to Remark~\ref{rem:effscope}).} If $f$ is only $\gamma$-H\"older, the bias in Lemma~\ref{lem:median} becomes $(k_n/n)^{\gamma}$. A window with $n^{1/2}\ll k_n\ll n^{1-1/(4\gamma)}$ then exists if and only if $\gamma>1/2$. Local polynomial median regression \citep{Chaudhuri1991} also satisfies~\ref{A:pilot} for smoother $f$, without moment conditions on the noise. For the GP pilot and the Laplace ML estimate, Theorem~\ref{thm:rough} replaces~\ref{A:sob} by $f\in C^\beta$ with $\beta>3/2$. We do not know whether $3/2$ is sharp; it is where our second-order bound $\sqrt n\|\Delta\|_n^2$ stops vanishing.

Finally, the data-driven arguments used for Theorem~\ref{thm:ecf} in practice do not change its limit.

\begin{lemma}[Data-driven arguments]\label{lem:datat}
Under the conditions of Theorem~\ref{thm:ecf}(ii) with, in addition, $\sqrt n\,h_n\to0$ (which holds under~\ref{A:design}), let $t_k=c_k/s_n$ with $0<c_1<c_2$ and $s_n=\operatorname{median}_j|D_j|$. Then $\sqrt m(\tilde\alpha_n-\alpha)\Rightarrow N(0,v^\ast)$, where $v^\ast$ is $v$ of Theorem~\ref{thm:ecf} evaluated at $t_k^\ast=c_k/s^\ast$, $s^\ast=\sigma\,\mathrm{median}|\eta_1|$. The variance $v^\ast$ depends on $\alpha$ only.
\end{lemma}

\begin{proposition}[Posterior moments]\label{prop:moments}
Under M2 with any $\alpha\in(0,2]$, the joint posterior of $f$ at $x_1,\dots,x_n,x_\ast$ exists and has finite moments of all orders.
\end{proposition}
\begin{proof}[Proof of Proposition~\ref{prop:moments}]
$p_\alpha$ is bounded by $p_\alpha(0)=\Gamma(1+1/\alpha)/\pi$, so the posterior density is at most \[\{p_\alpha(0)/\sigma\}^n\,\varphi(f,f_\ast;0,K^+)/p(y),\qquad 0<p(y)<\infty.\] The Gaussian bound has all moments.
\end{proof}
\ifssbuild By Proposition~\ref{prop:moments}, the posterior of $f$ has moments of all orders, so\else So\fi{} EI and UCB on the latent $f$ are well defined for every $\alpha$. By contrast, EI on a new \emph{noisy} observation is infinite for $\alpha\le1$, since $\E|y_\ast|=\infty$ there.

\subsection{Algorithm for symmetric \texorpdfstring{$\alpha$}{alpha}-stable noise}\label{sec:alg}
\ifssbuild\noindent\emph{Algorithm details.}\fi
\begin{enumerate}[label=(\alph*),leftmargin=2.2em]
\item Density: $\log p_\alpha$ is tabulated on $|z|\le60$ from Nolan's integral representation \citep{Nolan1997}, as implemented in \texttt{scipy.stats}\allowbreak\texttt{.levy\_stable}, on an $\alpha$-grid of step $0.01$, splined in $z$ and interpolated linearly in $\alpha$, with a continuous power-law extension beyond $60$. This is the general (dense) implementation, used for Table~\ref{tab:exact} and the multivariate BO problems. The $O(n)$ implementation introduced in Section~\ref{sec:cons}, used for Tables~\ref{tab:cons}, \ref{tab:cf} and~\ref{tab:boot} and for the one-dimensional BO fits, interpolates cubically in $\alpha$ instead, to obtain a smooth objective for optimization. An earlier table switched from Fourier inversion to a two-term tail series at $|z|=8$; the resulting jump (up to $0.04$ nats) spiked the first two derivatives there. It changed the fitted $\alpha$ negligibly (90\% of fits by less than $0.014$) but corrupted the curvature used by Laplace, so it is replaced. The theorems concern the exact density, so computed fits approximate the estimators they cover rather than being certified instances of them. As a check (\texttt{density\_sensitivity.py}), at 3{,}000 random points with off-grid $\alpha\in[0.8,1.98]$ and $|z|\le80$, the cubic table differs from \texttt{scipy.stats.levy\_stable} (used as the numerical reference) by at most $1.8\times10^{-4}$ nats for $|z|\le60$ and $3.8\times10^{-3}$ beyond, its $z$-derivative by at most $2.3\times10^{-4}$, and its $\alpha$-derivative by a median of $2.7\times10^{-5}$ (at most $0.11$); the linear table differs by at most $4.2\times10^{-3}$ nats, $2.4\times10^{-3}$ in its $z$-derivative and a median of $4.5\times10^{-3}$ (at most $0.65$) in its $\alpha$-derivative. Refitting the known-location ML estimator with the reference density on 40 samples of size 480 ($\alpha_0\in\{1.3,1.7\}$) changed $\hat\alpha$ by at most $1.1\times10^{-4}$, about $1/600$ of its standard error.
\item Posterior mode: Newton with line search, started from IRLS/EM. Because the posterior is not log-concave, this is a local mode, as for the Student-$t$ likelihood \citep{Vanhatalo2009,Jylanki2011}. The weights $w(z)=\psi(z)/z$, with $\psi=-(\log p_\alpha)'$, are exactly the EM weights of the normal scale mixture.
\item Marginal likelihood: Laplace approximation.
\item Predictive covariance: for prediction we use a Gaussian covariance obtained by replacing the likelihood Hessian curvature $W$ with the positive EM weights. This Fisher/EM covariance is positive definite and is used as a stable predictive approximation, including for the acquisition functions in Section~\ref{sec:bo}; it is distinct from the Hessian-based Laplace covariance used in the marginal-likelihood calculation.
\end{enumerate}

\subsection{Simulation: consistency of \texorpdfstring{$\hat\alpha$}{alpha-hat}}\label{sec:cons}
Design: $f(x)=x\sin(10x)+\cos(2x)$ on $[0,1.2]$, stratified inputs, noise $0.05\,\varepsilon$ with $\varepsilon\sim S\alpha S(1)$ simulated as in \citet{CMS1976}, and Mat\'ern-$5/2$ kernel. All fits in this section follow the standardization and parameter box of Theorem~\ref{thm:lapeff} (its global-mode and global-maximizer conditions are checked only empirically, Remark~\ref{rem:effscope}(a)): $y$ is centered by its median and scaled by its normalized median absolute deviation, the working $\alpha$ is restricted to $[0.8,1.98]$, and the Gaussian model is fitted separately. Laplace ML uses the $O(n)$ code described below, with two starts per data set (the difference estimate $\tilde\alpha_n$ with its scale estimate, and the fixed start $\alpha=1.95$), keeping the fit with the larger likelihood. There are 100 data sets for each $(\alpha_0,n)$.

\begin{table}[!htb]\centering\renewcommand{\baselinestretch}{1}\small
\begin{tabular}{@{}lrrrrrr@{}}\toprule
$\alpha_0$ & $n$ & median $\hat\alpha$ & mean $\hat\alpha$ & RMSE & at bound & $\|\hat f-f\|_n$\\\midrule
% --- table body (generated, inlined from tab_cons.tex) ---
1.3 & 30 & 1.65 & 1.65 & 0.452 & 0.28 & 0.0527\\
1.3 & 60 & 1.40 & 1.39 & 0.223 & 0.02 & 0.0372\\
1.3 & 120 & 1.31 & 1.33 & 0.138 & 0.00 & 0.0245\\
1.3 & 240 & 1.32 & 1.33 & 0.108 & 0.00 & 0.0189\\
1.3 & 1{,}920 & 1.30 & 1.30 & 0.032 & 0.00 & 0.0076\\\addlinespace
1.7 & 30 & 1.98 & 1.85 & 0.258 & 0.62 & 0.0480\\
1.7 & 60 & 1.77 & 1.78 & 0.185 & 0.29 & 0.0348\\
1.7 & 120 & 1.75 & 1.75 & 0.167 & 0.12 & 0.0234\\
1.7 & 240 & 1.72 & 1.71 & 0.103 & 0.02 & 0.0178\\
1.7 & 1{,}920 & 1.70 & 1.70 & 0.037 & 0.00 & 0.0072\\\addlinespace
2.0 & 30 & 1.98 & 1.97 & 0.063 & 0.96 & 0.0426\\
2.0 & 60 & 1.98 & 1.96 & 0.069 & 0.86 & 0.0327\\
2.0 & 120 & 1.98 & 1.97 & 0.038 & 0.87 & 0.0227\\
2.0 & 240 & 1.98 & 1.98 & 0.026 & 0.93 & 0.0167\\
2.0 & 1{,}920 & 1.98 & 1.98 & 0.020 & 0.99 & 0.0067\\\bottomrule
% --- end of table body ---

\end{tabular}
\caption{Consistency of $\hat\alpha$ in M2 (Laplace ML, 100 data sets per row). ``At bound'': share of fits with $\hat\alpha$ at the upper bound $1.98$ of the working $\alpha$. Last column: median of $\|\hat f-f\|_n$ at the posterior mode. Rows for other $n$ up to $7{,}680$ are in \texttt{ml\_unified\_summary.json}.}\label{tab:cons}
\end{table}

\begin{figure}[!htb]\centering\includegraphics[width=0.75\textwidth]{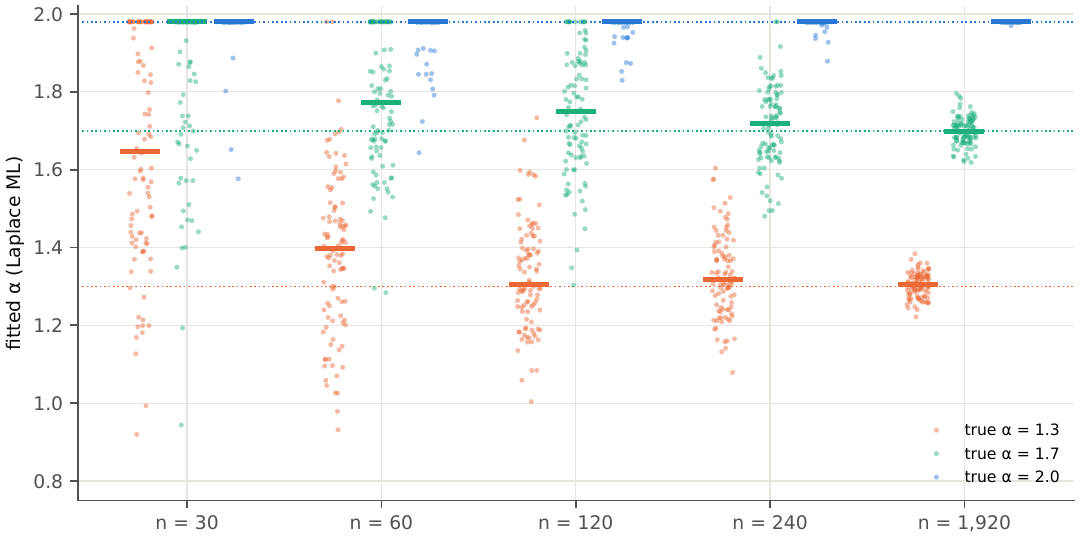}
\caption{Fitted $\alpha$ by sample size, 100 data sets per $n$; bars are medians, dotted lines the truth (for $\alpha_0=2$, the bound $1.98$).}\label{fig:cons}\par\smallskip{\small\noindent Alt text: Strip plot of the fitted tail index against sample size from 30 to 1,920, for true values 1.3, 1.7 and 2. The spread of the fitted values narrows as the sample size grows, and the medians approach the true values; for true value 2 almost all fits lie at the upper bound 1.98.\par}\end{figure}

\textbf{Reading.} Table~\ref{tab:cons} and Figure~\ref{fig:cons} show the results. In contrast to M1, where $\hat\alpha=2$ for every data set by Theorem~\ref{thm:gauss}, the M2 estimate moves toward the truth as $n$ grows. At small $n$ it is biased toward the Gaussian: with 30 points the heavy tail is often not visible in the sample, and $28\%$ ($\alpha_0=1.3$) and $62\%$ ($\alpha_0=1.7$) of the fits sit at the upper bound. By $n=120$ no fit with $\alpha_0=1.3$ is at the bound, and the RMSE falls from $0.45$ at $n=30$ to $0.14$ at $n=120$ and $0.03$ at $n=1{,}920$. With Gaussian noise ($\alpha_0=2$), 86--99\% of the fits are at the bound. The pilot error $\|\hat f-f\|_n$ at the mode falls like $n^{-0.45}$ to $n^{-0.47}$ from $n=30$ to $7{,}680$ for $\alpha_0\in\{1.3,1.7\}$, faster than the $n^{-2/5}$ guaranteed by Theorem~\ref{thm:gppilot}. Only one of the 1{,}800 fits with $\alpha_0<2$ had a fold at its optimum (at $n=60$). Section~\ref{sec:exact} shows that the small-$n$ bias is mostly a property of the likelihood itself.

\begin{figure}[!htb]\centering\includegraphics[width=\textwidth]{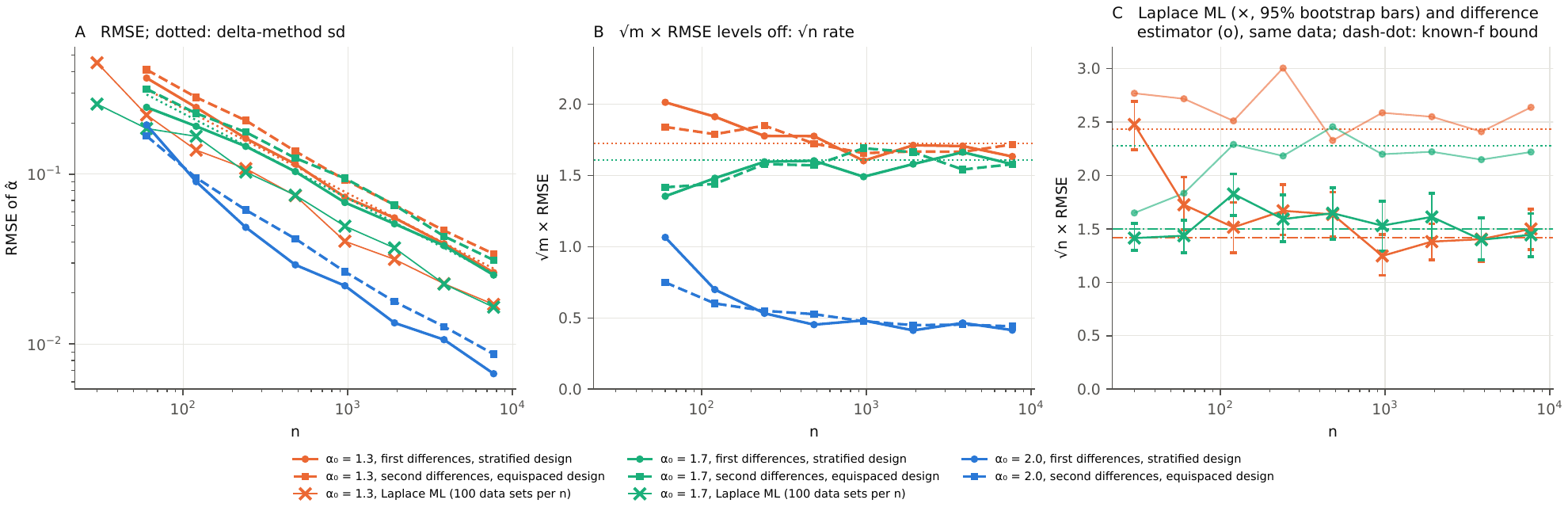}
\caption{Estimators of $\alpha$. A, B: difference-ECF estimator (Theorem~\ref{thm:ecf}), 500 replicates per point; dotted lines are the delta-method standard deviation. Crosses in A: Laplace ML started from $\tilde\alpha_n$, 100 data sets per point. C: $\sqrt n\times$RMSE of Laplace ML (crosses) and of $\tilde\alpha_n$ on the same data sets (circles). Dash-dot: known-$f$ bound; dotted: delta-method value for $\tilde\alpha_n$.}\label{fig:rate}\par\smallskip{\small\noindent Alt text: Three panels with sample size from about 30 to 8,000 on a log scale. Panel A shows the root mean squared error of the difference-based estimator decreasing linearly on the log-log scale for tail indices 1.3, 1.7 and 2. Panel B shows that this error, multiplied by the square root of the number of differences, levels off; for tail indices 1.3 and 1.7 it levels off near the delta-method values. Panel C shows that the error of Laplace maximum likelihood, multiplied by the square root of the sample size, is close to the known-function bound and below that of the difference-based estimator on the same data sets.\par}\end{figure}

\textbf{Rate study for the difference estimator.} Figure~\ref{fig:rate} uses the same $f$ and noise with $n$ from 60 to 7{,}680 and 500 replicates per cell. First differences are used on stratified designs and second differences on equispaced designs. The numbers below are for first differences. From $n\approx480$ onwards, $\sqrt m\times$RMSE is flat and within about 10\% of the delta-method value of Theorem~\ref{thm:ecf}, which is 1.72 for $\alpha_0=1.3$ and 1.61 for $\alpha_0=1.7$. The RMSE falls from $0.247$ at $n=120$ to $0.026$ at $n=7{,}680$ for $\alpha_0=1.3$, and the bias, which at small $n$ comes mostly from the signal remainder, vanishes ($+0.089$ at $n=120$, $+0.004$ at $n=1{,}920$).

For Gaussian noise the estimate is clipped at 2, which halves its mean squared error. For $n\ge1{,}920$ the observed $\sqrt m\times$RMSE is $0.41$--$0.47$, against $\sqrt{v/2}\approx0.42$.

The difference estimator costs $O(n)$ operations: about 0.25\,ms at $n=7{,}680$ and 3\,ms at $n=10^5$. So it suits large data sets, starting values, and checks on the likelihood-based estimate.

\textbf{Rate of the ML estimate.} The dense Laplace fit costs $O(n^3)$ per evaluation. For the Mat\'ern-$5/2$ kernel on one input, $f$ is the first coordinate of a three-dimensional linear stochastic differential equation. So $(K^{-1}+\mathrm{diag}\,c)^{-1}b$ and $\log|I+K\,\mathrm{diag}\,c|$ follow from a Kalman filter and Rauch--Tung--Striebel smoother in $O(n)$ operations \citep{HartikainenSarkka2010}. Both are block Gaussian elimination, so they stay exact when some $c_i<0$, as long as no pivot vanishes. Newton's method for the mode needs only these two operations, with $f^\top K^{-1}f$ tracked through $K^{-1}f=b-\mathrm{diag}(c)\,f$. On the data sets used for testing at $n=60$ and $240$ this agrees with the dense code to within $6\times10^{-4}$ nats; the difference comes from the jitter $10^{-8}I$ that the dense code adds to $K$. A fit at $n=7{,}680$ takes about 25\,s on one core.

The fits of Table~\ref{tab:cons} extend to $n$ from 30 to 7{,}680, with 100 data sets per cell. Here, unlike the general implementation of Section~\ref{sec:alg}, the density is interpolated cubically in $\alpha$, so that the likelihood is smooth in $\alpha$. The benchmark is the known-$f$ bound. If $f$ were known, $(\alpha,\log\sigma)$ would be estimated from $n$ iid residuals, with asymptotic standard deviation $\{[I^{-1}]_{\alpha\alpha}/n\}^{1/2}$ for the Fisher information $I$ \citep{DuMouchel1973}. Numerically, $[I^{-1}]_{\alpha\alpha}^{1/2}=1.42$ at $\alpha=1.3$ and $1.50$ at $\alpha=1.7$. A Monte Carlo check of the iid ML ($n=4{,}000$, 300 replicates, standard error about 4\%) gave 1.47 and 1.60.

The results are in Figure~\ref{fig:rate}C.
\begin{enumerate}[label=(\alph*),leftmargin=2.2em]
\item \emph{Rate.} The RMSE of the ML estimate scales approximately as $n^{-1/2}$ over the simulated range: $\sqrt n\times$RMSE stays between 1.25 and 1.67 for $n$ from 240 to 7{,}680 (1.52 and 1.83 at $n=120$). Its RMSE at $n=7{,}680$ is $0.017$ ($\alpha_0=1.3$) and $0.016$ ($\alpha_0=1.7$). The bias is at most $0.005$ in absolute value for $n\ge480$, against $+0.027$ ($\alpha_0=1.3$) and $+0.049$ ($\alpha_0=1.7$) at $n=120$.
\item \emph{Efficiency.} For $n\ge480$, $\sqrt n\times$RMSE is on average within 2\% of the known-$f$ bound (ratios $0.88$--$1.15$; ratios below 1 reflect Monte Carlo error; bootstrap standard errors of $\sqrt n\times$RMSE are 5--8\% of the value with 100 data sets). The 95\% bootstrap interval contains the bound in 9 of these 10 cells; the exception is $\alpha_0=1.3$, $n=480$, with interval $[1.43,1.84]$.
\item \emph{Against the difference estimator.} On the same data, $\tilde\alpha_n$ has mean squared error larger by a factor of 2.0--4.3 for $\alpha_0=1.3$ and 1.9--2.4 for $\alpha_0=1.7$ ($n\ge480$). Asymptotically the factors are $(2.44/1.42)^2=2.9$ and $(2.28/1.50)^2=2.3$, where $2.44=\sqrt2\times1.72$ and $2.28=\sqrt2\times1.61$ convert the delta-method values above from $\sqrt m$ to $\sqrt n$ scale ($m=n/2$).
\item \emph{Starting value.} The fixed start gave the larger likelihood in 12 and 8 of 100 data sets at $n=30$ and $60$ ($\alpha_0=1.3$; 6, 4 and 3 at $n=30$, 60, 120 for $\alpha_0=1.7$), by up to $5.2$ nats and with $\hat\alpha$ changing by up to $0.78$. For $n\ge240$ the two starts gave the same $\hat\alpha$ to within $0.004$ in every data set. So at small $n$ several starts are needed; at large $n$ the start $\tilde\alpha_n$ suffices.
\end{enumerate}
Why can the bound be reached when $f$ is unknown? The scores for $\alpha$ and $\sigma$ are even functions of the residual, so errors in $f$ affect $\hat\alpha$ only at second order. Theorem~\ref{thm:eff} makes this precise for a cross-fitted plug-in estimator, and Theorem~\ref{thm:lapeff} for the Laplace marginal-likelihood estimate, under conditions on the data and the parameter box that these fits meet.

\textbf{The cross-fitted estimator.} Table~\ref{tab:cf} checks Theorem~\ref{thm:eff} on fresh data sets from the same generator. With the GP pilot, which is covered by Corollary~\ref{cor:gppilot}, the cross-fitted estimator is close to the oracle with $f$ known from $n=1{,}920$ on. Its $\sqrt n\times$RMSE is $1.56$ and $1.47$ against the oracle's $1.46$ and $1.49$ at $n=1{,}920$ and $7{,}680$ for $\alpha_0=1.3$, and $1.47$ and $1.63$ against $1.49$ and $1.63$ for $\alpha_0=1.7$. Its Wald intervals cover $\alpha_0$ in 90--97\% of data sets. The pilot's own error behaves as Theorem~\ref{thm:gppilot} allows: the median of $\|\Delta\|_n$ falls from $0.038$ at $n=120$ to $0.0055$ at $n=7{,}680$ ($\alpha_0=1.3$), a log-log slope of $-0.47$ ($-0.46$ for $\alpha_0=1.7$), faster than the guaranteed $n^{-2/5}$; the sup-norm error falls at the same rate. None of the $1{,}600$ GP-pilot fits had a fold. Newton's method started from the true $f$ reached a different mode (sup-distance above $10^{-4}$) in 11 of 400 pilot fits at $n=120$, in 1 of 400 at $n=480$, and in none at $n\ge1{,}920$, in line with the uniqueness in Proposition~\ref{prop:infill}.

The running-median pilot (Lemma~\ref{lem:median}) converges more slowly. Its bias toward $\alpha=2$ falls from $+0.22$ at $n=120$ to $+0.013$ at $n=7{,}680$ ($\alpha_0=1.3$), but at $n=7{,}680$ its $\sqrt n\times$RMSE is still 17--26\% above the oracle's, and its coverage is $0.89$--$0.90$. This test function is demanding for a local smoother: its Lipschitz constant is about 240 times the noise scale, so the second-order term $\sqrt n\,\|\Delta\|_n^2$ of the proof vanishes only slowly; at $n=7{,}680$ the median of $\|\Delta\|_n$ is $0.013$, more than twice the GP pilot's. The marginal-likelihood estimate (last $\sqrt n\times$RMSE column) attains the bound at the same rate as the GP-pilot estimator, as Theorem~\ref{thm:lapeff} predicts.

\begin{table}[!htb]\centering\renewcommand{\baselinestretch}{1}\footnotesize\setlength{\tabcolsep}{3.5pt}
\begin{tabular}{@{}lrrcccccccc@{}}\toprule
& & & \multicolumn{4}{c}{$\sqrt n\times$RMSE of $\hat\alpha$} & \multicolumn{3}{c}{Wald 95\% coverage}\\\cmidrule(lr){4-7}\cmidrule(l){8-10}
$\alpha_0$ & $n$ & $k_n$ & median pilot & GP pilot & $f$ known & Laplace ML & median & GP & $f$ known\\\midrule
% --- table body (generated, inlined from tab_cf.tex) ---
1.3 & 120 & 5 & 2.85 & 2.04 & 1.48 & 1.52 & 0.65 & 0.86 & 0.94\\
1.3 & 480 & 11 & 2.79 & 1.75 & 1.37 & 1.64 & 0.64 & 0.91 & 0.95\\
1.3 & 1{,}920 & 25 & 2.33 & 1.56 & 1.46 & 1.38 & 0.79 & 0.90 & 0.95\\
1.3 & 7{,}680 & 59 & 1.88 & 1.47 & 1.49 & 1.50 & 0.89 & 0.97 & 0.92\\
\addlinespace
1.7 & 120 & 5 & 1.78 & 1.74 & 1.47 & 1.83 & 0.74 & 0.73 & 0.86\\
1.7 & 480 & 11 & 1.85 & 1.56 & 1.51 & 1.65 & 0.85 & 0.93 & 0.95\\
1.7 & 1{,}920 & 25 & 1.69 & 1.47 & 1.49 & 1.61 & 0.91 & 0.96 & 0.94\\
1.7 & 7{,}680 & 59 & 1.90 & 1.63 & 1.63 & 1.44 & 0.90 & 0.94 & 0.95\\
% --- end of table body ---

\bottomrule
\end{tabular}
\caption{The cross-fitted plug-in estimator of Theorem~\ref{thm:eff}, 100 fresh data sets per row. Median pilot: Lemma~\ref{lem:median} with $k_n$ the smallest odd integer that is at least $\max(5,\lceil0.15\,n^{2/3}\rceil)$. GP pilot: posterior mode of the GP with S$\alpha$S noise, fitted by Laplace ML on the other fold after median/MAD standardization, with working $\alpha\le1.98$ (Corollary~\ref{cor:gppilot}). $f$ known: the oracle ML estimator. Laplace ML: the marginal-likelihood estimate of Table~\ref{tab:cons} and Section~\ref{sec:cons}, on that study's own 100 data sets per row, which differ from the data sets of the other columns. The known-$f$ bound is $1.42$ ($\alpha_0=1.3$) and $1.50$ ($\alpha_0=1.7$); Monte Carlo standard errors, from the bootstrap over data sets, are $0.09$--$0.16$ for each $\sqrt n\times$RMSE ($5$--$8\%$ of its value); the binomial standard error $\{p(1-p)/100\}^{1/2}$ of a coverage $p$ is $0.02$--$0.05$ in this table.}\label{tab:cf}
\end{table}

\subsection{How accurate is the Laplace approximation?}\label{sec:exact}
We compared three objectives for $(\sigma_f,\ell,\sigma,\alpha)$ on the first 50 data sets of Section~\ref{sec:cons} at $n=30$ and $60$ for each $\alpha_0\in\{1.3,1.7,2\}$, under the same protocol (median/MAD standardization, $\alpha\le1.98$, the Gaussian model fitted separately).
\begin{enumerate}[label=(\alph*),leftmargin=2.2em]
\item \emph{Laplace}, as in Section~\ref{sec:alg}.
\item \emph{A variational lower bound.} Because $u\mapsto\log p_\alpha(\sqrt u)$ is convex for every normal scale mixture \citep{Palmer2005}, $\log p_\alpha(z)\ge\log p_\alpha(\xi)-\tfrac12w(\xi)(z^2-\xi^2)$ with $w=\psi(\xi)/\xi$. Integrating this Gaussian lower bound over $f$ gives a closed-form bound on $\log p(y)$. It is maximized over $\xi$ by EM, $\xi_i^2=\E_q z_i^2$. It is exact at $\alpha=2$ and never exceeds the true value, so it also checks the Monte Carlo estimates below.
\item \emph{The importance-sampling (IS) likelihood}, a Monte Carlo estimate of the marginal likelihood; it is not exact, and we call it the IS likelihood throughout. The proposal is a defensive mixture of multivariate $t_4$ laws centered at the Laplace mode and at the variational posterior. Weights are truncated at $\bar w\sqrt S$ \citep{Ionides2008}, and common random numbers make the estimate smooth enough to optimize. Each optimum was re-evaluated with an independent run of $S=40{,}000$. At every one of the 900 reported optima, the delta-method Monte Carlo standard error of the estimate is at most $0.05$ nats, and an independent estimate with $S=160{,}000$ draws differs by at most $0.19$ nats (median $0.007$). The effective sample size of the weights is above 100 at 95\% of the IS optima; at the few optima where it is small, the delta-method error understates the uncertainty, but the independent re-estimate still agrees. At the IS and Laplace optima we also ran annealed importance sampling (AIS; \citealp{Neal2001}) from a Gaussian-noise posterior with elliptical slice moves \citep{Murray2010}, with 64 chains and 1{,}000 temperatures; it does not use the Laplace mode. Wherever its own effective sample size was at least half the chains (425 of 600 runs), AIS agrees with IS to within $0.35$ nats (median $-0.003$). Where AIS degenerates it usually falls below IS, as a logarithm of collapsed AIS weights is biased downwards, so these runs do not indicate an IS error.
\end{enumerate}

\begin{table}[!htb]\centering\renewcommand{\baselinestretch}{1}\small
\begin{tabular}{@{}llccccc@{}}\toprule
& & \multicolumn{3}{c}{median $\hat\alpha$ (RMSE)} & \multicolumn{2}{c}{reject $\alpha=2$}\\\cmidrule(lr){3-5}\cmidrule(l){6-7}
$\alpha_0$ & $n$ & Laplace & VB bound & IS & Laplace & IS\\\midrule
1.3 & 30 & 1.65 (0.42) & 1.77 (0.47) & 1.45 (0.33) & 0.66 & 0.68\\
1.3 & 60 & 1.44 (0.23) & 1.57 (0.31) & 1.37 (0.22) & 0.94 & 0.94\\
1.7 & 30 & 1.98 (0.24) & 1.98 (0.25) & 1.98 (0.29) & 0.32 & 0.30\\
1.7 & 60 & 1.77 (0.17) & 1.87 (0.20) & 1.74 (0.18) & 0.52 & 0.56\\
2.0 & 30 & 1.98 (0.12) & 1.98 (0.03) & 1.98 (0.11) & 0.04 & 0.02\\
2.0 & 60 & 1.98 (0.11) & 1.98 (0.04) & 1.98 (0.12) & 0.06 & 0.04\\\bottomrule
\end{tabular}
\caption{Laplace, variational bound and IS ML. ``Reject $\alpha=2$'': share of data sets in which the likelihood-ratio statistic against the Gaussian-noise GP exceeds $2.71$, the nominal 5\% point of $\tfrac12\chi^2_0+\tfrac12\chi^2_1$. This nominal calibration is not justified at $\alpha=2$ (finding (iv)), so these are rejection frequencies; Table~\ref{tab:boot} gives the bootstrap-calibrated size and power. 50 data sets per row.}\label{tab:exact}
\end{table}

\begin{figure}[!htb]\centering\includegraphics[width=\textwidth]{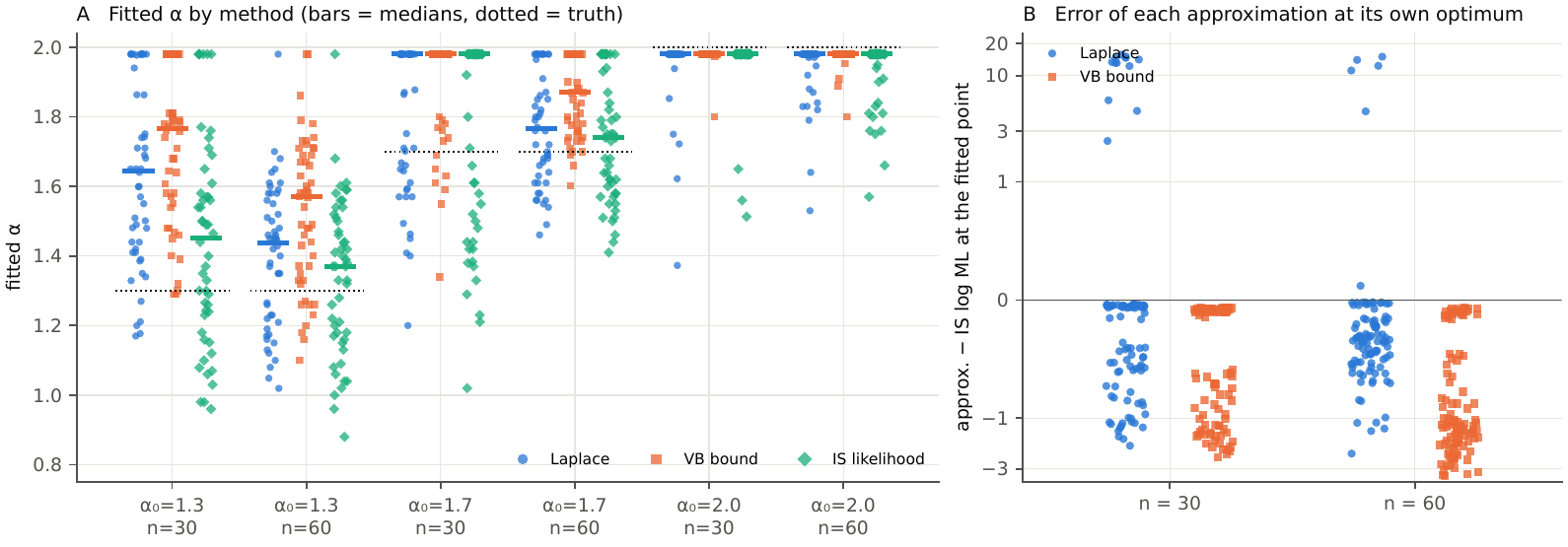}
\caption{A: fitted $\alpha$ under the three objectives. B: the approximation minus the IS estimate of the log marginal likelihood, at the point where each method's optimizer stopped (symmetric log scale).}\label{fig:exact}\par\smallskip{\small\noindent Alt text: Panel A is a strip plot of the fitted tail index under the Laplace approximation, the variational bound and importance sampling, for true values 1.3, 1.7 and 2 at sample sizes 30 and 60; the variational bound is pulled toward 2. Panel B shows, on a symmetric log scale, the error of the Laplace and variational approximations to the log marginal likelihood at their optima; the variational errors are all negative and a few Laplace errors are large and positive.\par}\end{figure}

\textbf{Findings} (Table~\ref{tab:exact}, Figure~\ref{fig:exact}).
(i) \emph{Most of the small-sample bias toward $\alpha=2$ is in the likelihood, not the approximation.} IS ML is also biased at $n=30$ (median $1.45$ for $\alpha_0=1.3$). Laplace adds $0.07$--$0.20$ at $\alpha_0=1.3$, and $0.03$ at $\alpha_0=1.7$, $n=60$. On Gaussian data all three medians are at the bound. The Laplace column differs slightly from Table~\ref{tab:cons} because of the subset and the optimizer (three Nelder--Mead starts on the dense code here).
(ii) \emph{Laplace occasionally fails badly.} In 20 of the 300 data sets, optimization of the Laplace criterion was drawn to the neighborhood of a fold of the posterior mode (Section~\ref{sec:fold}), where the criterion diverges: the optimizer stopped where the Laplace value exceeded the IS value by more than 3 nats (up to $15.7$), a number set by solver tolerance rather than by the model. At all 20 optima $\lambda_{\min}(I+K^{1/2}WK^{1/2})\le1.2\times10^{-4}$ (zero to working precision in 15), whereas $\lambda_{\min}\ge0.0026$ at every other Laplace optimum (\texttt{diag\_fold\_c2.py}). The failures are a small-$n$ phenomenon: 14 of the 20 are at $n=30$. \ifssbuild Supplementary Section~\ref{supp:fold}\else Section~\ref{sec:fold}\fi{} analyzes the three failures of an earlier, smaller run (64 data sets, standardized by the sample mean and standard deviation) in detail. Laplace approximations require care for non-log-concave likelihoods, where the posterior can be multimodal and some $W_{ii}<0$ \citep{Vanhatalo2009,Jylanki2011}. For accuracy problems of Laplace in GP classification, see \citet{KussRasmussen2005}. In the other 184 data sets with $\alpha_0<2$, Laplace understates the IS value (median error $-0.35$ nats). Over the 200 data sets with $\alpha_0<2$, using the Laplace optimum instead of the IS one costs a median of $0.08$ nats of IS log-likelihood, and at most $3.5$.
(iii) \emph{The variational bound is valid but biased toward the GP.} It never exceeded the IS value (largest difference $-0.06$ nats). Its gap grows as $\alpha$ falls, so it favors large $\alpha$: median $1.87$ when $\alpha_0=1.7$, $n=60$. It is useful as a certificate, not as an estimator.
(iv) \emph{Comparing $\alpha<2$ with the GP.} At the nominal cut-off of Table~\ref{tab:exact}, the IS likelihood ratio rejected $\alpha=2$ in $0.68$--$0.94$ of data sets at $\alpha_0=1.3$, in $0.30$--$0.56$ at $\alpha_0=1.7$, and in 3 of 100 Gaussian data sets. These are not sizes and powers. The stable family is non-regular at $\alpha=2$: computed as for the known-$f$ bound in Section~\ref{sec:cons}, the Fisher information for $\alpha$ is $0.50$, $0.89$, $1.43$ and $3.44$ at $\alpha=1.7$, $1.9$, $1.95$ and $1.985$, and appears to grow without bound. The latent GP and its hyperparameters are nuisance parameters. We therefore test
\begin{gather*}
H_0:\ \text{Gaussian noise (S$\alpha$S with }\alpha=2\text{)}\quad\text{against}\\ H_1:\ \text{S$\alpha$S noise with }\alpha\in[0.8,1.98],
\end{gather*}
with $\mathrm{LR}=2\{\sup_{H_1}\ell_{\rm Lap}-\sup_{H_0}\ell_{\rm G}\}$, where $\ell_{\rm Lap}$ is the Laplace log marginal likelihood of the S$\alpha$S-noise model and $\ell_{\rm G}$ the exact log marginal likelihood of the Gaussian-noise GP (to which Laplace reduces at $\alpha=2$). The working alternative is separated from the null, not nested: the upper bound $1.98$ is part of the definition of the test, $H_1$ does not contain $H_0$, and $\mathrm{LR}$ can be negative. So the boundary limit $\tfrac12\chi^2_0+\tfrac12\chi^2_1$ cannot be invoked. We calibrate $\mathrm{LR}$ by a parametric bootstrap under the Gaussian-noise GP fitted to the data. For a single data set, one draws $B$ bootstrap data sets (e.g.\ $B=499$) from the fitted GP, refits both models to each, and rejects when $\mathrm{LR}$ exceeds the $1-a$ quantile of the bootstrap statistics (\texttt{lr\_test\_bootstrap.py}; about 6 minutes on one core at $n=60$ with $B=499$). On the first data set of Table~\ref{tab:cons} at $n=60$ this gave $p=0.002$ for $\alpha_0=1.7$ and $p=0.10$ for $\alpha_0=2$. The validity of this bootstrap in a non-regular, separated comparison is not covered by our theory; we check it by Monte Carlo. To estimate its size and power we use the warp-speed method of \citet{Giacomini2013}: one bootstrap data set per Monte Carlo data set, the bootstrap critical value being the quantile of the bootstrap statistics within a cell (\texttt{lr\_bootstrap.py}). Table~\ref{tab:boot} gives the results on the data sets of Section~\ref{sec:cons}. The bootstrap test holds its level: its rejection rate under Gaussian noise is $3.8$--$6.4\%$ at the 5\% level (Monte Carlo standard error about $1\%$). The nominal cut-off $2.71$ is very conservative, rejecting in $0.8$--$2.6\%$, because the likelihood-ratio statistic is mostly near or below zero under the null; the bootstrap critical values are $0.03$--$1.00$. The bootstrap test is correspondingly more powerful at small $n$: at $\alpha_0=1.7$ it rejects in $37\%$ and $70\%$ of data sets at $n=30$ and $60$, against $26\%$ and $58\%$ for the nominal cut-off. Near $\alpha=2$ the likelihood is flat: in the 17 Gaussian data sets of Table~\ref{tab:exact} in which IS ML returned $\hat\alpha<1.975$, the gain over the GP was at most $1.9$ nats and below $0.9$ nats in 14.

\begin{table}[!htb]\centering\renewcommand{\baselinestretch}{1}\small
\begin{tabular}{@{}rccccccc@{}}\toprule
& \multicolumn{3}{c}{$\alpha_0=2$ (size)} & \multicolumn{2}{c}{$\alpha_0=1.7$ (power)} & \multicolumn{2}{c}{$\alpha_0=1.3$ (power)}\\\cmidrule(lr){2-4}\cmidrule(lr){5-6}\cmidrule(l){7-8}
$n$ & boot.\ crit. & bootstrap & nominal & bootstrap & nominal & bootstrap & nominal\\\midrule
30 & 0.03 & 0.042 & 0.026 & 0.37 & 0.26 & 0.64 & 0.58\\
60 & 0.32 & 0.064 & 0.014 & 0.70 & 0.58 & 0.96 & 0.94\\
120 & 0.93 & 0.048 & 0.018 & 0.86 & 0.81 & 1.00 & 1.00\\
240 & 1.00 & 0.038 & 0.008 & 0.98 & 0.97 & 1.00 & 1.00\\
\bottomrule
\end{tabular}
\caption{Likelihood-ratio test at the 5\% level of $H_0$: Gaussian noise ($\alpha=2$) against the separated working alternative $H_1$: S$\alpha$S noise with $\alpha\in[0.8,1.98]$ (Laplace likelihood ratio). Bootstrap: parametric bootstrap under the fitted GP, size and power estimated by the warp-speed method \citep{Giacomini2013}; ``boot.\ crit.'' is the bootstrap 5\% critical value under Gaussian noise. Nominal: cut-off $2.71$, the 5\% point of $\tfrac12\chi^2_0+\tfrac12\chi^2_1$. 500 data sets per cell for $\alpha_0=2$, 300 otherwise.}\label{tab:boot}
\end{table}

\emph{Recommendation.} Fit by Laplace and check the fold diagnostic $\lambda_{\min}$ of Section~\ref{sec:fold}, which needs one extra eigensolve; in this study the threshold $0.1$ flagged all 20 failures and one of the 280 other fits. Optionally also evaluate the IS likelihood once at the Laplace optimum (about one second at $n=60$). If $\lambda_{\min}<0.1$, or the IS value falls short of the Laplace value by more than one nat, refit on the IS objective.

\subsection{Why Laplace fails: folds of the posterior mode}\label{sec:fold}
Write $\psi_\theta(f)=\sum_i\log p(y_i\mid f_i)+\log\varphi(f;0,K_\theta)$ for the log posterior kernel, $W=-\nabla^2\sum_i\log p(y_i\mid f_i)$ (diagonal), and $H_\theta=K_\theta^{-1}+W$. The Laplace value at a local mode $\hat f_\theta$ is
\begin{gather*}
L_{\rm Lap}(\theta)=\log p(y\mid\hat f_\theta)-\tfrac12\hat f_\theta^\top K_\theta^{-1}\hat f_\theta-\tfrac12\log|B_\theta|,\\
B_\theta=I+K_\theta^{1/2}WK_\theta^{1/2}=K_\theta^{1/2}H_\theta K_\theta^{1/2},
\end{gather*}
and $B_\theta\succ0$ at a nondegenerate local mode. A \emph{fold} (saddle-node) is a hyperparameter value $\theta^\ast$ at which a local mode meets a saddle and disappears, so that $\lambda_{\min}(B_\theta)\to0$.

\begin{proposition}[Laplace diverges at a fold]\label{prop:fold}
Let $\theta_k\to\theta^\ast$ with $\hat f_{\theta_k}$ nondegenerate local modes, $\hat f_{\theta_k}\to f^\ast$, and $\lambda_{\min}(B_{\theta_k})\to0$. Then $L_{\rm Lap}(\theta_k)\to+\infty$, while $\log p(y\mid\theta_k)\to\log p(y\mid\theta^\ast)\in\R$.
\end{proposition}
\begin{proof}[Proof of Proposition~\ref{prop:fold}]
The first two terms of $L_{\rm Lap}$ are jointly continuous in $(f,\theta)$ and converge to a finite limit. $W$ is bounded, so the eigenvalues of $B_{\theta_k}$ are bounded above, and $\log|B_{\theta_k}|\le\log\lambda_{\min}(B_{\theta_k})+(n-1)\log\lambda_{\max}(B_{\theta_k})\to-\infty$.

The exact evidence is
\[
p(y\mid\theta)=\int\prod_ip(y_i\mid f_i)\,\varphi(f;0,K_\theta)\,df .
\]
It is positive because $p_\alpha>0$. On a neighborhood $U$ of $\theta^\ast$ where $K_\theta$ is uniformly positive definite and $\sigma$ is bounded away from $0$, the integrand is bounded by $C\exp(-|f|^2/2\Lambda)$ with $\Lambda=\sup_U\lambda_{\max}(K_\theta)$, using $p_\alpha\le p_\alpha(0)$ (Proposition~\ref{prop:moments}). Since $p_\alpha(z)$ is continuous in $(z,\alpha)$, dominated convergence gives continuity in $\theta$.
\end{proof}

So along the branch of modes that the inner solver tracks, the Laplace objective is unbounded above near any fold the hyperparameters can reach, and a local optimizer can climb toward the fold. For an isolated saddle-node crossed transversally, $\lambda_{\min}$ vanishes like the square root of the distance to the fold, so the error grows like $-\tfrac14\log(\text{distance})$; the one-observation example below behaves in this way. Along the paths to the three failing optima of an earlier, smaller run (64 data sets), $\lambda_{\min}$ decays faster, like (distance)$^{0.64\text{--}0.74}$, so the error grows like $-(0.32\text{--}0.37)\log(\text{distance})$. In both cases the divergence is only logarithmic in the distance, but it is unbounded. So on that branch the Laplace criterion has no finite maximum. Below, ``Laplace optimum'' means the point at which the optimizer stopped, and the size of the error there reflects solver tolerance. That Laplace approximations need care for non-log-concave likelihoods is well known: the posterior can be multimodal with some $W_{ii}<0$ \citep{Vanhatalo2009,Jylanki2011}, and \citet{HartmannVanhatalo2019} address the lack of concavity in a heteroscedastic Student-$t$ model with natural-gradient mode finding and a Laplace--Fisher approximation, which uses the Fisher information in place of the Hessian. What we add is the characterization of the failures as folds of the posterior mode (Proposition~\ref{prop:fold}), the sufficient condition for no fold (Proposition~\ref{prop:nofold}) and the eigenvalue diagnostic below.

\begin{proposition}[A sufficient condition for no fold]\label{prop:nofold}
Let $\ell_\alpha=\log p_\alpha$ and $c_\alpha=\sup_z\ell_\alpha''(z)$, so that $W_i=-\ell_\alpha''(z_i)/\sigma^2\ge-c_\alpha/\sigma^2$ with $z_i=(y_i-f_i)/\sigma$. Let $N=\{i:W_i<0\}$, $W_+=\mathrm{diag}(\max(W_i,0))$ and $\Sigma=(K^{-1}+W_+)^{-1}$. If
\[
\rho=\frac{c_\alpha}{\sigma^2}\,\lambda_{\max}(\Sigma_{NN})<1,
\]
then $H=K^{-1}+W\succ0$, so a mode at which $\rho<1$ cannot be at a fold.
\end{proposition}
\begin{proof}[Proof of Proposition~\ref{prop:nofold}]
$H=\Sigma^{-1}-D$ with $D=\mathrm{diag}(\max(-W_i,0))$, which is supported on $N$ and satisfies $D\preceq(c_\alpha/\sigma^2)I_N$. Now $H\succ0$ if and only if $I-\Sigma^{1/2}D\Sigma^{1/2}\succ0$. The nonzero eigenvalues of $\Sigma^{1/2}D\Sigma^{1/2}$ are those of $D_N^{1/2}\Sigma_{NN}D_N^{1/2}$, which are at most $(c_\alpha/\sigma^2)\lambda_{\max}(\Sigma_{NN})=\rho<1$. This proves the result.
\end{proof}

Folds therefore need residuals in the convex tail of $\ell_\alpha$, i.e.\ $|z_i|$ beyond the inflection point $z^\ast_\alpha$. Numerically $z^\ast_\alpha=1.00,1.73,2.25,2.84,3.68$ at $\alpha=1,1.3,1.5,1.7,1.9$, and there is none for Gaussian noise. Folds also need $f$ at those points to be poorly determined by the rest of the data, with $\lambda_{\max}(\Sigma_{NN})\ge\sigma^2/c_\alpha$. We find $c_\alpha=0.25,0.19,0.20,0.26,0.46$ at the same $\alpha$, so this threshold is about $2$--$5\,\sigma^2$. Since $\Sigma\preceq K$, a cruder sufficient condition is $c_\alpha\lambda_{\max}(K)<\sigma^2$. Then $H\succ0$ for every $f$, the log posterior is strictly concave, and there are no folds at all. Under infill with hyperparameters in a compact set, the global mode is eventually unique and nondegenerate, so its branch has no folds (Proposition~\ref{prop:infill}). The argument does not go through $N$: with $W$ evaluated at the true $f$, $\bar w=\E W_i>0$ and $P=I+\bar wK$, the normalized fluctuation $P^{-1/2}K^{1/2}(W-\bar wI)K^{1/2}P^{-1/2}$ has norm $O_p(n^{-1/3}\sqrt{\log n})$ by a matrix Bernstein inequality, because posterior variances are $O(n^{-2/3})$; the negative $W_i$ are averaged out. In our simulations folds are a sparse-data phenomenon (small $n$, short length-scales, or an outlier in a gap), and the largest $\rho$ falls with $n$ (below).

\begin{figure}[!htb]\centering\includegraphics[width=\textwidth]{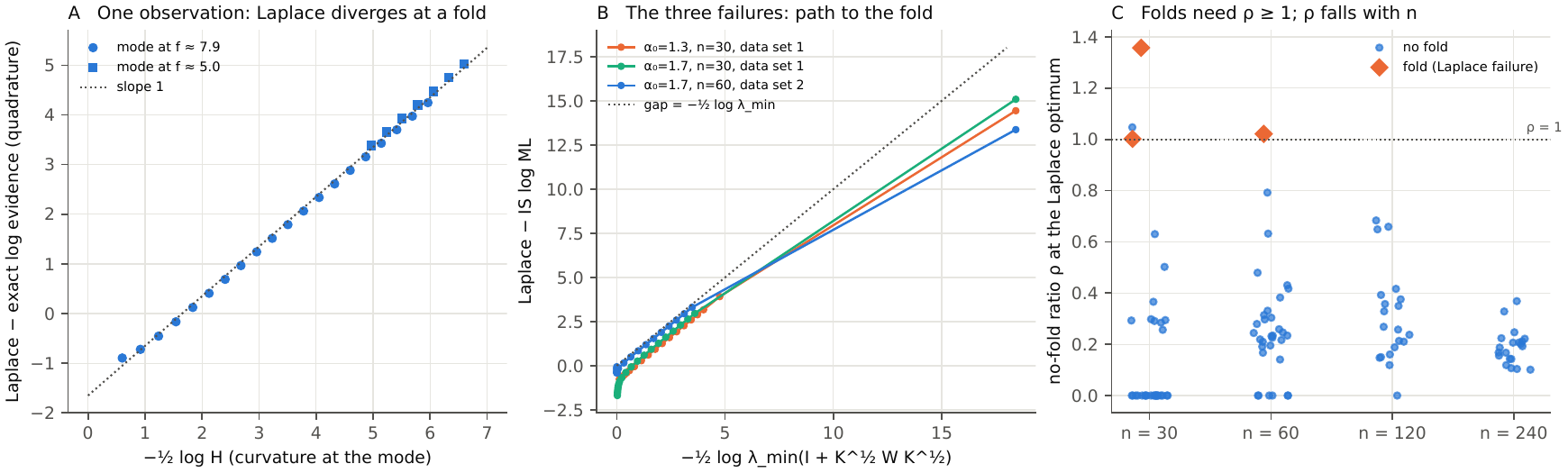}
\caption{Folds explain the Laplace failures. A: a single observation; Laplace minus exact log evidence (by quadrature) against $-\tfrac12\log H$ at the vanishing mode, for the two folds of the posterior. B: the three failing data sets of the earlier run of Section~\ref{sec:exact}, Laplace minus IS log-likelihood along the line from the IS optimum to the point where the Laplace optimizer stopped; at the last point of each curve $\lambda_{\min}$ is zero to working precision and is plotted at $10^{-16}$. C: the no-fold ratio $\rho$ of Proposition~\ref{prop:nofold} at the Laplace optima, by $n$.}\label{fig:fold}\par\smallskip{\small\noindent Alt text: Panel A shows the Laplace error for one observation lying on a line of slope one against minus one half of the log curvature at the vanishing mode. Panel B shows, for the three failing data sets, the Laplace minus importance-sampling log-likelihood growing along the path to the fold, close to the reference line minus one half of the log smallest eigenvalue. Panel C is a strip plot of the no-fold ratio at the Laplace optima by sample size; the three failures have ratio at least one and the ratio falls as the sample size grows.\par}\end{figure}

\textbf{Evidence (Figure~\ref{fig:fold}).} The detailed evidence uses the earlier run of the comparison of Section~\ref{sec:exact}: 64 data sets at $n=30$ and $60$, standardized by the sample mean and standard deviation, with $\alpha\le2$. In the run of Section~\ref{sec:exact}, all 20 failures also sit at folds.
\begin{enumerate}[label=(\alph*),leftmargin=2.2em]
\item \emph{The failures sit at folds.} At the Laplace optima of the earlier run, the three failures have $\lambda_{\min}(B)$ equal to zero to working precision; all 61 other fits have $\lambda_{\min}\ge0.94$. At these points the hyperparameters have just passed the fold, and the Newton iterate stopped next to a mode that no longer exists. So the size of the reported errors (13.4--15.1 nats) is set by solver tolerance; what is intrinsic is the divergence. On the path below, at $1-t=10^{-6}$ from the Laplace optimum, the error is already 3.0--3.9 nats.
\item \emph{The divergence is in the determinant.} Along the straight line from the IS optimum to each failing Laplace optimum, the IS log-likelihood falls smoothly, by 0.4--1.1 nats in all, and over the final approach ($1-t\le10^{-4}$) it is constant to two decimals. Over that approach the Laplace value rises by about $\tfrac12\log2$ each time $\lambda_{\min}$ halves, and $\lambda_{\min}$ falls like (distance)$^{0.64\text{--}0.74}$.
\item \emph{One observation shows the same.} With $f\sim N(0,v)$ and an outlier at $y=10$ ($\alpha=1.3$, $\sigma=1$), the curvature $H$ at the vanishing mode behaves like $|v-v^\ast|^{1/2}$, and the Laplace error tracks $-\tfrac12\log H$ without bound.
\item \emph{The no-fold condition holds.} All three folds have $\rho\ge1$. The largest $\rho$ among the Laplace optima falls with $n$: $1.36$, $1.02$, $0.68$ and $0.37$ at $n=30,60,120,240$. Standard Laplace refits of the earlier Table~\ref{tab:cons} data sets at $n=120$ and $240$ (10 data sets each for $\alpha_0=1.3,1.7$) had no folds ($\lambda_{\min}\ge0.956$). Their median $\hat\alpha$ was $1.27$, $1.34$, $1.75$ and $1.69$, close to Table~\ref{tab:cons}.
\item \emph{BO.} In the BO study of Section~\ref{sec:bostudy}, Laplace exceeded the IS check by more than 5 nats in 17.4\% of the 18{,}200 S$\alpha$S fits (5.4\% of 1{,}840 in the earlier 20-repetition run, which standardized by the mean and standard deviation). On Gaussian data the rate fell with $n$ (20.9\%, 9.6\% and 3.8\% for $n=8$--14, 15--22 and 23--30). With symmetric outliers it did not (21.3\%, 36.2\% and 33.2\%): outliers keep arriving, and some land in gaps. Both patterns are consistent with Proposition~\ref{prop:nofold}, although $\rho$ was not computed in the BO runs. The higher overall rate is plausibly due to median/MAD standardization, which places outliers further into the convex tail of $\ell_\alpha$ in standardized units; we have not verified this.
\end{enumerate}

\textbf{Remedies.}
\begin{enumerate}[label=(\alph*),leftmargin=2.2em]
\item \emph{Detect and refit} (recommended). $\lambda_{\min}(B)$ costs one extra symmetric eigensolve, and it flagged all three failures. Refit on the IS objective when it is small.
\item \emph{A guard} $\lambda_{\min}\ge\tau$ ($\tau=0.1$) repairs the value: at the guarded optima of the three failing data sets, Laplace is within 0.3--1.0 nat of the IS value. But the optimum then sits on the guard boundary, so the estimate is not repaired.
\item \emph{Replacing $W$ by the EM weights} in the determinant (a Fisher-type Laplace approximation; \citealp{HartmannVanhatalo2019} use the expected Fisher information, which is also positive) eliminates the fold-induced determinant singularity in this Fisher-type approximation, since the weights are positive; the underlying branch of posterior modes can still fold. It never exceeded the IS value (at most $+0.01$ nats over the 64 data sets). But, like the variational bound, which uses the same curvature, it pulls $\hat\alpha$ toward 2. Its median $\hat\alpha$ was $1.80$ and $1.60$ for $\alpha_0=1.3$ at $n=30,60$, against $1.51$ and $1.41$ for IS ML. At $n=120$ and $240$ it still exceeded standard Laplace by $0.05$--$0.12$.
\end{enumerate}

\section{Consequences for Bayesian optimization}\label{sec:bo}
In this section, we discuss the consequences of Sections~\ref{sec:m1} and~\ref{sec:m2} for BO, and report a repeated sequential BO study.
\subsection{Process-level surrogates}
By Corollary~\ref{cor:acq}, a TP or sub-Gaussian stable surrogate fitted by ML behaves in BO as follows:
\begin{enumerate}[label=(\alph*),leftmargin=2.2em]
\item it ranks candidates by PI exactly as the GP does;
\item its UCB is GP-UCB with a rescaled constant;
\item its EI is GP-EI with a fixed, data-free inflation in low-$z$ regions, which for the TP, and numerically for sub-Gaussian stable mixing, vanishes as $n$ grows (Remark~\ref{rem:gshape}).
\end{enumerate}
The tail parameter cannot adapt to the data, because its profile likelihood does not depend on them (Corollary~\ref{cor:profile}).

In a companion demonstration with a Mat\'ern-$5/2$ kernel and $\nu$ searched up to 1000, the fitted TP had $\hat\nu=1000$ in every data set. There were 90 synthetic data sets of $n=25$, with Gaussian noise, four gross outliers, or $t_{1.5}$ noise, and 60 NIFTY~50 data sets: noisy evaluations, over random historical windows of 2015--2026 index data, of the annualized excess return of a moving-average trend rule as a function of its look-back (an illustrative benchmark, not a financial application). This is Corollary~\ref{cor:tp} in practice: the profile is increasing in $\nu$ whatever the data.

\subsection{Observation-level surrogates}
By Proposition~\ref{prop:moments}, acquisitions on the latent $f$ are well defined under M2 for every $\alpha$. 
\emph{Companion demonstration for Section~\ref{sec:bo}.} In the companion demonstration we compared the GP with the GP-$t$ (M2 with Student-$t$ noise) on fixed data sets of $n=25$. The rule was to choose the heavy-tailed model when it has tail evidence (fitted $\nu<30$ and a log Bayes factor above 1) and better leave-one-out predictive density, overall and in the region of high posterior mean. The acquisition function was deliberately \emph{not} allowed to veto that choice. On fresh replicates the results were:
\begin{enumerate}[label=(\alph*),leftmargin=2.2em]
\item \emph{Gaussian noise:} the GP-$t$ was never selected.
\item \emph{Four outliers:} it was selected in 18 of 30 data sets. The median regret of the final recommendation was $0.0048$ under the rule, $0.0040$ with the GP-$t$ always, and $0.0297$ with the GP always.
\item \emph{A rule that also let the acquisition function veto the choice:} median regret $0.0150$.
\end{enumerate}

\subsection{A repeated sequential BO study}\label{sec:bostudy}
\textbf{Surrogates.} Five surrogates each run their own BO loop:
\begin{enumerate}[label=(\alph*),leftmargin=2.2em]
\item the GP;
\item the TP, with $\nu$ learned up to 1000;
\item the GP-$t$ (Student-$t$ noise, Laplace ML);
\item an RCGP-type robust GP: the conjugate generalized-Bayes posterior of \citet{Altamirano2024} with inverse-multiquadric weights $w_i=\beta\{1+(y_i-m)^2/c_{\rm R}^2\}^{-1/2}$, constant prior mean $m$ ($m=0$ after median/MAD standardization), $\beta=\sigma/\sqrt2$ with $\sigma$ the Gaussian noise standard deviation (the choice of \citealp{Altamirano2024}, under which $c_{\rm R}\to\infty$ recovers the GP), and $c_{\rm R}$ the empirical $0.95$ quantile of $|y_i-m|$; the kernel and noise hyperparameters are selected by leave-one-out predictive density, while $c_{\rm R}$ is set by the quantile rule, not by leave-one-out;
\item the GP with S$\alpha$S noise, fitted by Laplace ML with $\alpha\in[0.8,2]$ ($\alpha=2$ is the GP), from a warm start and one random start, with the safeguard of Section~\ref{sec:exact}.
\end{enumerate}
In the safeguard, one IS-likelihood evaluation ($S=3000$) is made at the Laplace optimum. If Laplace exceeds it by more than one nat, the model is refitted on the IS objective. (The BO runs used this check only, not the $\lambda_{\min}$ diagnostic.) The GP and the TP standardize the responses by the sample mean and standard deviation; the three robust surrogates use the median and MAD. All use an isotropic Mat\'ern-$5/2$ kernel.

\textbf{Problems.} There are nine maximization problems:
\begin{enumerate}[label=(\alph*),leftmargin=2.2em]
\item \emph{five one-dimensional problems} with $f(x)=x\sin(10x)+\cos(2x)$ on $[0,1.2]$ and noise
\begin{enumerate}[label=(\roman*),leftmargin=2.2em]
\item $0.05\,N(0,1)$ (``Gaussian'');
\item the same, plus a gross error $\pm U(0.5,1)$ with probability $0.15$ (``outliers'');
\item the same, but with the gross error always positive (``one-sided outliers''; asymmetric, so misspecified for every symmetric model);
\item $0.05$ S$\alpha$S$(1.3)$;
\item $0.05\,t_2$ (heavy-tailed and symmetric but not stable, so misspecified for the S$\alpha$S surrogate);
\end{enumerate}
\item \emph{Branin} on $[0,1]^2$ and \emph{Hartmann-6} on $[0,1]^6$, negated and standardized to unit standard deviation under uniform inputs ($f^\ast=1.05$ and $8.00$), with noise $0.1$ S$\alpha$S$(1.3)$;
\item the two constrained problems of \citet{Gardner2014} on $[0,6]^2$:
\begin{enumerate}[label=(\roman*),leftmargin=2.2em]
\item \emph{Gardner~1}: minimize $\cos(2x_1)\cos(x_2)+\sin(x_1)$ subject to\\ $\cos(x_1)\cos(x_2)-\sin(x_1)\sin(x_2)\le0.5$ (feasible set 66\% of the domain);
\item \emph{Gardner~2}: minimize $\sin(x_1)+x_2$ subject to $\sin(x_1)\sin(x_2)\le-0.95$ (feasible set 1.8\% of the domain).
\end{enumerate}
The objectives are negated and standardized as for Branin ($f^\ast=2.32$ and $1.47$), the noise is $0.1$ S$\alpha$S$(1.3)$ on the objective, and the constraint is observed without noise.
\end{enumerate}
An earlier version of the study included an illustrative NIFTY~50 trading-rule benchmark; we replaced it by the Gardner problems before running them. Under the analysis below it showed no significant difference between the S$\alpha$S surrogate and any other (all Holm-adjusted $p=1$ in the earlier 32-test family) (results in the accompanying package).

\textbf{Protocol.} The one-dimensional problems start from 8 stratified points and take 22 EI steps ($n=30$). Branin and the Gardner problems start from a 10-point Latin hypercube and take 30 steps, Hartmann-6 from 20 points and 40 steps. EI is computed on the latent $f$ (Proposition~\ref{prop:moments}), with the incumbent taken as the largest posterior mean at the data. In one dimension EI is maximized over a 401-point grid. In $d>1$ the candidates are 1{,}024 scrambled Sobol points plus 256 Gaussian perturbations (s.d.\ $0.05$) of the four data points with the largest posterior mean, redrawn at every step (for the constrained problems, the four feasible data points; perturbations are on the unit-cube scale). After each step the recommendation is the maximizer of the posterior mean over the candidates and the data, and we record its simple regret $f^\ast-f(\hat x)$. For the constrained problems the constraint is modeled by a separate GP, the same for all five surrogates, and the acquisition is constrained EI, EI times the posterior probability of feasibility \citep{Gardner2014}, with the incumbent the largest posterior mean at the feasible data; until a feasible point has been evaluated the acquisition is the probability of feasibility. The recommendation is the evaluated feasible point with the largest posterior mean, and until one exists the regret is that of the worst feasible point. Initial designs, noise and candidate sets use common random numbers across surrogates, so the comparisons are paired by repetition. There are 100 repetitions for each one-dimensional problem and 50 for each multivariate problem; these numbers were set by the compute budget.

\textbf{Analysis.} The analysis was specified in the summary script before the results were tabulated.
\begin{enumerate}[label=(\alph*),leftmargin=2.2em]
\item \emph{Primary outcome:} the simple regret after the last step.
\item \emph{Tests:} the S$\alpha$S surrogate against each of the other four on every problem, by the two-sided Wilcoxon signed-rank test on the paired differences (zero differences dropped). The 36 $p$-values are Holm-adjusted at family-wise level $0.05$.
\item \emph{Effect sizes:} the median paired difference (S$\alpha$S minus the other) with a 95\% paired-bootstrap percentile interval.
\item \emph{Tail risk:} $\mathrm{VaR}_{0.9}$ and $\mathrm{CVaR}_{0.9}$ (the mean of the worst 10\%) of the final regret over repetitions.
\item \emph{Robustness check (added after the primary analysis was fixed):} the exact paired sign test on wins against losses, Holm-adjusted over the same family, because final-regret differences are skewed and tied and need not be symmetric about their median (Supplementary Table~\ref{tab:signtest}).
\item \emph{Equivalence on Gaussian noise:} two one-sided tests at level $0.05$ for the mean paired difference with the GP, margin $\delta=0.025$ (about 1\% of the range of $f$); equivalence is declared if the 90\% paired-bootstrap interval lies inside $(-\delta,\delta)$.
\end{enumerate}
These results replace the 20-repetition study of an earlier draft. That study suggested a significant gain under S$\alpha$S noise ($p=0.009$, unadjusted); it does not survive here.

\begin{table}[!htb]\centering\renewcommand{\baselinestretch}{1}\scriptsize
\setlength{\tabcolsep}{1.5pt}
\begin{tabular*}{\linewidth}{@{\extracolsep{\fill}}lccccc@{}}\toprule
\multicolumn{6}{@{}l}{(a) Median final regret ($\mathrm{CVaR}_{0.9}$ of the final regret); bold: smallest median}\\
problem & GP & TP & GP-$t$ & RCGP-type & GP + S$\alpha$S noise\\\midrule
% --- table body (generated, inlined from tab_bo_a.tex) ---
Gaussian noise & \textbf{0.0010} (0.019) & \textbf{0.0010} (0.017) & 0.0027 (0.029) & 0.0017 (0.016) & 0.0018 (0.019)\\
15\% gross outliers & 0.0455 (0.117) & 0.0297 (0.136) & \textbf{0.0027} (0.086) & 0.0239 (0.144) & \textbf{0.0027} (0.107)\\
one-sided outliers$^\dagger$ & 0.0119 (0.111) & 0.0082 (0.145) & \textbf{0.0017} (0.058) & 0.0218 (0.147) & 0.0027 (0.193)\\
S$\alpha$S($1.3$) noise & 0.0100 (0.153) & 0.0100 (0.118) & \textbf{0.0039} (0.073) & 0.0108 (0.113) & 0.0059 (0.096)\\
$t_2$ noise$^\dagger$ & 0.0073 (0.073) & 0.0073 (0.073) & \textbf{0.0055} (0.069) & 0.0061 (0.058) & 0.0070 (0.069)\\
Branin, S$\alpha$S($1.3$) & 0.0468 (0.400) & 0.0472 (0.412) & 0.0492 (0.203) & 0.0433 (0.187) & \textbf{0.0396} (0.279)\\
Gardner~1, S$\alpha$S($1.3$) & 0.0564 (1.096) & 0.0641 (1.188) & \textbf{0.0488} (1.273) & 0.0873 (1.023) & 0.0575 (1.129)\\
Gardner~2, S$\alpha$S($1.3$) & 0.0085 (1.699) & 0.0086 (1.723) & 0.0091 (1.128) & 0.0098 (2.245) & \textbf{0.0071} (0.627)\\
Hartmann-6, S$\alpha$S($1.3$) & 0.68 (6.60) & 0.75 (6.70) & 0.64 (5.55) & 0.93 (5.35) & \textbf{0.62} (3.73)\\\bottomrule
% --- end of table body ---

\end{tabular*}

\medskip
\begin{tabular*}{\linewidth}{@{\extracolsep{\fill}}lcccc@{}}\toprule
\multicolumn{5}{@{}l}{(b) GP + S$\alpha$S noise against each surrogate: wins/losses, Holm-adjusted $p$; bold: significant}\\
problem & vs GP & vs TP & vs GP-$t$ & vs RCGP-type\\\midrule
Gaussian noise & 42/42, 1 & 39/46, 1 & 52/35, 1 & 37/55, 1\\
15\% gross outliers & $\mathbf{81/16}$, $2\cdot10^{-9}$ & $\mathbf{74/23}$, $2\cdot10^{-7}$ & 44/45, 1 & $\mathbf{75/21}$, $4\cdot10^{-7}$\\
one-sided outliers$^\dagger$ & $\mathbf{66/30}$, $2\cdot10^{-4}$ & $\mathbf{61/36}$, $5\cdot10^{-3}$$^\ddagger$ & 36/59, 1 & $\mathbf{69/27}$, $8\cdot10^{-7}$\\
S$\alpha$S($1.3$) noise & 53/42, 0.74 & 56/39, 0.50 & 43/48, 1 & 58/38, 1\\
$t_2$ noise$^\dagger$ & 47/47, 1 & 46/47, 1 & 40/53, 1 & 43/54, 1\\
Branin, S$\alpha$S($1.3$) & 28/22, 1 & 29/21, 1 & 28/22, 1 & 31/19, 1\\
Gardner~1, S$\alpha$S($1.3$) & 22/28, 1 & 24/26, 1 & 20/30, 1 & 30/20, 1\\
Gardner~2, S$\alpha$S($1.3$) & 24/14, 0.71 & 27/13, 0.25 & 23/10, 0.49 & 30/14, 0.11\\
Hartmann-6, S$\alpha$S($1.3$) & 35/15, 0.12 & 35/15, 0.12 & 29/20, 1 & $\mathbf{37/13}$, $5\cdot10^{-3}$\\\bottomrule
\end{tabular*}

\caption{Repeated BO study. $^\dagger$Misspecified for the S$\alpha$S surrogate. (a) Final simple regret over 100 paired repetitions (one-dimensional problems) or 50 (the others); Branin, Gardner~1, Gardner~2 and Hartmann-6 on the standardized scale ($f^\ast=1.05$, $2.32$, $1.47$ and $8.00$). (b) Paired comparisons by repetition, ties omitted; Wilcoxon signed-rank $p$-values Holm-adjusted over all 36 comparisons; the exact sign test agrees on significance in 35 of the 36 (Supplementary Table~\ref{tab:signtest}). $^\ddagger$The one discordant comparison: not significant under the sign test (Holm $p=0.40$).}\label{tab:bostudy}
\end{table}

\begin{table}[!htb]\centering\renewcommand{\baselinestretch}{1}\scriptsize
\setlength{\tabcolsep}{2.5pt}
\begin{tabular}{@{}lcccccc@{}}\toprule
\multicolumn{7}{@{}l}{Tail estimates after the last step (shares of runs) and safeguard statistics (shares of all S$\alpha$S fits)}\\
problem & \begin{tabular}[b]{@{}c@{}}TP: $\hat\nu$\\at bound\end{tabular} & \begin{tabular}[b]{@{}c@{}}TP regret\\$\equiv$ GP\end{tabular} & \begin{tabular}[b]{@{}c@{}}GP-$t$:\\$\hat\nu<30$\end{tabular} & \begin{tabular}[b]{@{}c@{}}S$\alpha$S: median $\hat\alpha$\\(share $\approx2$)\end{tabular} & refits & \begin{tabular}[b]{@{}c@{}}Laplace\\gap $>5$\end{tabular}\\\midrule
Gaussian noise & 1.00 & 90/100 & 0.11 & 2.00 (0.89) & 0.12 & 0.11\\
15\% gross outliers & 1.00 & 60/100 & 0.98 & 1.09 (0.01) & 0.34 & 0.31\\
one-sided outliers$^\dagger$ & 1.00 & 60/100 & 0.97 & 1.12 (0.03) & 0.33 & 0.29\\
S$\alpha$S($1.3$) noise & 1.00 & 80/100 & 0.82 & 1.54 (0.19) & 0.29 & 0.26\\
$t_2$ noise$^\dagger$ & 1.00 & 83/100 & 0.64 & 1.76 (0.42) & 0.20 & 0.19\\
Branin, S$\alpha$S($1.3$) & 1.00 & 13/50 & 0.80 & 1.35 (0.20) & 0.24 & 0.14\\
Gardner~1, S$\alpha$S($1.3$) & 1.00 & 25/50 & 0.66 & 1.79 (0.32) & 0.11 & 0.05\\
Gardner~2, S$\alpha$S($1.3$) & 1.00 & 36/50 & 0.84 & 1.47 (0.10) & 0.18 & 0.09\\
Hartmann-6, S$\alpha$S($1.3$) & 1.00 & 22/50 & 0.46 & 1.90 (0.36) & 0.10 & 0.03\\\bottomrule
\end{tabular}
\caption{Repeated BO study: tail estimates and safeguard statistics. $^\dagger$Misspecified for the S$\alpha$S surrogate. ``At bound'': $\hat\nu>990$; ``TP regret $\equiv$ GP'': identical regret curves over the whole run; ``share $\approx2$'': $\hat\alpha>1.975$; ``refits'': the IS check triggered a refit; ``Laplace gap $>5$'': Laplace exceeded the IS value by more than 5 nats at the Laplace optimum.}\label{tab:bosafe}
\end{table}

\begin{figure}[!htb]\centering\includegraphics[width=\textwidth]{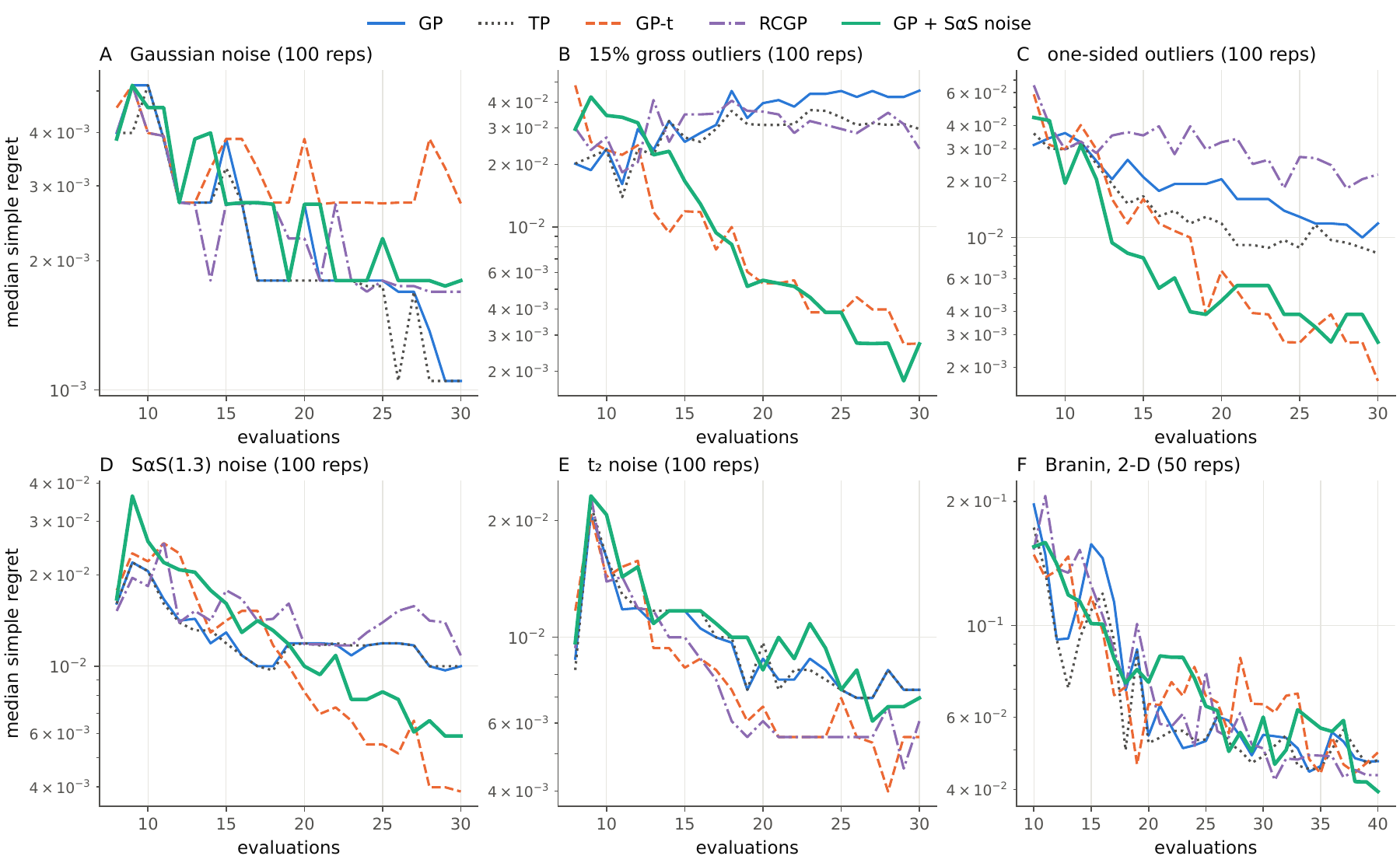}
\caption{Median simple regret of the recommendation against the number of evaluations (log scale), for six of the nine problems; Branin is on the standardized scale. The curves for Gardner~1, Gardner~2 and Hartmann-6 are in Supplementary Figure~\ref{fig:bostudysupp}.}\label{fig:bostudy}\par\smallskip{\small\noindent Alt text: Six panels of median simple regret on a log scale against the number of evaluations for five surrogates. Under Gaussian noise the curves are close, with the Gaussian process with Student-t noise ending somewhat higher. Under symmetric and one-sided gross outliers the surrogates with heavy-tailed noise, the Gaussian process with symmetric alpha-stable noise and the Gaussian process with Student-t noise, reach much lower regret than the Gaussian process, Student-t process and robust conjugate baseline. Under stable noise with index 1.3, t-distributed noise with two degrees of freedom and on Branin, the differences are smaller.\par}\end{figure}

\textbf{Findings} (Tables~\ref{tab:bostudy} and~\ref{tab:bosafe}).
\begin{enumerate}[label=(\alph*),leftmargin=2.2em]
\item \emph{The TP is the GP.} Its $\hat\nu$ was at the bound after all 700 runs, as Corollary~\ref{cor:tp} predicts. Its regret curve was identical to the GP's in 469 of the 700 runs, including 90 of the 100 Gaussian runs. The remaining differences come from the fixed EI inflation of Corollary~\ref{cor:acq}(c), not from learning about the tails; in $d>1$ they also propagate through the model-dependent candidate sets. (No TP--GP test was pre-specified.)
\item \emph{Gross outliers: observation-level tails help, but the GP-$t$ does as well.} With symmetric outliers the GP's regret \emph{grows} as it chases contaminated points (Figure~\ref{fig:bostudy}). The S$\alpha$S surrogate reduced the median final regret from $0.0455$ to $0.0027$ (81 wins, 16 losses; median paired difference $-0.030$, 95\% interval $[-0.045,-0.016]$) and was also significantly better than the TP and the RCGP-type GP. With one-sided outliers the pattern was the same (median $0.0027$ against $0.0119$ for the GP), except that the advantage over the TP is not significant under the sign test (Supplementary Table~\ref{tab:signtest}). In neither problem was it significantly different from the GP-$t$, which had the same or a smaller median.
\item \emph{The tail risk is not uniformly better.} On the outlier problem the S$\alpha$S surrogate's $\mathrm{CVaR}_{0.9}$ was close to the GP's ($0.107$ against $0.117$), despite a 17-fold smaller median. On one-sided outliers it was the worst of the five ($0.193$; GP-$t$ $0.058$). Nine of the ten worst S$\alpha$S runs there ended with $\hat\alpha<1.02$, five of them at the lower bound $0.8$; in three, the regret jumped from below $0.01$ to $0.2$--$0.7$ at the last step. For $\alpha<1$ the S$\alpha$S log-likelihood is far from concave, so the posterior mode, and with it the recommendation, can jump between modes. Across the one-dimensional problems, runs ending at $\hat\alpha=0.8$ had a larger median regret than the rest, except under symmetric outliers (equal medians). A lower bound $\alpha\ge1$ for BO is a natural restriction; we have not tested it.
\item \emph{Stable noise: no significant gain at these budgets.} Under the surrogate's own noise model, S$\alpha$S$(1.3)$, it had a smaller median than the GP in one dimension ($0.0059$ against $0.0100$; unadjusted $p=0.034$, Holm $0.74$), on Branin ($0.040$ against $0.047$; Holm $1$), on Gardner~2 ($0.0071$ against $0.0085$; 24 wins, 14 losses, unadjusted $p=0.031$, Holm $0.71$) and on Hartmann-6 ($0.62$ against $0.68$; 35 wins, 15 losses, unadjusted $p=0.0045$, Holm $0.12$). On Gardner~1 it did not ($0.058$ against $0.056$; Holm $1$). None of these comparisons with the GP survives the Holm adjustment. On Hartmann-6 it was significantly better than the RCGP-type GP only, and it had the smallest $\mathrm{CVaR}_{0.9}$ ($3.7$ against $6.6$ for the GP). On Gardner~2 its $\mathrm{CVaR}_{0.9}$ was also the smallest ($0.63$ against $1.70$); the tail there comes from runs that settled in the inferior one of the two feasible regions (final regret about $2.75$), one S$\alpha$S run against three for the GP. In one dimension the GP-$t$ had the smallest median, so the gain need not come from the stable model being correct.
\item \emph{No detectable cost when tails are absent.} On Gaussian noise the S$\alpha$S surrogate returned $\hat\alpha\approx2$ in 89 of 100 runs. Against the GP it had 42 wins, 16 ties and 42 losses. The mean paired difference was $0.0003$ with 90\% interval $[-0.0009,0.0014]$, inside the margin, so equivalence holds at $\delta=0.025$. This pre-specified margin is generous: it is 25 times the GP's median final regret ($0.0010$). Post hoc, the upper end of the 90\% interval, $0.0014$ (about $1.4$ times that median), is a one-sided 95\% upper confidence bound for the mean cost; any margin above $0.0014$ would have given equivalence.
\end{enumerate}
\emph{Further findings of the BO study.}
\begin{enumerate}[label=(\alph*),leftmargin=2.2em]\setcounter{enumi}{5}
\item \emph{Misspecified symmetric heavy tails ($t_2$): no differences.} No comparison was significant.
\item \emph{Constrained problems} (Supplementary Figure~\ref{fig:bostudysupp}). All five surrogates found a feasible point in every run of both Gardner problems; the constraint model, shared by all surrogates, did that work. The S$\alpha$S surrogate estimated $\hat\alpha<1.975$ in 68\% and 90\% of the runs (median $\hat\alpha$ $1.79$ and $1.47$), and no comparison was significant after adjustment.
\item \emph{The RCGP-type baseline.} It improved on the GP under symmetric outliers ($0.0239$ against $0.0455$) but had a larger median than the GP under S$\alpha$S$(1.3)$ noise and one-sided outliers in one dimension, on both Gardner problems and on Hartmann-6. Its weights are centered at a constant prior mean, so they also down-weight the largest genuine observations, which in maximization are the most informative. This result should not be read as a verdict on RCGP with a fitted mean function or on the method of \citet{Ezzerg2026}.
\item \emph{The safeguard matters in BO.} The IS check triggered a refit in 10\%--34\% of S$\alpha$S fits, most often with outliers. Laplace exceeded the IS value by more than 5 nats in 17.4\% of all 18{,}200 fits, the fold failure of Section~\ref{sec:fold} arising within the BO loop.
\end{enumerate}

\section{Concluding remarks}\label{sec:disc}
In this article, we showed that heavy tails can be put into a GP surrogate at two levels, and that only one of them is learnable from a single run. At the process level the mixing law is identifiable from the distribution of $y$ (Remark~\ref{rem:m1id}), but no estimator can be consistent for it over the model along one realization (Proposition~\ref{prop:noconsist}). The likelihood depends on the data only through the GP's quantities (Theorem~\ref{thm:factor}), and the Gaussian has the largest profile likelihood among all mixing laws (Theorem~\ref{thm:gauss}). The predictive is a data-free transformation of the GP predictive, and the standard PI, UCB and EI acquisitions inherit corresponding fixed transformations (Theorem~\ref{thm:pred}, Corollary~\ref{cor:acq}). At the observation level the tail index is identifiable and estimable from one realization (Theorem~\ref{thm:ecf}), and, for one-dimensional inputs under fixed-domain asymptotics with compact hyperparameter sets and $\alpha_0<2$, it can be estimated as efficiently as if the regression function were known (Theorem~\ref{thm:eff}), including by the global-mode Laplace marginal-likelihood estimate (Theorem~\ref{thm:lapeff}). The heavy tails then act where they should: they down-weight individual observations that the smooth signal cannot explain.

\paragraph{Scope of the process-level result.} It relies on a free amplitude and on a single realization. With a known amplitude, Proposition~\ref{prop:noconsist}(b) still rules out consistent estimation between non-degenerate members of the usual families; with an informative prior on $c$, or several independent realizations, $\theta$ enters the likelihood, but only through the draws of $cV$ (one per realization). Processes that are not scale mixtures fall outside the result. These include stable moving-average and harmonizable processes \citep{SamorodnitskyTaqqu1994}, whose estimability from one path is an interesting open problem.

\paragraph{What the BO study does and does not show.} The repeated study supports observation-level tails when the evaluations are contaminated by gross errors, and it shows no detectable cost on clean data. It does not show an advantage of the stable noise model specifically: the GP-$t$ was never significantly different from the S$\alpha$S surrogate and is cheaper to fit, and under S$\alpha$S noise itself the gains did not survive multiplicity adjustment at budgets of 30--60 evaluations. On the one-dimensional outlier problems the S$\alpha$S surrogate's worst runs came from fits with $\hat\alpha<1$, where the recommendation can jump between modes of the posterior; for BO a restriction $\alpha\ge1$, or acquisition and recommendation rules that account for multimodality, deserve study.

\paragraph{Computation.} For S$\alpha$S noise the Laplace approximation is usually adequate for estimating $\alpha$, and its global-mode version is asymptotically efficient (one-dimensional inputs, compact hyperparameter sets, $\alpha_0<2$). In small samples, however, hyperparameter optimization can be drawn toward folds of the branch of posterior modes that the solver tracks, where the Laplace criterion diverges while the marginal likelihood stays finite (Section~\ref{sec:fold}). A cheap eigenvalue diagnostic detects this. A single IS evaluation at the fitted point is a cheap safeguard. In the BO study it triggered a refit in up to a third of the fits. The variational bound is valid and useful as a certificate, but it favors the GP. Better approximations, such as expectation propagation or variational inference over the mixing variables, are natural next steps.

\paragraph{Open problems.}
\begin{enumerate}[label=(\alph*),leftmargin=2.2em]
\item Efficiency of the IS marginal-likelihood estimate, and of the Laplace estimate when the hyperparameter box grows with $n$; the sharp rate for the GP posterior mode (Sobolev entropy would give $n^{-5/12}$); extensions to inputs in $\mathbb R^d$, via nearest-neighbor differences and pilots.
\item Regret bounds for BO with symmetric $\alpha$-stable noise. Existing heavy-tailed analyses assume $1+\epsilon$ moments \citep{Bubeck2013,Chowdhury2019}, but symmetry allows median-type estimators that need no moments.
\item Whether the mode found by a practical solver (Newton from an IRLS start) is the global mode with probability tending to one, which would extend Theorem~\ref{thm:lapeff} from the global-mode criterion to the computed one.
\item Stein-type discrepancies to test the noise model on BO residuals.
\end{enumerate}

\paragraph{Supplementary material.} The Supplementary Material contains the proofs of the results of Section~\ref{sec:eff} (Section~\ref{app:proofs}), the regularity of the S$\alpha$S log-density (Section~\ref{app:reg}), the analysis of the GP posterior mode, including Theorem~\ref{thm:rough} (Section~\ref{app:gp}), the proofs of Proposition~\ref{prop:infill} and Theorem~\ref{thm:lapeff} (Section~\ref{app:lap}), and additional results for the BO study (Section~\ref{app:bo}). For the code and data, we refer to the Data availability statement below.

\section*{Data availability statement}
The Python code that reproduces all simulations, tables and figures of this article, together with the stored results and a smoke test, is available from the authors and will be deposited in a public GitHub repository archived at Zenodo, with a DOI. During review, the same code is supplied to the referees as a compressed archive. All data used in Sections~\ref{sec:m2} and~\ref{sec:bostudy} are simulated by this code. The companion demonstration of Section~\ref{sec:bo} also uses daily opening and closing values of the NIFTY~50 index (Yahoo Finance, ticker \texttt{\^{}NSEI}, 14 September 2015 to 21 September 2026), which are publicly available from that source.

\section*{Acknowledgements}
This work was supported by the Anusandhan National Research Foundation (ANRF), Government of India, under its ARG-MATRICS scheme (grant ANRF/\allowbreak ARGM/\allowbreak 2025/\allowbreak 000532/\allowbreak MTR). The authors used Claude (Anthropic), a large language model, to assist with writing and testing code, running and summarizing simulations, editing the text, and checking references. The authors checked all code, results, mathematical arguments and references, and take full responsibility for the content of this article.

% --- bibliography (processed .bbl, inlined) ---

% --- end of bibliography ---

\input{supplement-body.tex}
\end{document}

%% file: supplement-body.tex
\clearpage
\begingroup
\setcounter{section}{0}
\setcounter{theorem}{0}
\setcounter{proposition}{0}
\setcounter{lemma}{0}
\setcounter{corollary}{0}
\setcounter{remark}{0}
\setcounter{example}{0}
\setcounter{equation}{0}
\setcounter{footnote}{0}
\setcounter{figure}{0}
\setcounter{table}{0}
\setcounter{page}{1}
\renewcommand{\thesection}{S\arabic{section}}
\renewcommand{\thetheorem}{S\arabic{theorem}}
\renewcommand{\theproposition}{S\arabic{proposition}}\renewcommand{\thelemma}{S\arabic{lemma}}\renewcommand{\thecorollary}{S\arabic{corollary}}\renewcommand{\theremark}{S\arabic{remark}}\renewcommand{\theexample}{S\arabic{example}}
\renewcommand{\theequation}{S\arabic{equation}}
\renewcommand{\thefootnote}{S\arabic{footnote}}
\renewcommand{\thefigure}{S\arabic{figure}}
\renewcommand{\thetable}{S\arabic{table}}
\makeatletter
\def\@title{Supplementary Material for ``Heavy tails in Gaussian-process surrogates:\\ process-level profiling degeneracy versus\\ observation-level identifiability''}
\def\@author{Neelesh S.\ Upadhye and Rohit Kumar Mani\\[2pt]
\normalsize Department of Mathematics, Indian Institute of Technology Madras,\\ \normalsize Chennai 600036, India}
\def\@date{}
\arxivmaketitle
\thispagestyle{plain}
\makeatother
\noindent Theorems, propositions, corollaries, lemmas, remarks, sections, tables and figures without the prefix S refer to the main article; those with the prefix S are in this supplement.

\section{Proofs for Section~\ref{sec:eff}}\label{app:proofs}
In this section, we prove Theorem~\ref{thm:eff}, Lemma~\ref{lem:median} and Lemma~\ref{lem:datat}.

\begin{proof}[Proof of Theorem~\ref{thm:eff}]
Write $\tilde\varepsilon_i=\sigma_0\varepsilon_i=y_i-f(x_i)$ and $r_i=y_i-\hat f_i=\tilde\varepsilon_i-\Delta_i$. Let $B_1=\sup_{\Theta,r}|\partial_r\ell_\theta(r)|$. Let $B_2$ and $B_3$ be the bounds in~\ref{A:reg} for $\partial_r^2 s_\theta$ and $\partial_r\partial_\theta s_\theta$ on $U$.

\emph{Consistency.} Let $M_n(\theta)=n^{-1}\sum_i\ell_\theta(r_i)$, $M_n^0(\theta)=n^{-1}\sum_i\ell_\theta(\tilde\varepsilon_i)$ and $M(\theta)=\E\,\ell_\theta(\tilde\varepsilon_1)$. By the mean-value theorem and Cauchy--Schwarz, $\sup_\Theta|M_n-M_n^0|\le B_1n^{-1}\sum_i|\Delta_i|\le B_1\|\Delta\|_n\to0$ in probability. The map $\theta\mapsto\ell_\theta(r)$ is continuous, $\Theta$ is compact, and the envelope in~\ref{A:reg} is integrable. So $\sup_\Theta|M_n^0-M|\to0$ almost surely by the uniform law of large numbers \citep[Ch.~5]{suppbib-vanderVaart1998}. $M$ has a unique maximizer at $\theta_0$, because the S$\alpha$S family is identifiable in $(\alpha,\sigma)$ and the Kullback--Leibler divergence is positive. The argmax argument \citep[Ch.~5]{suppbib-vanderVaart1998} gives $\hat\theta\to\theta_0$ in probability.

\emph{Expansion.} With probability tending to one, $\hat\theta$ is interior to $U$ and $\Psi_n(\hat\theta)=0$, where $\Psi_n(\theta)=n^{-1}\sum_is_\theta(r_i)$. Shrinking $U$ if necessary, we take it to be a closed ball, so that it is convex and contains the segment from $\theta_0$ to $\hat\theta$. By the fundamental theorem of calculus along that segment, $0=\Psi_n(\theta_0)+\bar J_n(\hat\theta-\theta_0)$, where $\bar J_n=\int_0^1J_n\{\theta_0+u(\hat\theta-\theta_0)\}\,du$ and $J_n(\theta)=n^{-1}\sum_i\partial_\theta s_\theta(r_i)$. (A single intermediate point common to both rows need not exist; the integrated Jacobian avoids it.)

\emph{The score at $\theta_0$.} Write $s=s_{\theta_0}$ and $\dot s=\partial_rs$. By Taylor's theorem in $r$,
\[
\Psi_n(\theta_0)=\frac1n\sum_is(\tilde\varepsilon_i)-\frac1n\sum_i\dot s(\tilde\varepsilon_i)\Delta_i+R_n,\qquad \|R_n\|\le\tfrac12B_2\|\Delta\|_n^2 .
\]
The first term satisfies $\sqrt n\,n^{-1}\sum_is(\tilde\varepsilon_i)\Rightarrow N(0,I(\theta_0))$ by the central limit theorem, since $\E s(\tilde\varepsilon_1)=0$ and $\E\|s\|^2<\infty$.

For the second term, let $F_1,F_2$ be the folds and $\mathcal G_j$ the $\sigma$-field generated by the data of the fold other than $F_j$. Conditionally on $\mathcal G_j$, the $\Delta_i$ with $i\in F_j$ are fixed, and the $\tilde\varepsilon_i$ with $i\in F_j$ are iid and independent of $\mathcal G_j$. Since $p_\alpha$ is symmetric, $\ell_\theta$ is even in $r$, so $s$ is even and $\dot s$ is odd. With $\tilde\varepsilon_i$ symmetric, $\E\dot s(\tilde\varepsilon_i)=0$. Hence $T_j=n^{-1}\sum_{i\in F_j}\dot s(\tilde\varepsilon_i)\Delta_i$ satisfies $\E(T_j\mid\mathcal G_j)=0$ and
\[
\E(\|T_j\|^2\mid\mathcal G_j)=n^{-2}\textstyle\sum_{i\in F_j}\Delta_i^2\,\E\|\dot s(\tilde\varepsilon_1)\|^2\ \le\ n^{-1}\|\Delta\|_n^2\,\sup_r\|\dot s(r)\|^2 .
\]
By the conditional Chebyshev inequality, $\sqrt n\,T_j=O_p(\|\Delta\|_n)=o_p(1)$. Finally, $\sqrt n\|R_n\|\le\tfrac12B_2\sqrt n\|\Delta\|_n^2=o_p(1)$ by~\ref{A:pilot}.

\emph{The Jacobian.} Uniformly on $U$, $\|J_n(\theta)-n^{-1}\sum_i\partial_\theta s_\theta(\tilde\varepsilon_i)\|\le B_3\|\Delta\|_n\to0$ in probability. The second matrix converges to $\E\,\partial_\theta s_\theta(\tilde\varepsilon_1)$ uniformly on $U$ by the uniform law of large numbers, using the envelope in~\ref{A:reg}, and the limit is continuous and equals $-I(\theta_0)$ at $\theta_0$. Since $\hat\theta\to\theta_0$, every point $\theta_0+u(\hat\theta-\theta_0)$, $u\in[0,1]$, lies in a shrinking neighborhood of $\theta_0$, so $\sup_{u\in[0,1]}\|J_n\{\theta_0+u(\hat\theta-\theta_0)\}+I(\theta_0)\|\to0$ and hence $\bar J_n\to-I(\theta_0)$ in probability.

Combining, $\sqrt n(\hat\theta-\theta_0)=I(\theta_0)^{-1}n^{-1/2}\sum_is(\tilde\varepsilon_i)+o_p(1)\Rightarrow N(0,I(\theta_0)^{-1})$. This proves the result.
\end{proof}

\begin{proof}[Proof of Lemma~\ref{lem:median}]
\emph{Bias.} The $k=k_n$ inputs of the other fold that are nearest to $x_i$ in rank lie within $2k+2$ ranks of $x_i$, hence within distance $(2k+2)C_D/n$ by~\ref{A:design}. So $|f(x_j)-f(x_i)|\le\delta_n=L(2k+2)C_D/n$ for each of them. The median is monotone and translation-equivariant, so $|\mathrm{med}_j\{f(x_j)+\tilde\varepsilon_j\}-f(x_i)-\mathrm{med}_j\tilde\varepsilon_j|\le\delta_n$. Hence $\Delta_i^2\le2\delta_n^2+2\mu_i^2$, where $\mu_i$ is the median of $k$ iid copies of $\tilde\varepsilon_1$.

\emph{Second moment of the median.} Let $G(t)=P(0<\tilde\varepsilon_1\le t)$ and $q(t)=P(\tilde\varepsilon_1>t)=\tfrac12-G(t)$. For $t>0$, $\{\mu_i>t\}$ is the event that at least $(k+1)/2$ of the $k$ copies exceed $t$. By Hoeffding's inequality \citep{suppbib-Hoeffding1963}, its probability is at most $\exp\{-2kG(t)^2\}$. It is also at most $\binom{k}{(k+1)/2}q(t)^{(k+1)/2}\le(4q(t))^{k/2}$.

The S$\alpha$S density is continuous and positive, and $q(t)\le C_\alpha t^{-\alpha_0}$. Choose $t_0$ with $4C_\alpha t_0^{-\alpha_0}\le\tfrac12$, and let $g_0>0$ be the minimum of the density of $\tilde\varepsilon_1$ on $[0,t_0]$, so that $G(t)\ge g_0t$ there. By symmetry,
\begin{align*}
\E\mu_i^2=2\int_0^\infty2t\,P(\mu_i>t)\,dt&\le4\int_0^{t_0}t\,\exp(-2kg_0^2t^2)\,dt\\&\qquad+4\int_{t_0}^\infty t\,(4C_\alpha t^{-\alpha_0})^{k/2}\,dt\\
&\le\frac1{kg_0^2}+\frac{4t_0^2\,2^{-k/2}}{\alpha_0k/2-2}
\end{align*}
for $k>4/\alpha_0$. The second term is exponentially small in $k$. Hence $\E\mu_i^2\le C/k$ for $k\ge k_0(\alpha_0)$, and $\E\|\Delta\|_n^2\le2\delta_n^2+2C/k$.

\emph{Rate.} By Markov's inequality, $\|\Delta\|_n=O_p(k_n/n+k_n^{-1/2})$, which is $o_p(n^{-1/4})$ under the stated window conditions. Each $\hat f_i$ uses only the other fold, by construction.
\end{proof}

\begin{proof}[Proof of Lemma~\ref{lem:datat}]
Fix $0<a<c_1/(2s^\ast)$ and $b>2c_2/s^\ast$, and let $\mathcal T=[a,b]$.

\emph{Signal remainder, uniformly in $t$.} With $\bar\phi(t)=m^{-1}\sum_j\cos(t\sigma\eta_j)$,
\[
\cos(tD_j)=\cos(t\sigma\eta_j)\cos(t\Delta_j)-\sin(t\sigma\eta_j)\sin(t\Delta_j).
\]
Using $|\Delta_j|\le Lh_n$, this gives $\sup_{\mathcal T}|\hat\phi-\bar\phi|\le bLh_n+\tfrac12b^2L^2h_n^2$. So $\sqrt m\sup_{\mathcal T}|\hat\phi-\bar\phi|\to0$, because $\sqrt n\,h_n\to0$.

\emph{The scale.} $\big||D_j|-\sigma|\eta_j|\big|\le Lh_n$, and the sample median is 1-Lipschitz in the sup-norm. So $|s_n-\mathrm{median}_j\,\sigma|\eta_j||\le Lh_n$. The sample median of the iid $\sigma|\eta_j|$ converges almost surely to $s^\ast$, because their density is positive at $s^\ast$. Hence $t_k=c_k/s_n\to t_k^\ast$ almost surely, and $t_k\in\mathcal T$ eventually.

\emph{Equicontinuity.} Let $X=\sigma\eta_1$ and $\mathcal F=\{x\mapsto\cos(tx):t\in\mathcal T\}$. If $t,t'\in\mathcal T$ and $|t-t'|\le\delta$, then $|\cos(tx)-\cos(t'x)|\le\min(2,\delta|x|)$. So a $\delta$-grid of $\mathcal T$ yields $O(1/\delta)$ brackets of $L_2(P)$-size at most $2\{\E\min(4,\delta^2X^2)\}^{1/2}$. For $\alpha<2$, since $P(|X|>u)\le Cu^{-\alpha}$,
\[
\E\min(4,\delta^2X^2)=\int_0^4P(|X|>\sqrt u/\delta)\,du\le C\delta^\alpha\int_0^4u^{-\alpha/2}du ,
\]
and for $\alpha=2$ it is at most $\delta^2\E X^2$. So the bracketing numbers grow polynomially in $1/\epsilon$. Hence the bracketing entropy integral
\[
\int_0^1\{\log N_{[\,]}(\varepsilon,\mathcal F,L_2(P))\}^{1/2}\,d\varepsilon
\]
is finite. Since $\mathcal F$ has the bounded envelope $1$, the bracketing Donsker theorem \citep[Theorem~2.5.6, p.~130]{suppbib-vanderVaartWellner1996} implies that $\mathcal F$ is $P$-Donsker. Hence $Z_m(t)=\sqrt m\{\bar\phi(t)-\psi(t)\}$ is asymptotically equicontinuous on $\mathcal T$, and $Z_m(t_k)-Z_m(t_k^\ast)\to0$ in probability.

\emph{Conclusion.} Note that $\log(t_1/t_2)=\log(c_1/c_2)$ does not depend on $s_n$, and that $g\{\psi(c_1/s),\psi(c_2/s)\}=\alpha$ for every $s>0$. Hence
\[
\tilde\alpha_n-\alpha=g\{\hat\phi(t_1),\hat\phi(t_2)\}-g\{\psi(t_1),\psi(t_2)\}.
\]
By the two previous steps, $\sqrt m\{\hat\phi(t_k)-\psi(t_k)\}=Z_m(t_k^\ast)+o_p(1)$, which converges jointly to $N(0,\Sigma^\ast)$, where $\Sigma^\ast$ is $\Sigma$ of Theorem~\ref{thm:ecf} at $t^\ast$. Also, $g$ is continuously differentiable near $(\psi(t_1^\ast),\psi(t_2^\ast))\in(0,1)^2$, and $\psi(t_k)\to\psi(t_k^\ast)$. The delta method gives the limit $N(0,\nabla g^\top\Sigma^\ast\nabla g)$. Finally, $\psi(t_k^\ast)=\exp\{-2(c_k/\mathrm{median}|\eta_1|)^\alpha\}$ does not depend on $\sigma$.
\end{proof}

\section{Regularity of the S\texorpdfstring{$\alpha$}{alpha}S log-density}\label{app:reg}
In this section, we collect the regularity properties of the S$\alpha$S log-density used in the proofs.

Throughout, $p_\alpha$ is the S$\alpha$S density with characteristic function $\exp\{-|t|^\alpha\}$ (Nolan's S0 and S1 parameterizations coincide when $\beta=0$), $L(\alpha,z)=\log p_\alpha(z)$, $c_\alpha=\Gamma(1+\alpha)\sin(\pi\alpha/2)/\pi>0$, and $A\subset(0,2)$ is compact. Constants $C$ may depend on $A$ and on the orders of differentiation, and change from line to line.

\begin{lemma}[Tail expansion]\label{lem:tail}
$(x,\alpha)\mapsto p_\alpha(x)$ is $C^\infty$ on $\R\times(0,2)$, even in $x$, positive, and strictly decreasing in $|x|$. For all integers $j,m\ge0$ there is $C$ such that, for all $\alpha\in A$ and $x\ge2$,
\[
\big|\partial_x^j\partial_\alpha^m\{p_\alpha(x)-c_\alpha x^{-1-\alpha}\}\big|\le C\,x^{-1-2\alpha-j}(\log x)^m .
\]
\end{lemma}
\begin{proof}
\emph{Smoothness, positivity.} $p_\alpha(x)=\pi^{-1}\int_0^\infty\cos(tx)\exp\{-t^\alpha\}dt$. Differentiating $j$ times in $x$ and $m$ times in $\alpha$ under the integral produces integrands bounded by $C t^j(1+t)^{2m}(1+|\log t|)^m\exp(1-t^{a})$, $a=\min A$, uniformly on $A$ (note $\exp\{-t^\alpha\}\le \exp\{1-t^a\}$ for all $t\ge0$), which are integrable; so $p$ is $C^\infty$. Since $\lambda\mapsto \exp\{-\lambda^{\alpha/2}\}$ is the Laplace transform of the positive $(\alpha/2)$-stable law $\mu_\alpha$, $\exp\{-|t|^\alpha\}=\int_0^\infty \exp\{-vt^2\}\mu_\alpha(dv)$, so $p_\alpha(x)=\int_0^\infty(4\pi v)^{-1/2}\exp\{-x^2/(4v)\}\mu_\alpha(dv)$ \citep{suppbib-SamorodnitskyTaqqu1994}: a Gaussian scale mixture, hence even, positive and strictly decreasing in $|x|$.

\emph{Contour rotation.} Fix $x>0$ and $\varphi=\min\{\pi/2,\pi/(2\max A)\}$, so that $\sin\varphi>0$ and $\alpha\varphi\le\pi/2$ on $A$. The function $F(t)=\exp(ixt-t^\alpha)$ (principal branch) is analytic in the sector $0<\arg t<\varphi$. On the arc $t=R\exp(i\omega)$: for $\omega\le\varphi/2$, $\alpha\omega\le\pi/4$ and $|F|\le \exp\{-R^\alpha/\sqrt2\}$; for $\omega\in[\varphi/2,\varphi]$, $|F|\le \exp\{-xR\sin(\varphi/2)\}$ because $\cos(\alpha\omega)\ge0$. So the arc integral vanishes as $R\to\infty$, $F$ is bounded near $0$, and Cauchy's theorem gives
\begin{align*}
\int_0^\infty \exp(ixt-t^\alpha)\,dt&=\exp(i\varphi)\int_0^\infty \exp\{ixs\exp(i\varphi)-w(\alpha,s)\}\,ds,\\
w(\alpha,s)&=s^\alpha \exp(i\alpha\varphi),\qquad \operatorname{Re}w\ge0 .
\end{align*}

\emph{Expansion.} Let $r_N(w)=\exp\{-w\}-\sum_{k=0}^N(-w)^k/k!$. For $\operatorname{Re}w\ge0$ and $q\le N+1$, $r_N^{(q)}=(-1)^qr_{N-q}$ and the integral form of Taylor's remainder gives $|r_N^{(q)}(w)|\le|w|^{N+1-q}$. Since $\operatorname{Re}\{-i\exp(i\varphi)\}=\sin\varphi>0$, $\int_0^\infty s^{k\alpha}\exp\{ixs\exp(i\varphi)\}ds=\Gamma(k\alpha+1)\{x \exp\{i(\varphi-\pi/2)\}\}^{-k\alpha-1}$. Collecting terms, the factors $\exp\{i\varphi\}$ cancel, and taking real parts,
\begin{multline*}
\pi p_\alpha(x)=\sum_{k=1}^N\frac{(-1)^{k+1}}{k!}\Gamma(k\alpha+1)\sin\frac{k\pi\alpha}2\,x^{-k\alpha-1}+\operatorname{Re}E_N(\alpha,x),\\
E_N=\exp\{i\varphi\}\int_0^\infty \exp\{ixs\exp(i\varphi)\}r_N(w(\alpha,s))\,ds .
\end{multline*}
This is the classical series \citep{suppbib-Zolotarev1986,suppbib-Nolan2020}; the $k=1$ term is $\pi c_\alpha x^{-1-\alpha}$. What we need is control of the remainder and of its derivatives, uniformly in $\alpha$.

\emph{Derivatives of the remainder.} Differentiation in $x$ brings down $\{is\exp(i\varphi)\}^j$. In $\alpha$, $\partial_\alpha^pw=w(\log s+i\varphi)^p$, so by Fa\`a di Bruno's formula $\partial_\alpha^m\{r_N(w)\}$ is a sum of terms $r_N^{(q)}(w)$, $1\le q\le m$, times products of $q$ factors $\partial_\alpha^{p}w$ with $\sum p=m$. Taking $N\ge\max(m,1)$, each term is bounded by $C|w|^{N+1}(|\log s|+2)^m=Cs^{(N+1)\alpha}(|\log s|+2)^m$. Differentiation under the integral is justified by these bounds, and
\begin{align*}
|\partial_x^j\partial_\alpha^mE_N|&\le C\int_0^\infty s^{(N+1)\alpha+j}(|\log s|+2)^m\exp(-xs\sin\varphi)\,ds\\
&\le C\,x^{-(N+1)\alpha-j-1}(\log x)^m\qquad(x\ge2),
\end{align*}
by the substitution $s=v/x$. The terms $k=2,\dots,N$ are smooth in $\alpha$, and their $\partial_x^j\partial_\alpha^m$ derivatives are $O(x^{-k\alpha-1-j}(\log x)^m)$. With $N\ge\max(m,1)$ every term other than $k=1$ is $O(x^{-2\alpha-1-j}(\log x)^m)$, uniformly on $A$.
\end{proof}

\begin{lemma}[The log-density]\label{lem:logtail}
For all $j\ge1$, $m\ge0$: $\sup_{\alpha\in A,z\in\R}(1+|z|)^j|\partial_z^j\partial_\alpha^mL(\alpha,z)|<\infty$. For $m\ge0$, $|\partial_\alpha^mL(\alpha,z)|\le C\{1+\log(1+|z|)\}$, and $\partial_\alpha^mL$ is bounded for $m\ge2$. Moreover $|L(\alpha,z)-\log c_\alpha+(1+\alpha)\log(1+|z|)|\le C$, and, as $|z|\to\infty$, $z\,\partial_zL(\alpha,z)\to-(1+\alpha)$ and $\partial_\alpha L(\alpha,z)+\log|z|\to(\log c_\alpha)'$, uniformly on $A$.
\end{lemma}
\begin{proof}
By evenness take $z\ge2$. Write $p_\alpha(z)=c_\alpha z^{-1-\alpha}\{1+\rho(\alpha,z)\}$. By Leibniz's rule, $\partial_z^j\partial_\alpha^m\rho$ is a sum of products of a derivative of $p_\alpha-c_\alpha z^{-1-\alpha}$, of order $O(z^{-1-2\alpha-j_1}(\log z)^{m_1})$ by Lemma~\ref{lem:tail}, and a derivative of $c_\alpha^{-1}z^{1+\alpha}$, of order $O(z^{1+\alpha-j_2}(\log z)^{m_2})$ since $c_\alpha^{-1}$ is smooth on $(0,2)$. Hence $\partial_z^j\partial_\alpha^m\rho=O(z^{-\alpha-j}(\log z)^m)$. For $z\ge z_0$, $|\rho|\le\frac12$, and Fa\`a di Bruno's formula gives the same bound for $\log(1+\rho)$. Now
\[
L(\alpha,z)=\log c_\alpha-(1+\alpha)\log z+\log\{1+\rho(\alpha,z)\} .
\]
For $j\ge1$, $\partial_z^j\partial_\alpha^m\{-(1+\alpha)\log z\}$ is $O(z^{-j})$ (and zero for $m\ge2$), and $O(z^{-\alpha-j}(\log z)^m)=O(z^{-j})$. For $j=0$ the claims follow in the same way. On $|z|\le z_0$ all derivatives are bounded, since $p$ is smooth and positive (Lemma~\ref{lem:tail}).
\end{proof}

\begin{proof}[Proof of Proposition~\ref{prop:reg}]
Write $\varsigma=\log\sigma$ and $z=r\exp(-\varsigma)$, so $\ell_\theta(r)=-\varsigma+L(\alpha,z)$. For a function $g(\alpha,z)$, $\partial_\varsigma\{g\{\alpha,r\exp(-\varsigma)\}\}=-(z\partial_zg)(\alpha,z)$ and $\partial_r\{g\{\alpha,r\exp(-\varsigma)\}\}=\exp\{-\varsigma\}(\partial_zg)(\alpha,z)$. Hence, by induction, $\partial_r^j\partial_\varsigma^k\partial_\alpha^m\ell_\theta(r)$ is $\exp\{-j\varsigma\}$ times a finite linear combination of the functions $z^a\partial_z^{a+j}\partial_\alpha^mL(\alpha,z)$, $0\le a\le k$, plus the constant $-1$ when $(j,k,m)=(0,1,0)$.

(i) If $j\ge1$, each term satisfies $|z^a\partial_z^{a+j}\partial_\alpha^mL|\le C|z|^a(1+|z|)^{-a-j}\le C(1+|z|)^{-j}$ by Lemma~\ref{lem:logtail}, and $(1+|z|)^{-1}\le C(1+|r|)^{-1}$ because $\sigma$ ranges over a compact subset of $(0,\infty)$.

(ii) If $j=0$, the terms with $a\ge1$ are bounded, and the term $a=0$ is $\partial_\alpha^mL$, which is $O(1+\log(1+|z|))$. Also $|L(\alpha,z)|\le C\{1+\log(1+|z|)\}$.

(iii) The tails of $p_{\alpha}$ (Lemma~\ref{lem:tail}) give $\E\{\log(1+|\sigma\varepsilon|)\}^q<\infty$ for all $q$, so $I(\theta)$ is finite by (ii). For $\theta$ in a compact neighborhood $U$ and $|\beta|\le2$, $|\partial_\theta^\beta p_\theta(r)|\le Cp_\theta(r)\{1+\log(1+|r|)\}^2$ by (ii), and $\sup_Up_\theta(r)\le C(1+|r|)^{-1-\min_U\alpha}$ by Lemma~\ref{lem:logtail}. So $\int p_\theta=1$ can be differentiated twice under the integral, which gives the information identities; continuity of $I$ follows by dominated convergence. If $v^\top I(\theta)v=0$ then $v_1\partial_\alpha\ell_\theta(r)+v_2\partial_\varsigma\ell_\theta(r)=0$ for almost every, hence (by continuity) every, $r$. As $|r|\to\infty$, $\partial_\varsigma\ell_\theta(r)=-1-z\partial_zL\to\alpha\ne0$ while $\partial_\alpha\ell_\theta(r)=\partial_\alpha L\to-\infty$ (Lemma~\ref{lem:logtail}). So $v_1=0$, and then $v_2=0$.

Finally, \ref{A:reg} asks for a bounded $\partial_r\ell_\theta$ on $\Theta$ and bounded $\partial_r^j\partial_\theta^\beta\ell_\theta$ for $j\in\{1,2\}$ on $U$, given by (i); for an integrable envelope of $\ell_\theta$ and square-integrable envelopes of its $\theta$-derivatives up to order~2, given by (ii); and for the identities and nonsingularity, given by (iii).
\end{proof}

\section{The GP posterior mode: proof of Theorem~\ref{thm:gppilot}}\label{app:gp}
In this section, we study the posterior mode of the GP with S$\alpha$S noise and prove Theorem~\ref{thm:gppilot}, Corollary~\ref{cor:gppilot} and Theorem~\ref{thm:rough}.

Write $\tilde\varepsilon_i=\sigma_0\varepsilon_i$, $\rho_\theta=-\ell_\theta$, $h=g-f$, $h_i=h(x_i)$, and, for $i\in J$,
\begin{gather*}
\phi_{\theta,i}(t)=\rho_\theta(\tilde\varepsilon_i-t)-\rho_\theta(\tilde\varepsilon_i),\qquad D_\theta(t)=\E\,\phi_{\theta,i}(t),\\
V_\theta(h)=\frac1N\sum_{i\in J}\{\phi_{\theta,i}(h_i)-D_\theta(h_i)\},
\end{gather*}
$Q(h)=N^{-1}\sum_{i\in J}\min(h_i^2,1)$, and $\|h\|_N^2=N^{-1}\sum_{i\in J}h_i^2$. By Proposition~\ref{prop:reg} applied to the compact $\Theta'$, there are constants with $|\rho_\theta'|\le L_\rho$, $|\rho_\theta''|\le R_2$, $|\rho_\theta'''|\le R_3$ and $|\partial_\theta\rho_\theta'|\le R_4$, and (Lemma~\ref{lem:logtail}) $|\rho_\theta(x)-(1+\alpha)\log(1+|x|)|\le C$. Since $|\log(1+|e-t|)-\log(1+|e|)|\le\log(1+|t|)$, this gives, for all $e$, $t$ and $\theta\in\Theta'$,
\begin{equation}\label{eq:Phi}
|\phi_{\theta,i}(t)|\le\Phi(t):=C_1\min(|t|,1)\{1+\log(1+|t|)\}.
\end{equation}

\begin{lemma}[Kernel facts]\label{lem:rkhs}
Uniformly over $\kappa\in\mathcal K$: (a) $c_0\|g\|_{H^3}\le\|g\|_{\mathcal H_\kappa}\le C_0\|g\|_{H^3}$; (b) $\|g\|_\infty+\|g'\|_\infty+\|g''\|_\infty+\sup_{x\ne x'}|g''(x)-g''(x')|/|x-x'|^{1/2}\le C\|g\|_{H^3}$; (c) $\log N(\epsilon,\{\|g\|_{H^3}\le R\},\|\cdot\|_\infty)\le A_0(R/\epsilon)^{2/5}$; (d) $k_\kappa(0)-k_\kappa(u)\le C u^2$.
\end{lemma}
\begin{proof}
(a) The Mat\'ern-$5/2$ spectral density is a multiple of $s^2\ell(1+\ell^2\omega^2/5)^{-3}$, which lies between two multiples of $(1+\omega^2)^{-3}$ uniformly on $\mathcal K$. So the RKHS on $\R$ is $H^3(\R)$ with uniformly equivalent norms, and the RKHS on $[0,T]$ carries the restriction (minimal-extension) norm, which is equivalent to the $H^3[0,T]$ norm by the extension theorem \citep[Ch.~10]{suppbib-Wendland2005}. (b) is the Sobolev embedding on an interval, with $|g''(x)-g''(x')|\le|x-x'|^{1/2}\|g'''\|_{L_2}$ by Cauchy--Schwarz. (c) follows from (b) and the entropy bound for H\"older balls of smoothness $5/2$ on an interval \citep[Theorem~2.7.1]{suppbib-vanderVaartWellner1996}. (d) holds because $k_\kappa$ is twice differentiable at $0$.
\end{proof}

\begin{lemma}[Excess risk]\label{lem:curv}
There is $\kappa_0>0$ such that $\kappa_0\min(t^2,1)\le D_\theta(t)\le\tfrac12R_2t^2$ for all $t\in\R$ and $\theta\in\Theta'$.
\end{lemma}
\begin{proof}
$D_\theta$ is finite and continuous in $(t,\theta)$ by~\eqref{eq:Phi}. Since $\rho_\theta$ is even with $\rho_\theta'>0$ on $(0,\infty)$ (Lemma~\ref{lem:tail}), $\rho_\theta(x)=\rho_\theta(0)+\int_0^\infty\rho_\theta'(v)\mathbf 1\{v<|x|\}dv$, and Fubini's theorem gives
\[
D_\theta(t)=\int_0^\infty\rho_\theta'(v)\{P(|\tilde\varepsilon_1|\le v)-P(|\tilde\varepsilon_1-t|\le v)\}dv .
\]
Let $p$ be the density of $\tilde\varepsilon_1$ and $G_v(t)=\int_{-v}^vp(x+t)dx$. Then $G_v'(t)=p(v+t)-p(v-t)<0$ for $t>0$, because $p$ is even and strictly decreasing in $|x|$. So the bracket is positive for $t\ne0$, $v>0$, and $D_\theta(t)>0$ for $t\ne0$. Next, $D_\theta'(0)=\E\rho_\theta'(\tilde\varepsilon_1)=0$ by symmetry, and $D_\theta''(t)=\E\rho_\theta''(\tilde\varepsilon_1-t)$ is continuous in $(t,\theta)$ with $|D_\theta''|\le R_2$; this gives the upper bound. Integrating by parts, $D_\theta''(0)=\int_\R\rho_\theta'(x)\{-p'(x)\}dx>0$, since both factors are odd and positive on $(0,\infty)$. By compactness of $\Theta'$ and continuity there are $t_1,d_0>0$ with $D_\theta''\ge d_0$ on $[-t_1,t_1]\times\Theta'$, so $D_\theta(t)\ge\frac12d_0t^2$ there. Finally $\inf\{D_\theta(t):|t|\ge t_1,\theta\in\Theta'\}>0$: $D$ is continuous and positive, and along any sequence $(t_k,\theta_k)$ with $|t_k|\to\infty$ and (after extraction) $\theta_k\to\theta^\ast$, Fatou's lemma gives $\liminf_kD_{\theta_k}(t_k)\ge\int_0^\infty\rho_{\theta^\ast}'(v)P(|\tilde\varepsilon_1|\le v)dv>0$.
\end{proof}

\begin{lemma}[Maximal inequality]\label{lem:max}
For $U\ge U_0:=1+\|f\|_{H^3}$ and $q\in(0,1]$ let $\mathcal G(U,q)=\{g:\|g\|_{H^3}\le U,\ Q(g-f)\le q\}$. Then
\[
\E\sup_{\theta\in\Theta',\,g\in\mathcal G(U,q)}|V_\theta(g-f)|\le C\,N^{-1/2}\log^2(2+U)\,\big\{U^{1/5}q^{2/5}+q^{1/2}\big\}.
\]
\end{lemma}
\begin{proof}
By the symmetrization inequality for independent processes (the proof of Lemma~2.3.1 of \citealp{suppbib-vanderVaartWellner1996} applies verbatim), the left side is at most $2\E\sup|Z|$, with $Z(\theta,h)=N^{-1}\sum_{i\in J}\xi_i\phi_{\theta,i}(h_i)$ and $\xi_i$ iid Rademacher, independent of the data. Condition on the data. By Hoeffding's inequality, $Z$ is sub-Gaussian for the metric $d((\theta,h),(\theta',h'))=N^{-1}\{\sum_{i\in J}(\phi_{\theta,i}(h_i)-\phi_{\theta',i}(h'_i))^2\}^{1/2}$. We have $|\phi_{\theta,i}(t)-\phi_{\theta,i}(t')|\le L_\rho|t-t'|$, and $|\phi_{\theta,i}(t)-\phi_{\theta',i}(t)|\le\min(R_4|t|\,|\theta-\theta'|,2\Phi(t))\le C_2\Phi(t)\min\{1,(1+|t|)|\theta-\theta'|\}$. On $\mathcal G(U,q)$, $\max_i|h_i|\le H_U:=C(U+U_0)$ by Lemma~\ref{lem:rkhs}(b), and by~\eqref{eq:Phi}, $d((\theta,h),(\theta,0))\le\Delta_U:=C_1N^{-1/2}\{1+\log(1+H_U)\}q^{1/2}$. Hence
\[
d((\theta,h),(\theta',h'))\le L_\rho N^{-1/2}\|h-h'\|_\infty+C_2(1+H_U)\Delta_U|\theta-\theta'| ,
\]
and the covering numbers at scale $\eta\le2\Delta_U$ are at most \[\exp\{A_0(2L_\rho U'/(\eta\sqrt N))^{2/5}\}\times\{C(1+H_U)\Delta_U/\eta\}^2,\] with $U'=U+U_0$ (Lemma~\ref{lem:rkhs}(c); translation by $f$ does not change covering numbers). The point $(\theta,0)$ belongs to the class and $Z(\theta,0)=0$. Dudley's entropy integral \citep[Section~2.2]{suppbib-vanderVaartWellner1996} then gives
\begin{multline*}
\E_\xi\sup|Z|\le C\int_0^{2\Delta_U}\Big\{\Big(\frac{U'}{\eta\sqrt N}\Big)^{1/5}+\sqrt{\log\frac{C(1+H_U)\Delta_U}{\eta}}\Big\}d\eta\\
\le C\Big\{\Big(\frac{U'}{\sqrt N}\Big)^{1/5}\Delta_U^{4/5}+\Delta_U\sqrt{\log(2+H_U)}\Big\},
\end{multline*}
which is the claim, since $\log(2+H_U)\le C\log(2+U)$.
\end{proof}

\begin{lemma}[From the empirical norm to the sup norm]\label{lem:interval}
Let $h\in C^1[0,T]$ with $\|h'\|_\infty\le K_1$, let $H=\|h\|_\infty$ and $\tau=\min(H,1)$, and let $J$ have spacings at most $C_J/N$, including those to the endpoints. Then $Q(h)\ge\frac{\tau^2}{4}\{\min(\tau/(2K_1),T)/C_J-1/N\}$.
\end{lemma}
\begin{proof}
Let $|h(x^\ast)|=H$. On an interval of length at least $\min(\tau/(2K_1),T)$ containing $x^\ast$, $|h|\ge H-\tau/2\ge\tau/2$, so $\min(h^2,1)\ge\tau^2/4$ there. The interval contains at least $N\min(\tau/(2K_1),T)/C_J-1$ points of $J$.
\end{proof}

\begin{proof}[Proof of Theorem~\ref{thm:gppilot}]
\emph{Existence.} $\rho_\theta\ge\log\sigma-\log p_\alpha(0)$ is bounded below on $\Theta'$, constants belong to $\mathcal H_\kappa$, and by the representer theorem $Q_\psi$ reduces to a continuous, coercive function of $(g(x_i))_{i\in J}\in\R^N$. So minimizers exist, and a measurable selection $\hat f_\psi$ exists by the measurable selection theorem for minimizers of normal integrands. The suprema below are measurable because the processes are continuous in $(\theta,h)$ on separable classes.

Write $\lambda_N(U)=\log^2N\,\log^4(2+U)$, and let $C$ denote constants that do not depend on $N$ or $M_0$ below.

\emph{Peeling.} Let $U_k=2^kU_0$ ($k\ge0$) and $q_l=2^{-l}$ ($0\le l\le l_N$, with $2^{-l_N}\le N^{-2}$). Apply Lemma~\ref{lem:max} to each $\mathcal G(U_k,q_l)$ and Markov's inequality with thresholds $M_0(k+1)^2(l+1)^2$ times the bound. Outside an event of probability at most $\sum_{k,l}\{M_0(k+1)^2(l+1)^2\}^{-1}\le3/M_0$, every $\theta\in\Theta'$ and every $g$ with $\|g\|_{H^3}=U$ and $Q(g-f)=q$ satisfy
\begin{multline}\label{eq:modulus}
|V_\theta(g-f)|\le CM_0\lambda_N(\bar U)N^{-1/2}\big\{\bar U^{1/5}(q+N^{-2})^{2/5}+(q+N^{-2})^{1/2}\big\},\\ \bar U=\max(U,U_0).
\end{multline}
We also take $\eta\le M_0$ (in the approximate case $\eta=O_p(1)$). We work on this event; all remaining steps are deterministic and hold for every $\psi\in\Psi$ and every minimizer.

\emph{Basic inequality.} Let $\hat h=\hat f_\psi-f$, $\hat U=\|\hat f_\psi\|_{H^3}$ and $\hat Q=Q(\hat h)$. Since $Q_\psi(\hat f_\psi)\le Q_\psi(f)+\eta$ ($\eta=0$ for a minimizer), and $\sum_{i\in J}\{\rho_\theta(\tilde\varepsilon_i-\hat h_i)-\rho_\theta(\tilde\varepsilon_i)\}=\sum_{i\in J}D_\theta(\hat h_i)+NV_\theta(\hat h)$, Lemma~\ref{lem:curv} and Lemma~\ref{lem:rkhs}(a) (with $\|\hat f_\psi-c\|_{\mathcal H_\kappa}\ge c_0(\hat U-M\|1\|_{H^3})$ and $\|f-c\|_{\mathcal H_\kappa}\le C_0U_0+C_0M\|1\|_{H^3}$) give
\begin{equation}\label{eq:basic}
\kappa_0\hat Q+c\,\hat U^2/N\ \le\ C(1+\eta)/N+|V_\theta(\hat h)| .
\end{equation}

\emph{Step 1: a crude bound on $\hat U$.} With $\hat Q\le1$, \eqref{eq:modulus} and~\eqref{eq:basic} give $\hat U^2\le C(1+\eta)+CM_0\lambda_N(\bar U)N^{1/2}\bar U^{1/5}$, so $\hat U\le C(M_0\log^6N)^{5/9}N^{5/18}$ for large $N$.

\emph{Step 2: $\hat Q\to0$.} Dropping the second term on the left of~\eqref{eq:basic} and using $\hat Q+N^{-2}\le2\hat Q$ unless $\hat Q\le N^{-2}$, $\hat Q\le C[N^{-1}(1+\eta)+(M_0\lambda_NN^{-1/2}\hat U^{1/5})^{5/3}+(M_0\lambda_N)^2N^{-1}]$, so by Step~1, $\hat Q\le C_{M_0}N^{-20/27}\log^{c}N$.

\emph{Step 3: the sup norm.} By Lemma~\ref{lem:rkhs}(b), $\|\hat h'\|_\infty\le K_1:=C(\hat U+U_0)\le C_{M_0}N^{5/18}\log^cN$. If $\|\hat h\|_\infty\ge1$, Lemma~\ref{lem:interval} gives $\hat Q\ge\frac14\{1/(2K_1C_J)-1/N\}\ge cN^{-5/18}\log^{-c}N$, which contradicts Step~2 for large $N$. So $\|\hat h\|_\infty<1$, and then Lemma~\ref{lem:interval} with $\tau=\|\hat h\|_\infty$ gives $\|\hat h\|_\infty^3\le8C_JK_1(\hat Q+1/N)\to0$.

\emph{Step 4: the rate.} With $\|\hat h\|_\infty<1$, $\hat Q=\delta^2$ with $\delta=\|\hat h\|_N$, and~\eqref{eq:basic} with~\eqref{eq:modulus} becomes
\[
\kappa_0\delta^2+c\hat U^2/N\le C(1+\eta)/N+CM_0\lambda_NN^{-1/2}\big\{\bar U^{1/5}\delta^{4/5}+\bar U^{1/5}N^{-4/5}+\delta+N^{-1}\big\},
\]
with $\lambda_N\le C\log^6N$ by Step~1. The left side is at most five times the largest of the five terms on the right. If it is the term in $\bar U^{1/5}\delta^{4/5}$, then $\delta^{6/5}\le C\lambda_NN^{-1/2}\bar U^{1/5}$ and $\hat U^{9/5}\le C\lambda_NN^{1/2}\delta^{4/5}$; eliminating $\delta$ gives $\hat U\le C\lambda_NN^{1/10}$ and then $\delta\le C\lambda_NN^{-2/5}$. The other cases give $\delta\le C\lambda_NN^{-1/2}$ and $\hat U\le C\lambda_N$. Finally Step~3 with $\hat Q=\delta^2$ and $K_1\le C\lambda_NN^{1/10}$ gives $\|\hat h\|_\infty^3\le C\lambda_N^3N^{-7/10}$. Since $M_0$ is arbitrary and $N\asymp n$, the three rates follow, uniformly over $\psi$ and over all (approximate) minimizers. This proves the result.
\end{proof}

\begin{proof}[Proof of Corollary~\ref{cor:gppilot}]
Let $J$ be the fold used for $\hat\psi$, $\hat h=\hat f_{\hat\psi}-f$, and $x$ a point of the other fold between consecutive points $x_a<x<x_b$ of $J$, with $x=wx_a+(1-w)x_b$. Then
\[
\hat f(x)-f(x)=w\hat h(x_a)+(1-w)\hat h(x_b)+\{wf(x_a)+(1-w)f(x_b)-f(x)\},
\]
and the last term is at most $(x_b-x_a)^2\|f''\|_\infty/8=O(n^{-2})$, since $f\in H^3\subset C^2$. Beyond the end points the error is at most $|\hat h(x_{\rm end})|+\|f'\|_\infty C_D/n$. With rank-parity folds each point of $J$ is a neighbor of at most two points of the other fold, so $\|\Delta\|_n^2\le C(\|\hat h\|_N^2+n^{-2})$. Theorem~\ref{thm:gppilot}, applied with $\psi=\hat\psi$ (it is uniform over $\Psi$), gives the rate; the same holds with the folds swapped. Each $\hat f_i$ is a function of the other fold's data only. (Evaluating $\hat f_{\hat\psi}$ at $x_i$ directly also works, since $|\hat h(x)-\hat h(x_a)|\le2C_D\|\hat h'\|_\infty/n=O_p(n^{-9/10}\log^6n)$; the sup-norm rate alone would not suffice.)
\end{proof}

\subsection*{Proof of Theorem~\ref{thm:rough}}
\emph{Sample paths.} Let $Z$ be a centered Mat\'ern-$5/2$ process with spectral density $S(\omega)\propto(1+\ell^2\omega^2/5)^{-3}$. Since $\int\omega^4S(\omega)d\omega<\infty$, $Z$ has mean-square derivatives of order two, and $Z''$ is a stationary Gaussian process with spectral density $\omega^4S(\omega)\le C(1+\omega^2)^{-1}$. Hence $\E\{Z''(x)-Z''(x')\}^2=2\int\{1-\cos(\omega h)\}\omega^4S(\omega)d\omega\le C|h|$ for $h=x-x'$, and by Gaussianity $\E|Z''(x)-Z''(x')|^{2p}\le C_p|h|^p$ for every $p$. Kolmogorov's continuity theorem gives a modification $\tilde Z_2$ of $Z''$ that is H\"older of every order below $1/2$. Then $\tilde Z(x)=Z(0)+xZ'(0)+\int_0^x\int_0^s\tilde Z_2(u)\,du\,ds$ is in $C^\beta[0,T]$ for every $\beta<5/2$, and $\tilde Z(x)=Z(x)$ almost surely for each $x$, because the mean-square integrals of $Z''$ agree almost surely with the pathwise integrals of $\tilde Z_2$; so $\tilde Z$ is a modification of $Z$.

\emph{The competitor.} Extend $f$ to $F\in C^\beta(\R)$ with compact support (Whitney's extension theorem, \citealp[Ch.~VI, \S2]{suppbib-Stein1970}, followed by multiplication with a smooth cut-off). Let $K$ be smooth, supported in $[-1,1]$, with $\int K=1$ and $\int yK(y)dy=\int y^2K(y)dy=0$, and let $g_b=F\ast K_b$ on $[0,T]$, $K_b=b^{-1}K(\cdot/b)$, $b\le1$. Taylor's formula to order $k=\lceil\beta\rceil-1\le2$ with H\"older remainder and the vanishing moments give $\|g_b-f\|_\infty\le Cb^\beta$. Since $\int K^{(j)}=0$ for $j\ge1$, $g_b'''(x)=\int\{F^{(k)}(x-y)-F^{(k)}(x)\}K_b^{(3-k)}(y)dy$, so $\|g_b'''\|_\infty\le Cb^{\beta-k}b^{-(3-k)}=Cb^{\beta-3}$. In the same way $\|g_b^{(j)}\|_\infty\le Cb^{\min(0,\beta-j)}\le Cb^{\beta-3}$ for $j\le2$. Hence $\|g_b\|_{H^3}\le Cb^{\beta-3}$.

\emph{Changes to the proof of Theorem~\ref{thm:gppilot}.} (i) Lemma~\ref{lem:max} holds with $U_0=1$ and $H_U=C(U+\|f\|_\infty)$: the class need not contain $h=0$, and Dudley's bound is applied with an arbitrary anchor $(\theta_1,h_1)$ in the class, for which $\E_\xi|Z(\theta_1,h_1)|\le\Delta_U$; this adds a term of the same order as the second. The same applies to Lemma~\ref{lem:se}, with anchor term $Cn^{-1/2}\delta$. (ii) Comparing with $g_b$ instead of $f$, and writing $e=g_b-f$,
\[
\kappa_0\hat Q+c\hat U^2/N\le C(1+\eta)/N+\frac{\|g_b-c\|_{\mathcal H_\kappa}^2}{2N}+\frac1N\sum_iD_\theta(e_i)+|V_\theta(e)|+|V_\theta(\hat h)| .
\]
Here $\|g_b-c\|^2_{\mathcal H_\kappa}\le Cb^{2(\beta-3)}$, $N^{-1}\sum_iD_\theta(e_i)\le\frac12R_2\|e\|_\infty^2\le Cb^{2\beta}$ (Lemma~\ref{lem:curv}), and $\sup_\theta|V_\theta(e)|=O_p(N^{-1/2}b^\beta)$, by Hoeffding's inequality and chaining over $\theta$ alone ($e$ is deterministic), using $|\phi_{\theta,i}(t)|\le L_\rho|t|$ and $|\phi_{\theta,i}(t)-\phi_{\theta',i}(t)|\le R_4|t|\,|\theta-\theta'|$. With $b=N^{-1/6}$ these three terms are $O_p(\varepsilon_N)$, $\varepsilon_N=N^{-\beta/3}$, because $\beta\le3$. (iii) Step~1 gives $\hat U^2\le C+CN\varepsilon_N+CM_0\lambda_NN^{1/2}\bar U^{1/5}$, so $\hat U\le C\lambda_N\max(N^{5/18},N^{(3-\beta)/6})=C\lambda_NN^{5/18}$ since $\beta>3/2$. Step~2 gives $\hat Q\le C_{M_0}\lambda_N^c\max(N^{-20/27},N^{-\beta/3})$. Step~3 uses Lemma~\ref{lem:interval} with $\|\hat h'\|_\infty\le C(\hat U+\|f'\|_\infty)$ ($\beta>1$); the contradiction argument needs $\beta/3>5/18$, and then $\|\hat h\|_\infty\to0$. In Step~4 the right side acquires the term $C\varepsilon_N$; if it dominates, $\delta^2\le C\varepsilon_N$ and $\hat U^2\le CN\varepsilon_N$, and otherwise the earlier cases apply. This gives the first two rates, and Step~3 with $\hat Q=\delta^2$ gives the sup-norm rate.

\emph{Consequences.} In Corollary~\ref{cor:gppilot} the interpolation error of $f$ is at most $C(x_b-x_a)^{\min(\beta,2)}=O(n^{-\min(\beta,2)})$, so $\|\Delta\|_n=O_p(n^{-\min(\beta/6,2/5)}\log^6n)=o_p(n^{-1/4})$ because $\beta>3/2$. Proposition~\ref{prop:infill} needs only $\sup_\psi\|\hat f_\psi-f\|_\infty\to0$. In the proof of Theorem~\ref{thm:lapeff}, take $u_n=n^{\max\{(3-\beta)/6,1/10\}}\log^7n$ and $\delta_n=n^{-\min(\beta/6,2/5)}\log^7n$. Then $\sqrt n\,\delta_n^2\to0$ because $\beta>3/2$, $u_n^{1/5}\delta_n^{4/5}\to0$ because $(3-\beta)/30<4\beta/30$ and $1/50<8/25$, and $u_n^2/n\to0$; the rest of the proof is unchanged. Approximate minimizers are measured against $\inf_gQ_\psi(g)$, because $Q_\psi(f)$ may be infinite; the proof compares with $g_b$ and uses only $Q_\psi(\hat f_\psi)\le Q_\psi(g_b)+\eta$.

\begin{lemma}[From conditional to unconditional statements]\label{lem:uncond}
Let $f$ be a random function independent of the noise, and let $A_n$ and $X_n$ be events and random vectors that are jointly measurable in $(f,\text{noise})$, so that $p_n(f)=P(A_n\mid f)$ and $P(|X_n|>Mr_n\mid f)$ are measurable in $f$ (Fubini). (For the events and suprema over $\psi$ used below, measurability holds because they can be taken over a countable dense subset of $\Psi$, by continuity in $\psi$.)
(i) If $P(A_n\mid f)\to1$ for almost every $f$, then $P(A_n)\to1$.
(ii) If $X_n=O_p(r_n)$ conditionally on $f$ for almost every $f$, i.e.\ $h_M(f)=\limsup_nP(|X_n|>Mr_n\mid f)\to0$ as $M\to\infty$, then $X_n=O_p(r_n)$.
(iii) If $P(X_n\le t\mid f)\to G(t)$ for almost every $f$ and every $t\in\R^d$, with $G$ continuous, then $P(X_n\le t)\to G(t)$ for every $t$.
\end{lemma}
\begin{proof}
(i) and (iii): $P(A_n)=\E\,p_n(f)$, and $0\le p_n\le1$, so bounded convergence applies. (ii): by the reverse Fatou lemma, $\limsup_nP(|X_n|>Mr_n)\le\E\,h_M(f)$, which tends to $0$ as $M\to\infty$ by bounded convergence.
\end{proof}
Applied with $f$ the Mat\'ern-$5/2$ path, (i) gives the statements that hold with probability tending to one (Proposition~\ref{prop:infill}), (ii) the rates of Theorem~\ref{thm:rough} and Corollary~\ref{cor:gppilot}, and (iii) the asymptotic normality of Theorems~\ref{thm:eff} (with the GP pilot) and~\ref{thm:lapeff}, whose limit $N(0,I(\theta_0)^{-1})$ does not depend on $f$.

\section{The Laplace marginal likelihood: proofs of Proposition~\ref{prop:infill} and Theorem~\ref{thm:lapeff}}\label{app:lap}
In this section, we prove Proposition~\ref{prop:infill} and Theorem~\ref{thm:lapeff}.

Now $J=\{1,\dots,n\}$ and $N=n$. We keep the notation of Section~\ref{app:gp}. Let $\bar w_\theta=\E\rho_\theta''(\tilde\varepsilon_1)=D_\theta''(0)$. By the proof of Lemma~\ref{lem:curv}, $w_{\min}=\inf_{\Theta'}\bar w_\theta>0$, and $\bar w_\theta\le R_2$. For $\kappa\in\mathcal K$ and $w>0$ write $K=K_\kappa$ (positive definite, since the inputs are distinct), $P=I+wK$, $A=K^{1/2}P^{-1/2}$, so that $A^2=KP^{-1}$ and $\|A^2\|\le1/w$, and $\mathrm{lev}_i=(KP^{-1})_{ii}$. By the representer theorem, minimizing $Q_\psi$ over $\mathcal H_\kappa$ is equivalent to minimizing $q_\psi(v)=\sum_i\rho_\theta(y_i-v_i)+\frac12(v-c\mathbf1)^\top K^{-1}(v-c\mathbf1)$ over $v\in\R^n$, with $v=(g(x_i))_i$; the Hessian of $q_\psi$ is $K^{-1}+W(v)$ with $W(v)=\mathrm{diag}\{\rho_\theta''(y_i-v_i)\}$.

\begin{lemma}[Posterior variances]\label{lem:lev}
Uniformly over $\kappa\in\mathcal K$ and $w\in[w_{\min},R_2]$: $\max_i\mathrm{lev}_i\le Cn^{-2/3}$, $\operatorname{tr}(KP^{-1})\le Cn^{1/3}$ and $\log|I+wK|\le Cn^{1/3}$.
\end{lemma}
\begin{proof}
Let $Z$ be a centered GP with covariance $k_\kappa$, and $e_j$ iid $N(0,1/w)$ independent of $Z$. Then $\mathrm{lev}_i=(K-K(K+w^{-1}I)^{-1}K)_{ii}$ is the conditional variance of $Z(x_i)$ given $(Z(x_j)+e_j)_j$, which is the smallest mean squared error of any linear predictor. Predicting by the average of $Z(x_j)+e_j$ over the $m$ inputs nearest to $x_i$, which lie within $mC_D/n$ of $x_i$, and using Jensen's inequality and Lemma~\ref{lem:rkhs}(d),
\[
\mathrm{lev}_i\le\frac1m\sum_j\E\{Z(x_i)-Z(x_j)\}^2+\frac1{wm}\le C\Big(\frac mn\Big)^2+\frac1{wm}\qquad(1\le m\le n).
\]
With $m=\lceil n^{2/3}\rceil$ this gives the first two claims. For the third, $\frac{d}{du}\log|I+uK|=\operatorname{tr}\{K(I+uK)^{-1}\}$. For $u\ge1/n$, the display with $m=\lceil n^{2/3}u^{-1/3}\rceil\le n$ gives $\operatorname{tr}\{K(I+uK)^{-1}\}\le Cn^{1/3}u^{-2/3}$; for $u<1/n$ the trace is at most $\operatorname{tr}K\le Cn$. Integrating over $u\in[0,w]$ gives $\log|I+wK|\le C+Cn^{1/3}w^{1/3}$.
\end{proof}

\begin{lemma}[Concentration of the curvature]\label{lem:bern}
Let $D^0_\theta=\mathrm{diag}\{\rho_\theta''(\tilde\varepsilon_i)-\bar w_\theta\}$ and $A_{\theta,\kappa}$ be $A$ with $w=\bar w_\theta$. Then $\sup_{\theta\in\Theta',\kappa\in\mathcal K}\|A_{\theta,\kappa}D^0_\theta A_{\theta,\kappa}\|=O_p(n^{-1/3}\sqrt{\log n})$.
\end{lemma}
\begin{proof}
Fix $(\theta,\kappa)$, and let $a_i=Ae_i$. Then $AD^0A=\sum_iX_i$ with $X_i=D^0_{\theta,i}\,a_ia_i^\top$ independent, $\E X_i=0$, $\|X_i\|\le2R_2\|a_i\|^2=2R_2\,\mathrm{lev}_i\le Cn^{-2/3}$, and $\|\sum_i\E X_i^2\|\le4R_2^2\max_i\mathrm{lev}_i\,\|\sum_ia_ia_i^\top\|=4R_2^2\max_i\mathrm{lev}_i\|A^2\|\le Cn^{-2/3}$, by Lemma~\ref{lem:lev}. The matrix Bernstein inequality \citep[Theorem~1.4]{suppbib-Tropp2012}, applied to $\pm\sum_iX_i$, gives $P(\|AD^0A\|\ge t)\le2n\exp\{-\tfrac12t^2/(Cn^{-2/3}+Cn^{-2/3}t)\}$, which is at most $n^{-30}$ for $t=C_5n^{-1/3}\sqrt{\log n}$ and $C_5$ large. Take a grid of mesh $n^{-6}$ in the compact $\Theta'\times\mathcal K\subset\R^4$, with at most $Cn^{24}$ points; by the union bound the claim holds on the grid. Between grid points: $\|D^0_\theta-D^0_{\theta'}\|\le C|\theta-\theta'|$ because $\partial_\theta\rho_\theta''$ is bounded; $\|A^2-A'^2\|\le Cn(|\theta-\theta'|+|\kappa-\kappa'|)$ by the resolvent identity, since $A^2=w^{-1}\{I-(I+wK)^{-1}\}$ and the entries of $K_\kappa$ are Lipschitz in $\kappa$; and $\|A-A'\|\le\|A^2-A'^2\|^{1/2}$. (To see this: for positive semi-definite $X,Y$, let $\lambda\ge0$ be an eigenvalue of $X^{1/2}-Y^{1/2}$ of largest modulus (else swap $X,Y$), with unit eigenvector $u$. Since $X-Y=(X^{1/2}-Y^{1/2})X^{1/2}+Y^{1/2}(X^{1/2}-Y^{1/2})$, $u^\top(X-Y)u=\lambda(u^\top X^{1/2}u+u^\top Y^{1/2}u)\ge\lambda^2$, because $u^\top X^{1/2}u=\lambda+u^\top Y^{1/2}u\ge\lambda$.) With $\|A\|\le w_{\min}^{-1/2}$ and $\|D^0\|\le2R_2$, the oscillation between neighboring grid points is $O(n^{-5/2})$.
\end{proof}

\begin{proof}[Proof of Proposition~\ref{prop:infill}]
Let $\eta_0=w_{\min}/(4R_3)$ and let $E_n$ be the event on which $\sup\|AD^0A\|\le\frac14$ (Lemma~\ref{lem:bern}) and $\sup_{\psi}\sup_x|\hat f_\psi(x)-f(x)|\le\eta_0/2$ for every minimizer (Theorem~\ref{thm:gppilot}); $P(E_n)\to1$. Let $\mathcal C=\{v\in\R^n:\max_i|v_i-f(x_i)|\le\eta_0\}$, a convex set that does not depend on $\psi$. For $v\in\mathcal C$, write $W(v)=\bar w_\theta I+D^0_\theta+E(v)$ with $|E_i(v)|=|\rho_\theta''(\tilde\varepsilon_i-(v_i-f(x_i)))-\rho_\theta''(\tilde\varepsilon_i)|\le R_3\eta_0$, so $\|AE(v)A\|\le R_3\eta_0\|A^2\|\le\frac14$. With $P=I+\bar w_\theta K$,
\[
I+K^{1/2}W(v)K^{1/2}=P^{1/2}\{I+A(D^0_\theta+E(v))A\}P^{1/2}\succeq\tfrac12P\succeq\tfrac12I .
\]
Hence on $E_n$ the Hessian $K^{-1}+W(v)=K^{-1/2}\{I+K^{1/2}W(v)K^{1/2}\}K^{-1/2}$ is positive definite on $\mathcal C$, and $q_\psi$ is strictly convex there. A global minimizer exists because $q_\psi$ is continuous and coercive ($\rho_\theta$ is bounded below and the RKHS term is quadratic). Every global minimizer lies in the interior of $\mathcal C$, so there is exactly one, and $\lambda_{\min}\ge\frac12$ there. A differentiable function that is strictly convex on the convex set $\mathcal C$ has at most one stationary point in $\mathcal C$, its minimizer over $\mathcal C$; so $q_\psi$ has no other stationary point in $\mathcal C$. Since $q$ is smooth in $(v,\psi)$ and its Hessian is nonsingular at the minimizer, the implicit function theorem gives a $C^1$ branch of critical points through it (one-sided at the boundary of $\Psi$, after extending $q$ smoothly to a neighborhood of $\Psi$). For $\psi'$ near $\psi$ the branch stays in the interior of $\mathcal C$, where the critical point of the strictly convex $q_{\psi'}$ is its unique minimizer over $\mathcal C$, hence its global minimizer. So $\psi\mapsto\hat f_\psi$ is $C^1$. This proves the result.
\end{proof}

\begin{lemma}[The log-determinant]\label{lem:logdet}
On $E_n$, with $W_\psi=W(\hat v_\psi)$, $\sup_\psi|\log|I+K^{1/2}W_\psi K^{1/2}||\le Cn^{1/3}$ and $\sup_\psi\|\partial_\theta\log|I+K^{1/2}W_\psi K^{1/2}|\|\le Cn^{1/3}$.
\end{lemma}
\begin{proof}
Write $D=W_\psi-\bar w_\theta I$, so $|D_i|\le2R_2$ and $\|ADA\|\le\frac12$ on $E_n$. Then $\log|I+K^{1/2}W_\psi K^{1/2}|=\log|P|+\log|I+ADA|$. The first term is $O(n^{1/3})$ by Lemma~\ref{lem:lev}. For the second, $|\log(1+\mu)|\le2|\mu|$ for $|\mu|\le\frac12$, so it is at most $2\|ADA\|_{\rm trace}\le2\sum_i|D_i|\|a_i\|^2\le4R_2\operatorname{tr}(KP^{-1})=O(n^{1/3})$.

For the derivative, let $M=(K^{-1}+W_\psi)^{-1}$. On $E_n$, $K^{-1}+W_\psi\succeq\frac12(K^{-1}+\bar w_\theta I)$, so $0\prec M\preceq2KP^{-1}$: $M_{ii}\le2\,\mathrm{lev}_i\le Cn^{-2/3}$, $\operatorname{tr}M\le Cn^{1/3}$ and $\|M\|\le2/w_{\min}$. Now $\partial_\theta\log|I+KW_\psi|=\operatorname{tr}(M\,\partial_\theta W_\psi)$, and with $r_i=y_i-\hat v_i$, $\partial_\theta W_{\psi,ii}=(\partial_\theta\rho_\theta'')(r_i)-\rho_\theta'''(r_i)\,\partial_\theta\hat v_i$. Differentiating the stationarity condition $K^{-1}(\hat v-c\mathbf1)=\rho_\theta'(r)$ gives $\partial_\theta\hat v=Mb$ with $b_i=(\partial_\theta\rho_\theta')(r_i)$, $|b_i|\le R_4$. Hence
\begin{multline*}
|\operatorname{tr}(M\,\partial_\theta W_\psi)|\le\sup|\partial_\theta\rho''_\theta|\operatorname{tr}M+R_3\max_iM_{ii}\sum_i|(Mb)_i|\\
\le Cn^{1/3}+Cn^{-2/3}\sqrt n\,\|M\|\,\sqrt n\,R_4=O(n^{1/3}).\qedhere
\end{multline*}
\end{proof}

\begin{lemma}[Stochastic equicontinuity]\label{lem:se}
Let $\dot s_\theta=\partial_rs_\theta$. For $u\ge U_0$ and $\delta\in(0,1]$,
\begin{multline*}
\E\sup\Big\{\Big\|\frac1n\sum_{i=1}^n\dot s_\theta(\tilde\varepsilon_i)\,(g-f)(x_i)\Big\|:\theta\in\Theta',\ \|g\|_{H^3}\le u,\ \|g-f\|_n\le\delta\Big\}\\
\le Cn^{-1/2}\big\{u^{1/5}\delta^{4/5}+\delta\sqrt{\log(2+u)}\big\}.
\end{multline*}
\end{lemma}
\begin{proof}
$\dot s_\theta$ is odd and bounded (Proposition~\ref{prop:reg}(i)) and $\tilde\varepsilon_i$ is symmetric, so each summand has mean zero, for every $\theta$; and $|\dot s_\theta(e)h_i-\dot s_{\theta'}(e)h'_i|\le B|h_i-h'_i|+B'|\theta-\theta'||h'_i|$. By Hoeffding's inequality the process is sub-Gaussian for the metric $Cn^{-1/2}(\|h-h'\|_n+\delta|\theta-\theta'|)$ on the class, which contains the points $h=0$ where the process vanishes. Dudley's entropy integral with Lemma~\ref{lem:rkhs}(c) gives the bound, exactly as in the proof of Lemma~\ref{lem:max}, without symmetrization.
\end{proof}

\begin{proof}[Proof of Theorem~\ref{thm:lapeff}]
Let $u_n=n^{1/10}\log^7n$ and $\delta_n=n^{-2/5}\log^7n$. By Theorem~\ref{thm:gppilot} and Proposition~\ref{prop:infill}, with probability tending to one, $E_n$ holds, and every $\psi\in\Psi$ has $\|\hat f_\psi\|_{H^3}\le u_n$ and $\|\Delta_\psi\|_n\le\delta_n$, where $\Delta_\psi=\hat f_\psi-f$ at the inputs. We work on this event. Then $L$ is finite and continuous on the compact $\Theta\times\mathcal K$, so a maximizer exists.

\emph{Consistency.} Uniformly over $\Theta\times\mathcal K$,
\begin{multline*}
\Big|\frac1nL(\theta,\kappa)-\frac1n\sum_i\ell_\theta(\tilde\varepsilon_i)\Big|\le L_\rho\|\Delta_\psi\|_n+\frac{\|\hat f_\psi-c\|_{\mathcal H_\kappa}^2}{2n}+\frac{|\log|I+K^{1/2}W_\psi K^{1/2}||}{2n}\\
\le L_\rho\delta_n+\frac{Cu_n^2}n+Cn^{-2/3}\to0,
\end{multline*}
by Lemma~\ref{lem:logdet}. As in the proof of Theorem~\ref{thm:eff}, $n^{-1}\sum_i\ell_\theta(\tilde\varepsilon_i)\to M(\theta)$ uniformly on $\Theta$, and $M$ has a unique, well-separated maximizer at $\theta_0$. Since $L(\hat\theta,\hat\kappa)\ge L(\theta_0,\hat\kappa)$, $M(\hat\theta)\ge M(\theta_0)-o_p(1)$, whatever $\hat\kappa$ is. Hence $\hat\theta\to\theta_0$ in probability.

\emph{Score.} With probability tending to one $\hat\theta$ is interior to $\Theta$, so $\partial_\theta L(\hat\theta,\hat\kappa)=0$. The first two terms of $L$ are $-q_\psi(\hat v_\psi)$, and $\nabla_vq_\psi(\hat v_\psi)=0$, so by the envelope theorem their $\theta$-derivative is $\sum_is_\theta(y_i-\hat v_{\psi,i})$. With $\hat\psi=(\hat\theta,\hat\kappa,\hat c)$ and $\hat\Delta=\Delta_{\hat\psi}$, Lemma~\ref{lem:logdet} and a second-order Taylor expansion in the residual give
\[
\frac1n\sum_is_{\hat\theta}(\tilde\varepsilon_i)=\frac1n\sum_i\dot s_{\hat\theta}(\tilde\varepsilon_i)\hat\Delta_i+O(\|\hat\Delta\|_n^2)+O(n^{-2/3}).
\]
Here $\sqrt n\|\hat\Delta\|_n^2\le\sqrt n\,\delta_n^2\to0$. The first term on the right is bounded by the supremum in Lemma~\ref{lem:se} with $(u,\delta)=(u_n,\delta_n)$, which is $O_p(n^{-1/2}u_n^{1/5}\delta_n^{4/5}+n^{-1/2}\delta_n\sqrt{\log n})=O_p(n^{-1/2}n^{-3/10}\log^7n)=o_p(n^{-1/2})$. So $n^{-1}\sum_is_{\hat\theta}(\tilde\varepsilon_i)=o_p(n^{-1/2})$.

\emph{Conclusion.} This is the estimating equation of the known-$f$ maximum-likelihood estimator, up to $o_p(n^{-1/2})$. Expanding with the integrated Jacobian, as in the proof of Theorem~\ref{thm:eff}, and using $n^{-1}\sum_i\partial_\theta s_\theta(\tilde\varepsilon_i)\to-I(\theta)$ uniformly on a neighborhood of $\theta_0$ (Proposition~\ref{prop:reg}), gives $\sqrt n(\hat\theta-\theta_0)=I(\theta_0)^{-1}n^{-1/2}\sum_is_{\theta_0}(\tilde\varepsilon_i)+o_p(1)$. This proves the result.

The plug-in estimator $\arg\max_\theta\sum_i\ell_\theta(y_i-\hat f_{\hat\psi}(x_i))$ with an uncross-fitted pilot and any $\hat\psi\in\Psi$ is handled by the same argument, without the log-determinant.
\end{proof}

\section{Additional results for the BO study}\label{app:bo}
In this section, we give additional results for the BO study of Section~\ref{sec:bostudy}. Figure~\ref{fig:bostudysupp} shows the median regret curves for the two constrained problems and Hartmann-6, which are omitted from Figure~\ref{fig:bostudy} for legibility.

\begin{figure}[h]\centering\includegraphics[width=\textwidth]{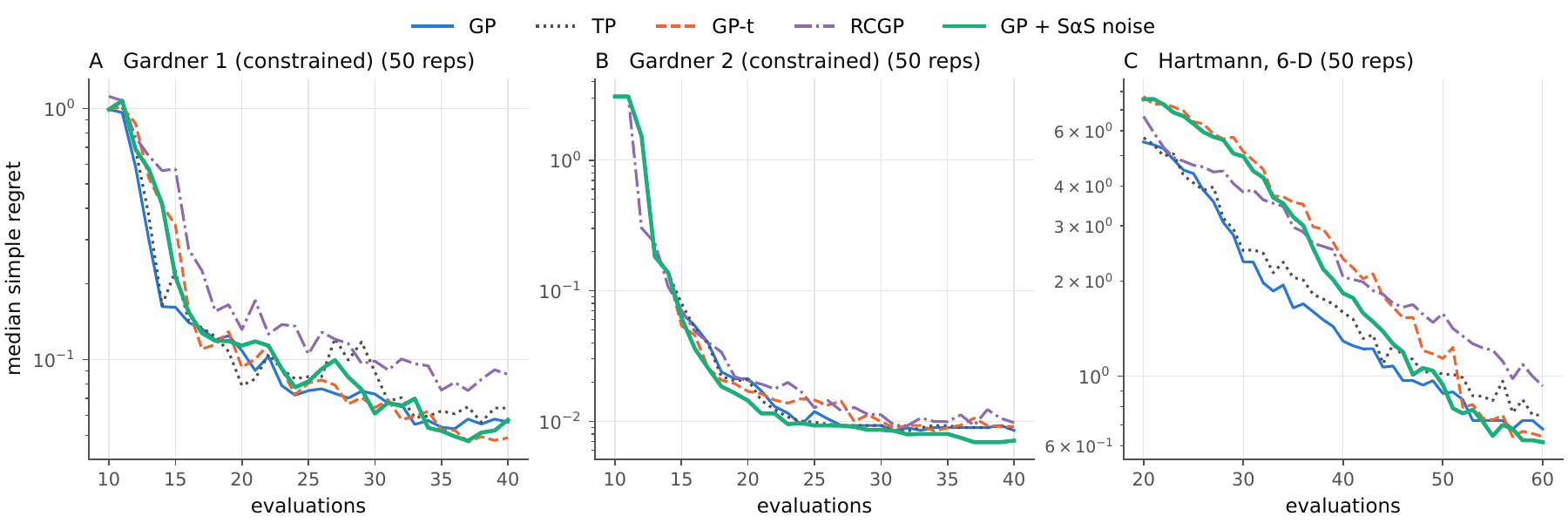}
\caption{Median simple regret against the number of evaluations (log scale) for Gardner~1, Gardner~2 and Hartmann-6, on the standardized scale; for the Gardner problems the recommendation is the best evaluated feasible point.}\label{fig:bostudysupp}\par\smallskip{\small\noindent Alt text: Three panels of median simple regret on a log scale against the number of evaluations for five surrogates on Gardner 1, Gardner 2 and Hartmann-6. On both Gardner problems the curves are close to each other, except that the robust conjugate baseline ends higher on Gardner 1. On Hartmann-6 the Gaussian process is lowest for most of the run, and the robust conjugate baseline ends highest.\par}\end{figure}

\textbf{Sign test.} Final-regret differences are skewed and often tied, and under one-sided contamination they need not be symmetric about their median, which the Wilcoxon signed-rank test assumes. As a robustness check added after the pre-specified analysis, Table~\ref{tab:signtest} repeats every comparison of Table~\ref{tab:bostudy}(b) with the exact two-sided paired sign test on wins against losses (ties dropped), Holm-adjusted over the same 36 comparisons (\texttt{summarize\_bo\_c4.py}). The two tests agree on significance in 35 of the 36 comparisons. The exception is the S$\alpha$S surrogate against the TP under one-sided outliers (61 wins, 36 losses), significant under the signed-rank test (Holm $p=0.005$) but not under the sign test (Holm $p=0.40$).

\begin{table}[h]\centering\renewcommand{\baselinestretch}{1}\scriptsize
\setlength{\tabcolsep}{2pt}
\resizebox{\linewidth}{!}{\begin{tabular}{@{}lcccccccccccc@{}}\toprule
& \multicolumn{3}{c}{vs GP} & \multicolumn{3}{c}{vs TP} & \multicolumn{3}{c}{vs GP-$t$} & \multicolumn{3}{c}{vs RCGP-type}\\\cmidrule(lr){2-4}\cmidrule(lr){5-7}\cmidrule(lr){8-10}\cmidrule(l){11-13}
problem & W/L & $p_{\rm W}$ & $p_{\rm S}$ & W/L & $p_{\rm W}$ & $p_{\rm S}$ & W/L & $p_{\rm W}$ & $p_{\rm S}$ & W/L & $p_{\rm W}$ & $p_{\rm S}$\\\midrule
Gaussian noise & 42/42 & $1$ & $1$ & 39/46 & $1$ & $1$ & 52/35 & $1$ & $1$ & 37/55 & $1$ & $1$\\
15\% gross outliers & 81/16 & $\mathbf{2\cdot10^{-9}}$ & $\mathbf{4\cdot10^{-10}}$ & 74/23 & $\mathbf{2\cdot10^{-7}}$ & $\mathbf{7\cdot10^{-6}}$ & 44/45 & $1$ & $1$ & 75/21 & $\mathbf{4\cdot10^{-7}}$ & $\mathbf{9\cdot10^{-7}}$\\
one-sided outliers & 66/30 & $\mathbf{2\cdot10^{-4}}$ & $\mathbf{1\cdot10^{-2}}$ & 61/36 & $\mathbf{5\cdot10^{-3}}$ & $0.40$ & 36/59 & $1$ & $0.61$ & 69/27 & $\mathbf{8\cdot10^{-7}}$ & $\mathbf{7\cdot10^{-4}}$\\
S$\alpha$S($1.3$) noise & 53/42 & $0.74$ & $1$ & 56/39 & $0.50$ & $1$ & 43/48 & $1$ & $1$ & 58/38 & $1$ & $1$\\
$t_2$ noise & 47/47 & $1$ & $1$ & 46/47 & $1$ & $1$ & 40/53 & $1$ & $1$ & 43/54 & $1$ & $1$\\
Branin & 28/22 & $1$ & $1$ & 29/21 & $1$ & $1$ & 28/22 & $1$ & $1$ & 31/19 & $1$ & $1$\\
Gardner~1 & 22/28 & $1$ & $1$ & 24/26 & $1$ & $1$ & 20/30 & $1$ & $1$ & 30/20 & $1$ & $1$\\
Gardner~2 & 24/14 & $0.71$ & $1$ & 27/13 & $0.25$ & $0.92$ & 23/10 & $0.49$ & $0.88$ & 30/14 & $0.11$ & $0.61$\\
Hartmann-6 & 35/15 & $0.12$ & $0.20$ & 35/15 & $0.12$ & $0.20$ & 29/20 & $1$ & $1$ & 37/13 & $\mathbf{5\cdot10^{-3}}$ & $\mathbf{0.03}$\\\bottomrule
\end{tabular}}
\caption{GP + S$\alpha$S noise against each other surrogate: wins/losses in final regret, and Holm-adjusted $p$-values of the Wilcoxon signed-rank test ($p_{\rm W}$, pre-specified) and of the exact paired sign test ($p_{\rm S}$, robustness check). Bold: Holm-adjusted $p<0.05$.}\label{tab:signtest}
\end{table}

\clearpage
% --- bibliography (processed .bbl, inlined) ---

% --- end of bibliography ---

\clearpage
\endgroup